\documentclass[twocolumn,showpacs,prb,aps,amsmath,amssymb,superscriptaddress]{revtex4}

\usepackage{graphicx}
\usepackage{bm}

\begin{document}

\title{
Vector chiral and multipolar orders in
the spin-$1/2$ frustrated ferromagnetic chain \\
in magnetic field
}
\author{Toshiya Hikihara}
\affiliation{Department of Physics, Hokkaido University,
Sapporo 060-0810, Japan}
\author{Lars Kecke}
\affiliation{Condensed Matter Theory Laboratory, RIKEN,
Wako, Saitama 351-0198, Japan}
\affiliation{Institut f{\"u}r theoretische Physik, 
Universit{\"a}t Ulm, 89069 Ulm, Germany.}
\author{Tsutomu Momoi}
\author{Akira Furusaki}
\affiliation{Condensed Matter Theory Laboratory, RIKEN,
Wako, Saitama 351-0198, Japan}

\date{\today}

\begin{abstract}
We study the one-dimensional spin-$1/2$ Heisenberg chain with competing
ferromagnetic nearest-neighbor $J_1$ and antiferromagnetic
next-nearest-neighbor $J_2$ exchange couplings in the
presence of magnetic field.
We use both numerical approaches (the density matrix renormalization
group method and exact diagonalization) and effective field-theory approach,
and obtain the ground-state phase diagram for wide parameter range of
the coupling ratio $J_1/J_2$.
The phase diagram is rich and has a variety of phases,
including the vector chiral phase, the nematic phase,
and other multipolar phases.
In the vector chiral phase, which appears in relatively weak 
magnetic field,
the ground state exhibits long-range order (LRO) of vector chirality
which spontaneously breaks a parity symmetry.
The nematic phase shows a quasi-LRO of antiferro-nematic spin correlation,
and arises as a result of formation of two-magnon bound states
in high magnetic fields.
Similarly, the higher multipolar phases, such as triatic ($p=3$) 
and quartic ($p=4$) phases,
are formed through binding of $p$ magnons near the saturation fields,
showing quasi-LRO of antiferro-multipolar spin correlations.
The multipolar phases cross over to spin density wave phases
as the magnetic field is decreased, before encountering a phase
transition to the vector chiral phase at a lower field.
The implications of our results to quasi-one-dimensional frustrated magnets
(e.g., LiCuVO$_4$) are discussed.
\end{abstract}

\pacs{
75.10.Jm, 
75.10.Pq,
75.40.Cx
}
\maketitle

\section{Introduction}\label{sec:intro}

There is resurgence of theoretical interest in
the one-dimensional frustrated ferromagnetic Heisenberg model
in magnetic field,\cite{Chubukov1991,CabraHP2000,HeidrichMHV2006,DmitrievK2006,KuzianD2007,KeckeMF2007,VekuaHMH2007,Furukawa2008,Katsura2008}
\begin{equation}
\mathcal{H} = J_1 \sum_l {\bm s}_l \cdot {\bm s}_{l+1}
+ J_2 \sum_l {\bm s}_l \cdot {\bm s}_{l+2}
-h \sum_l s^z_l,
\label{eq:Ham}
\end{equation}
where the nearest-neighbor exchange is ferromagnetic $J_1<0$,
the competing next-nearest-neighbor exchange is antiferromagnetic $J_2>0$,
and ${\bm s}_l$ is a spin-$\frac12$ operator on the site $l$.
The model has recently attracted much attention as it is considered
to describe magnetic properties of quasi-one-dimensional
edge-sharing chain
cuprates, such as
Rb$_2$Cu$_2$Mo$_3$O$_{12}$ (Ref.~\onlinecite{Hase2004})
and LiCuVO$_4$ (Ref.~\onlinecite{Enderle2005}).
In particular, there have been intensive experimental studies of
LiCuVO$_4$,
exploring an unusual
phase transition in magnetic field from a spiral-ordered phase
to a modulated-collinear-ordered phase\cite{Banks2007,Buttgen2007}
and a multiferroic behavior.\cite{Naito2007,Yasui2007,Schrettle2007}

From a theoretical point of view, the $J_1$-$J_2$ spin chain (\ref{eq:Ham})
is of special interest
as it is the simplest of the frustrated quantum spin models
and provides a good testing ground to look
for exotic quantum phases induced by frustration.
The theoretical studies over the past several decades
have mostly considered
the case where both couplings are antiferromagnetic,
$J_1 > 0 $ and $J_2> 0$.
It has been established that 
in zero magnetic field the ground state of 
the antiferromagnetic $J_1$-$J_2$
spin chain undergoes a phase transition from a critical phase 
with gapless excitations for $J_2 < J_{\rm 2c} = 0.2411 J_1$ 
to a gapped phase with spontaneous dimerization for $J_2 > J_{\rm 2c}$ 
as $J_2$ increases.\cite{MajumdarG1969A,MajumdarG1969B,Haldane1982,JullienH1983,OkamotoN1992,Eggert1996,WhiteA1996}
It has also been revealed that the model exhibits cusp singularities 
and a 1/3-plateau in the magnetization curve\cite{OkunishiHA1999,OkunishiT2003}
as well as a vector chiral order in the case of anisotropic exchange
couplings\cite{NersesyanGE1998,KaburagiKH1999,HikiharaKK2001}
or under magnetic
field.\cite{KolezhukV2005,McCulloch2007,Okunishi2008,HikiharaMFK}

In this paper we concentrate on the ferromagnetic case ($J_1<0$) of
the $J_1$-$J_2$ spin chain (\ref{eq:Ham}) in magnetic field
which partially polarizes spins to the $+z$ direction.
We show that the ground-state phase diagram in the case is a zoo of
exotic quantum phases, using 
the numerical density-matrix renormalization
group (DMRG) method,\cite{White1992,White1993,Schollwock2005,Hallberg2006}
exact-diagonalization method,
and effective field theories.
We find a phase with long-range vector chiral order
and phases with various kinds of multipolar spin correlations,
most of which have not been known to appear in this model.

Let us briefly review established results from previous studies
on the ferromagnetic $J_1$-$J_2$ spin chain and introduce
our new findings.

In zero field the ground state is ferromagnetic
for $J_1/J_2<-4$ and spin singlet for $-4<J_1/J_2<0$;
the nature of the spin singlet ground state is not well
understood.
The ground state manifold has extensive degeneracy
at the phase boundary
$J_1/J_2=-4$.\cite{HamadaKNN1988,TonegawaH1989}
In magnetic field
the spins order with a helical magnetic structure
\begin{equation}
\bm{s}_l/s=
(\sin\theta^c\cos\phi_l^c,\sin\theta^c\sin\phi_l^c,\cos\theta^c)
\label{eq:helical}
\end{equation}
in the classical limit ($s=|\bm{s}|\gg1$),
with a pitch angle
\begin{equation}
\phi^c=
\phi_{l+1}^c-\phi_l^c=\pm\arccos(-J_1/4J_2)
\end{equation}
and a canting angle
\begin{equation}
\theta^c=\arccos[4hJ_2/s(J_1+4J_2)^2],
\end{equation}
when $-4<J_1/J_2<0$.
One might expect that this helical magnetic
order should be completely destroyed by
quantum fluctuations in the quantum limit $s=1/2$.
It is important to notice, however,
that a part of the broken
symmetries in the classical helical spin configuration
may remain to be spontaneously broken even in the quantum limit.
Indeed, the chirality (the sign of $\phi^c$) of the helical spin
configuration is $Z_2$-valued and can be broken
in (1+1) dimensions.
The chirality can be measured with the vector chiral order parameter
\begin{equation}
\kappa_l^{(n)}=(\bm{s}_l\times\bm{s}_{l+n})^z
=s_l^xs_{l+n}^y-s_l^ys_{l+n}^x
\label{kappa^n_l}
\end{equation}
with $n=1$ and $2$;
its classical value is $\kappa_l^{(n)}=s^2\sin^2\theta^c\sin(n\phi^c)$.
In this paper we show, for the first time, that
the vector chirality $\kappa_l^{(n)}$ is long-range ordered
in the weak-field regime of the phase diagram of the ferromagnetic
$J_1$-$J_2$ model.\cite{AFcase}
We also show that the vector chiral
order parameters satisfy the relation
\begin{equation}
J_1\langle\kappa_l^{(1)}\rangle
+2J_2\langle\kappa_l^{(2)}\rangle
=0,
\label{eq:bloch relation}
\end{equation}
where $\langle\cdots\rangle$ denotes average in the ground state.
This implies that the spin current, $J_{ij}(\bm{s}_i\times\bm{s}_j)^z$,
flowing on the link connecting the sites $i$ and $j$ ($J_{ij}=J_1$ or $J_2$)
is confined and circulating in each triangle made of
the three neighboring sites.
Incidentally, we note that the classical helical configuration
(\ref{eq:helical}) satisfies Eq.\ (\ref{eq:bloch relation}).

The vector chirality (\ref{kappa^n_l}) is an antisymmetric product
of two spin-$\frac12$ operators.
This is an example of the $\bm{p}$-type nematic operator introduced by
Andreev and Grishchuk.\cite{Andreev1984}
In this paper, we shall reserve the term ``nematic'' for symmetric products
(termed $\bm{n}$-type in Ref.~\onlinecite{Andreev1984})
and call the antisymmetric product (\ref{kappa^n_l})
the vector chirality.
Examples of what we call nematic operators are
\begin{equation}
Q_{x^2-y^2}=s^x_is^x_j-s^y_is^y_j,
\qquad
Q_{xy}=s^x_is^y_j+s^y_is^x_j,
\label{Q}
\end{equation}
which can be thought of as members of quadrupolar spin operators.

Interestingly enough, the phase diagram of the ferromagnetic
$J_1$-$J_2$ spin chain has a Tomonaga-Luttinger (TL) liquid phase
with quasi-long-range antiferro-nematic order
$Q_{--}=Q_{x^2-y^2}-iQ_{xy}=s^-_is^-_j$,
where $s^-_j=s^x_j-is^y_j$: As first pointed out 
by Chubukov,\cite{Chubukov1991}
this nematic order is realized due to pairing of
two magnon excitations.
The paired magnons are the low-energy excitations of the
TL liquid with the nematic quasi-long-range order.
This was confirmed recently by numerical calculation
of nematic correlation function at $J_1/J_2=-1$.\cite{VekuaHMH2007}
In this paper we explore wider region of
the parameter space and show that the nematic TL liquid
phase occupies a large part of the phase diagram.

One can generalize the quadrupolar spin orders to higher
multipolar orders.
For example, one can define ocutupolar triatic order\cite{MomoiSS2006}
$O_{---}=O_{x^3-3xy^2}+iO_{y^3-3x^2y}=s^-_is^-_js^-_k$ with
\begin{subequations}
\begin{eqnarray}
O_{x^3-3xy^2}\!\!&=&\!\!
s^x_is^x_js^x_k
-s^x_is^y_js^y_k-s^y_is^x_js^y_k-s^y_is^y_js^x_k,
\\
O_{y^3-3x^2y}\!\!&=&\!\!
s^y_is^y_js^y_k
-s^y_is^x_js^x_k-s^x_is^y_js^x_k-s^x_is^x_js^y_k,
\quad
\end{eqnarray}
\end{subequations}
and, similarly, the hexadecapolar order
$H_{----}=H_{x^4-6x^2y^2+y^4}-iH_{x^3y-xy^3}=s^-_is^-_js^-_ks^-_l$,
which we dub the ``quartic" order, and so on.
In fact, signatures of binding of three or four magnons are
found in recent numerical studies of magnetization
curves\cite{CabraHP2000,HeidrichMHV2006}
and of multimagnon instabilities at a saturation field.\cite{KeckeMF2007}
In this paper we establish the existence of TL liquid phases with
the triatic and quartic orders through the DMRG calculation of
correlation functions.
It is interesting to note that quasi-long-range molecular
superfluid phases (called trionic and quartetting phases),
similar to the above-mentioned
triatic and quartic phases, have recently
been found in a model of one-dimensional multicomponent
fermionic cold atoms.\cite{RappZHH2007,CapponiRLABW2008,RouxCLA2008}

This paper is organized as follows.
In Sec.\ \ref{sec:phasediagram} we present the ground-state
phase diagram and briefly describe properties of the phases
newly identified in the present work.
These are vector chiral, nematic, incommensurate nematic,
triatic, quartic phases, and spin density wave phases (SDW$_2$ and SDW$_3$).
This section gives a summary of the main results of the paper.
In Sec.\ \ref{sec:multimagnon} we consider formation of multimagnon
bound states which destabilizes the fully polarized state.
The results of this consideration allow us to determine phases
emerging just below the saturation field.
In Sec.\ \ref{sec:magcurve} we study magnetization process
of the model (\ref{eq:Ham})
for several values of the ratio $J_1/J_2$,
and find a transition from a single-spin flip process to a multispin
flip process.
The transition point is identified as the boundary of the vector
chiral phase in the phase diagram.
The remaining sections present detailed analysis of
correlation functions in each phase,
which we calculate using the DMRG method.
In Sec.\ \ref{sec:VC}, we consider the vector chiral phase.
After a brief review of bosonization approach due to Kolezhuk
and Vekua\cite{KolezhukV2005} which is valid for $|J_1| \ll J_2$,
numerical results of the DMRG calculation are presented.
In Sec.\ \ref{sec:Nem} we discuss the nematic phase.
We introduce a hard-core bose gas of two-magnon bound states
as an effective theory for the nematic phase.
We find good agreement between the theory and numerics of various
correlation functions in the nematic phase.
In Sec.\ \ref{sec:incommensurate} we show our numerical results 
for the incommensurate nematic phase, which exhibits 
quasi-long-range order of the nematic correlation with
an incommensurate wave number.
In Sec.\ \ref{sec:TriQua} we apply the hard-core boson theory
to the triatic and quartic phases.
We show that these phases
can be understood as TL liquids of hard-core bosons which
correspond to three- and four-magnon bound states, respectively,
just as the nematic phase is a TL liquid of two-magnon bound states.
We conclude with some remarks in Sec.\ \ref{sec:conclusion}.
The relation (\ref{eq:bloch relation}) is derived
in the Appendix.

\section{Phase diagram}
\label{sec:phasediagram}

\begin{figure}
\begin{center}
\includegraphics[width=85mm]{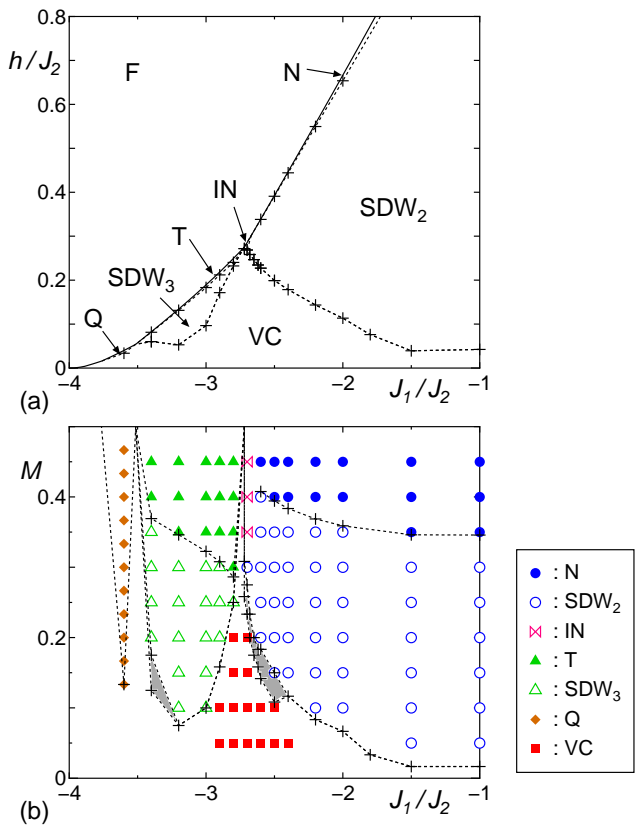}
\caption{
(Color online)
Magnetic phase diagram of the spin-1/2 zigzag chain with ferromagnetic
$J_1$ and antiferromagnetic $J_2$
(a) in the $J_1/J_2$ versus $h/J_2$ plane
and (b) in the $J_1/J_2$ versus $M$ plane.
Crosses show the transition and crossover points
obtained from the magnetization curves and correlation functions.
In (a), symbols ``VC", ``N", ``IN'', ``T", ``Q", and ``F" indicate
the vector chiral ($\Delta S^z_{\rm tot} = 1$),
nematic ($\Delta S^z_{\rm tot} = 2$),
incommensurate nematic ($\Delta S^z_{\rm tot} = 2$),
triatic ($\Delta S^z_{\rm tot} = 3$),
quartic ($\Delta S^z_{\rm tot} = 4$), and
ferromagnetic (fully polarized) phases, respectively.
Here $\Delta S^z_\mathrm{tot}$ is the unit of changes in the total
$S^z_\mathrm{tot}=\sum_l s^z_l$ when the magnetic field $h$ is swept.
There are also two kinds of spin-density wave phases:
SDW$_2$ ($\Delta S^z_{\rm tot} = 2$) and
SDW$_3$ ($\Delta S^z_{\rm tot} = 3$),
which are related to the nematic and triatic phases, respectively.
The solid curve shows the saturation field $h_{\rm s}$
and dotted lines are the guide for the eye.
In (b), symbols indicate parameter points for which
their ground-state phase is identified by analysis of correlation functions.
Shaded regions in (b) correspond to the magnetization jump
at the first-order transition;
see Sec.\ \ref{sec:magcurve}.
}
\label{fig:phasediagram}
\end{center}
\end{figure}

\begin{figure*}
\includegraphics[width=180mm]{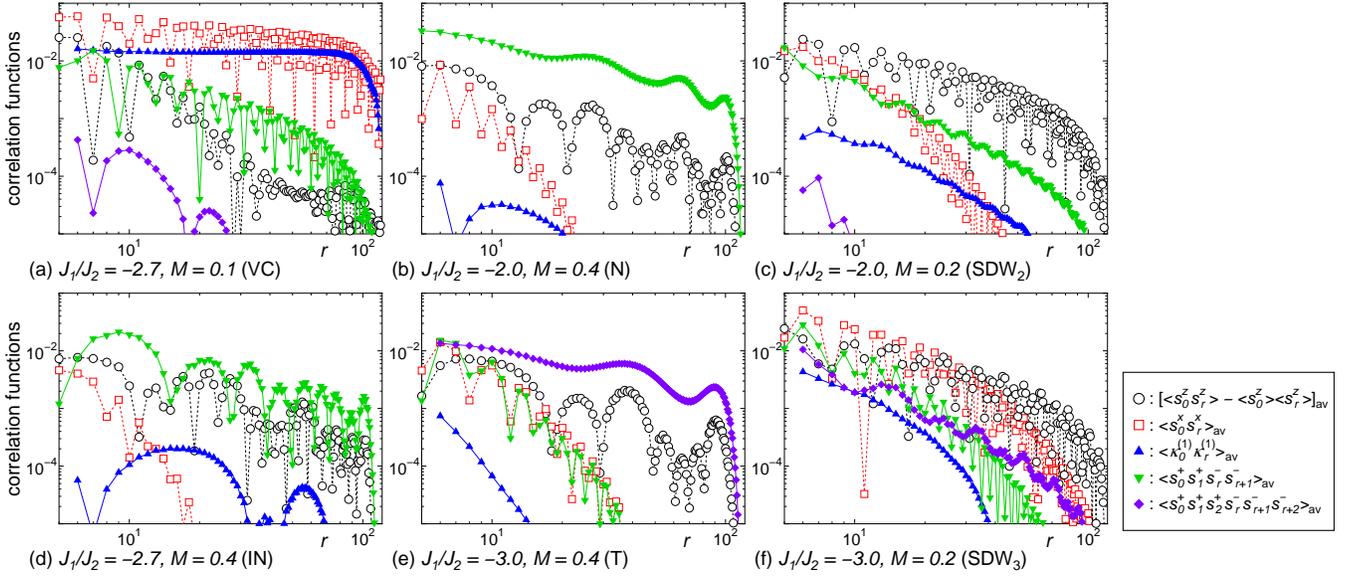}
\caption{
Typical behaviors of various correlation functions in
(a) the vector chiral (VC) phase,
(b) the nematic (N) phase,
(c) the SDW$_2$ phase,
(d) the incommensurate nematic (IN) phase,
(e) the triatic (T) phase, and
(f) the SDW$_3$ phase.
Absolute values of
spatially averaged correlation functions are plotted.
(For the averaging procedure, see Sec.\ \ref{sec:numerics-VC}.)
In (b) and (d), the triatic correlation function
$\langle s^+_l s^+_{l+1} s^+_{l+2} s^-_{l'} s^-_{l'+1} s^-_{l'+2} \rangle$
is smaller than $10^{-5}$.
}
\label{fig:correlators in various phases}
\end{figure*}

The ground-state phase diagram obtained in the present
work is summarized in Fig.\ \ref{fig:phasediagram}
in the planes of (a) $J_1/J_2$ versus $h/J_2$
and (b) $J_1/J_2$ versus the magnetization per site $M$.
The phase diagram contains (at least) eight phases:
vector chiral (VC) phase, nematic (N) phase, incommensurate
nematic (IN) phase, triatic (T) phase, quartic (Q) phase,
two kinds of spin density wave phases (SDW$_2$ and SDW$_3$),
and ferromagnetic (F) phase.
Brief explanation of these phases is given below.
More detailed discussions on each phase will be given
in Secs.~\ref{sec:VC}--\ref{sec:TriQua}.
Figure \ref{fig:correlators in various phases} shows
typical spatial dependence of various correlation functions in
these phases.

\textit{Ferromagnetic phase}:
In the ferromagnetic phase, spins are fully polarized, $M=1/2$.
This phase is stable when $J_1/J_2<-4$ or when large enough magnetic
field is applied.
We will discuss in Sec.~\ref{sec:multimagnon}
magnetic instabilities along the phase boundary of the ferromagnetic
phase.

\begin{figure}[ht]
\begin{center}
\includegraphics[width=60mm]{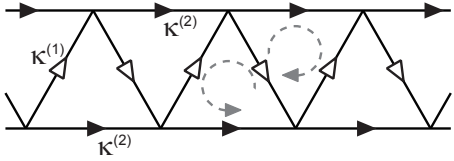}
\caption{
Schematic picture of the vector chiral order.
The arrows on bonds indicate breaking of the parity symmetry
by the vector chiral order
$\kappa_l^{(n)}=(\bm{s}_l\times\bm{s}_{l+n})^z$,
whose expectation values obeys the relation
$J_1\langle\kappa^{(1)}\rangle+2J_2\langle\kappa^{(2)}\rangle=0$.
The circulation of the $s^z$ spin current, shown by the dashed arrows, 
is alternating,
and there is no net spin current flow.
}
\label{fig:VCflow}
\end{center}
\end{figure}

\textit{Vector chiral phase}:
The vector chiral phase appears in small magnetic field.
This phase is characterized by long-range order of
the vector chiral correlation (\ref{kappa^n_l}).
The ground state breaks a $Z_2$ symmetry,
as Eq.\ (\ref{kappa^n_l}) indicates that the
parity about a bond center
is broken spontaneously.
We can also regard this
$Z_2$ symmetry breaking as choosing one of the two possible
directions of circulation of spontaneous $s^z$-spin current flow.
A schematic picture of the vector chiral order and circulating
spin current in the $Z_2$-symmetry
broken state is shown
in Fig.~\ref{fig:VCflow}, where the spin chain is drawn as
a zigzag ladder.
Numerical evidence for the long-range order will be presented in
Sec.~\ref{sec:VC}.
Another important feature of the vector chiral phase is
that the transverse spin correlation $\langle s^x_0 s^x_l\rangle$
is incommensurate with the lattice and stronger than
the longitudinal correlation
$\langle s^z_0 s^z_l\rangle - \langle s^z_0 \rangle \langle s^z_l\rangle$.

\textit{Nematic/SDW$_2$ phases}:
At higher magnetic field up to the saturation field,
the nematic/SDW$_2$
phases\cite{Andreev1984,Chubukov1991,VekuaHMH2007,KeckeMF2007}
appear at $J_1/J_2\gtrsim-2.7$.
These phases are a TL liquid of hard-core bosons which are actually
two-magnon bound states with total momentum $k=\pi$.
The boson creation operator $b^\dagger_l$ corresponds to
$s^-_ls^-_{l+1}$, and the boson density
$n_l=b^\dagger_lb^{}_l\propto\frac12-s^z_l$.
Since breaking a two-magnon bound state costs a finite binding energy,
the transverse spin correlation $\langle s^+_0 s^-_l\rangle$
is short-ranged, where $s^+_0=s^x_0+is^y_0$.
Being a TL liquid, the ground state exhibits power-law decaying
correlations of the single-boson propagator,
$\langle b^{}_0b^\dagger_l\rangle\propto
\langle s^+_0s^+_1s^-_ls^-_{l+1}\rangle$,
and the density fluctuations,
$\langle n_0 n_l\rangle-\langle n_0\rangle\langle n_l\rangle \propto
\langle s^z_0s^z_l\rangle - \langle s^z_0 \rangle \langle s^z_l\rangle$.
When the boson propagator decays slower than the density-density
correlation, it is appropriate to call this phase the (spin) nematic phase.
In the opposite case when the latter incommensurate 
density correlation is
dominant, we call this phase the spin density wave (SDW$_2$) phase.
The SDW$_2$ phase is extended to the antiferromagnetic side $J_1>0$ across
the decoupled-chain limit $J_1=0$; it is called even-odd phase
in Ref.~\onlinecite{OkunishiT2003}.
The boundary between the SDW$_2$ phase and the nematic phase is
shown by a dotted line in Fig.~\ref{fig:phasediagram}.

\begin{figure}
\begin{center}
\includegraphics[width=65mm]{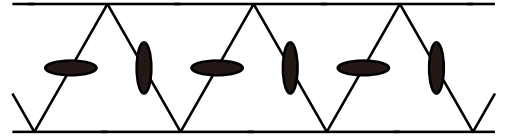}
\caption{
Schematic picture of antiferro-nematic quasi-long-range order
in the nematic phase.
Ellipses represent directors of the nematic order on each bond.
}
\label{fig:nematic}
\end{center}
\end{figure}

In the semiclassical picture we can
write $s^-_l=e^{-i\phi_l}$, where $\phi_l$ is
the angle of the two-dimensional vector $(s^x_l,s^y_l)$
measured from the positive $x$ direction,
$0\le\phi_l<2\pi$.
The product $s^-_ls^-_{l+1}=e^{-i(\phi_l+\phi_{l+1})}$ can
be represented by the vector
$\bm{N}_{l+\frac12}=(\cos\Phi_{l,2},\sin\Phi_{l,2})$ with
$\Phi_{l,2}=-(\phi_l+\phi_{l+1})/2$.
We now realize that we need to identify
$\bm{N}_{l+\frac12}$ with $-\bm{N}_{l+\frac12}$ because
of the physical identification
$(\phi_l,\phi_{l+1})\equiv(\phi_l+2\pi,\phi_{l+1})
\equiv(\phi_l,\phi_{l+1}+2\pi)$.
We can thus consider $\bm{N}_{l+\frac12}$ as a director representing
the nematic order.
We will show in Sec.~\ref{sec:Nem} that the nematic phase
has antiferro-nematic quasi-long-range order of the director, as shown
schematically in Fig.~\ref{fig:nematic}.
The ground state is not dimerized in this phase, as opposed to the initial
proposal of Chubukov.\cite{Chubukov1991}

\begin{figure}
\begin{center}
\includegraphics[width=65mm]{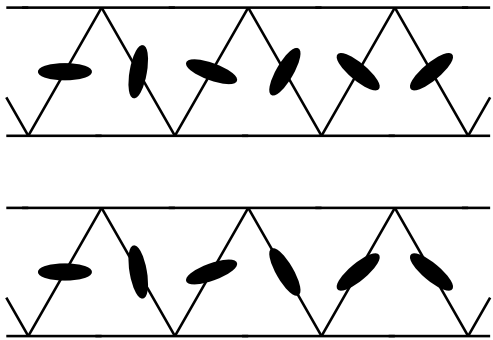}
\caption{
Schematic pictures of incommensurate nematic quasi-long-range order
in the incommensurate \emph{chiral} nematic phase.
Ellipses represent directors of the nematic order on each bond.
The numerical results in Sec.\ \ref{sec:incommensurate}
indicate that this chiral
symmetry is not broken in the incommensurate nematic phase;
the ground state is given by equal superposition of the upper
and lower configurations.
}
\label{fig:ICnematic}
\end{center}
\end{figure}

\textit{Incommensurate nematic phase}:
The incommensurate nematic phase occupies a very small
region in the phase diagram.
This phase has quasi-long-range order of the nematic correlation with
an incommensurate wave number.
The correlation is due to two-magnon bound states with
momentum $k=\pi+\delta$ and
$\pi-\delta$, instead of $k=\pi$ in the nematic phase.
Schematic pictures of the incommensurate nematic order are
depicted in Fig.~\ref{fig:ICnematic}, where the upper and lower
pictures represent the nematic correlation with wave number
$k=\pi+\delta$ and $\pi-\delta$, respectively.
If the densities of paired magnons
with $k=\pi+\delta$ and $\pi-\delta$
are different, one of the two correlation patterns
in Fig.~\ref{fig:ICnematic} becomes dominant,
and the $Z_2$ chiral symmetry is broken spontaneously,
as suggested by Chubukov.\cite{Chubukov1991}
However, we have found
no signature of long-range order of the chiral
correlation in our numerical calculation, which we discuss in
Sec.~\ref{sec:incommensurate}.

\begin{figure}
\begin{center}
\includegraphics[width=65mm]{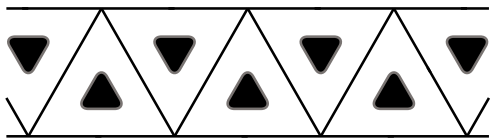}
\caption{
Schematic picture of antiferro-triatic quasi-long-range order in the triatic phase.
Solid triangles represent spin structure of the triatic order formed by
three $s=1/2$ spins on each plaquette.
}
\label{fig:triatic}
\end{center}
\end{figure}

\textit{Triatic and SDW$_3$ phases}:
The triatic phase exists below the saturation field and
next to the incommensurate nematic phase.
The triatic/SDW$_3$ phases are a TL liquid of
bosons which represent three-magnon
bound states with total momentum $k=\pi$.
In analogy with the nematic phase, the triatic order has a simple
semiclassical picture.
Writing the bound three magnons as
$s_l^- s_{l+1}^- s_{l+2}^-=e^{-i(\phi_l+\phi_{l+1}+\phi_{l+2})}
=e^{3i\Phi_{l,3}}$,
we may consider the triatic order as ordering of the angle
$\Phi_{l,3}=-(\phi_l+\phi_{l+1}+\phi_{l+2})/3$,
which has the property $\Phi_{l,3}\equiv\Phi_{l,3}+2\pi/3$.
A schematic picture of the triatic ordered state is shown in
Fig.~\ref{fig:triatic}.
In the triatic phase, correlation functions probing three-magnon
bound states, such as
$\langle s_0^+s_1^+s_2^+s_l^-s_{l+1}^-s_{l+2}^-\rangle$,
exhibit quasi-long-range order (power-law decay).
In contrast, both the transverse spin correlation $\langle s^+_0s^-_l\rangle$
and the nematic correlation $\langle s^+_0s^+_1s^-_ls^-_{l+1}\rangle$
are short-ranged, because of a finite energy cost for breaking
a three-magnon bound state.
The longitudinal spin correlation
$\langle s_0^zs_l^z\rangle-\langle s^z_0\rangle\langle s^z_l\rangle$
shows algebraic decay, as $s^z_l$ is related to the boson density.
When the most slowly decaying correlation is the longitudinal
spin correlation, we call the phase the SDW$_3$ phase.
The boundary between the triatic phase and the SDW$_3$ phase
is shown by a dotted curve in Fig.~\ref{fig:phasediagram}.
The detailed discussion of the correlation functions will be given
in Sec.~\ref{sec:TriQua}.

\textit{Quartic phase}:
The quartic phase is a TL liquid phase of
four-magnon bound states with momentum $k=\pi$.
Its properties can be easily deduced by straightforward generalization
from the triatic phase.

\section{Multimagnon instability}\label{sec:multimagnon}

We begin our study of the phase diagram
by examining instabilities of the fully polarized state.
To that end, we numerically calculate energy dispersion of low-energy
excitations with a small number of magnons (down spins).
The analysis presented here extends the result reported in our previous
study.\cite{KeckeMF2007}

Inside the ferromagnetic phase in magnetic field,
there is a finite energy gap between the
ground state and excited states.
With decreasing the magnetic field, the gap becomes smaller and
eventually vanishes at the boundary of the ferromagnetic phase.
We define the saturation field $h_\mathrm{s}$ as the magnetic field
$h$ at which a branch of excitations first becomes gapless
as the field $h$ is reduced.
In the ferromagnetic $J_1$-$J_2$ spin chain,
the excitation mode that becomes gapless (soft) at $h=h_\mathrm{s}$ is
\textit{a multimagnon bound state}.
Below the saturation field the soft multimagnon bound states
proliferate.
As a result the ground state can change into a TL liquid
with the correlation that is represented by the soft,
bound multimagnon mode.
It is therefore important to find out which branch of multimagnon bound
states is the soft mode.

We calculate energy of $p$-magnon excitations
using the method we introduced
in Ref.\ \onlinecite{KeckeMF2007}.
The number of magnons $p$ and the total momentum $k$
are good quantum numbers of the Hamiltonian (\ref{eq:Ham}).
We thus expand eigenstates in the sector of $p$ magnons with the basis
\begin{equation}
|p, k; \{ r_1,\ldots,r_{p-1} \} \rangle =
\frac{1}{\sqrt{\Omega}}
\sum_{l=1}^{\Omega}
\prod_{n=1}^p e^{ikl_n/p}s^-_{l_n}
|{\rm FM}\rangle,
\label{eq:p-magbasis}
\end{equation}
where
\begin{equation}
l_n=l+\sum_{i=1}^{n-1} r_i.
\end{equation}
Here
$|{\rm FM}\rangle$ is the fully polarized state
($s_i^+|\mathrm{FM}\rangle=0$),
$\Omega$ the system size
taken to be $\Omega \to \infty$,
and $r_i$ $(1 \le i \le p-1)$ is the distance
between the $i$-th and $(i+1)$-th magnons.
The periodic boundary condition is imposed in this calculation.
We take $r_i$ to be in the range of $1 \le r_i \le r_{\rm max}$,
where $r_{\max}$ is chosen so that
wave function vectors can be stored in the computer memory.
(The finite value of $r_{\max}$ limits the accuracy of energy calculations.
For tightly bound magnons, errors caused by this
approximation can be made negligibly small, on the order of
exponentially decaying tails of their wave function.)

\begin{table}
\caption{
Number of magnons, $p$, and total momentum $k$ of the
multimagnon bound states which become gapless
at the saturation field.
}
\label{tab:multimagnon}
\begin{ruledtabular}
\begin{tabular}{ccc}
parameter range & $p$ & $k$ \\
\hline
 $-2.669 < J_1/J_2 < 0$      & 2  & $\pi$  \\
 $-2.720 < J_1/J_2 < -2.669$ & 2  & $\pi\pm\delta$ ($\delta>0$) \\
 $-3.514 < J_1/J_2 < -2.720$ & 3  & $\pi$  \\
 $-3.764 < J_1/J_2 < -3.514$ & 4  & $\pi$  \\
 $-3.888 < J_1/J_2 < -3.764$ & 5  & $\pi$  \\
 $-3.917 < J_1/J_2 < -3.888$ & 6  & $\pi$  \\
 $-4<      J_1/J_2 < -3.917$ & 7  & $\pi$  \\
\end{tabular}
\end{ruledtabular}
\end{table}

We numerically diagonalize the Hamiltonian matrix expressed
in the restricted Hilbert space ($r_i\le r_{\max}$) and
obtain the lowest energy
as a function of the total momentum $k$ for each $p$ magnon sector.
In this way we obtain energy dispersion of $p$-magnon bound states.
In our previous study\cite{KeckeMF2007} we calculated energy dispersion
of multimagnon excitations for up to $p=4$.
Here we extend the calculation to include  more magnons ($p_{\max}=8$),
taking the maximum distance $r_{\max}$ to be at least $42/(p-1)$.
We thereby identify soft multimagnon modes and determine
the saturation field $h_\mathrm{s}$ at each value of the
ratio $J_1/J_2$ ($-4<J_1/J_2<0$).

Table\ \ref{tab:multimagnon} summarizes
the number of magnons, $p$, and
the momentum $k$ of the multimagnon modes ($p\le8$)
which become gapless at the saturation field.
We note that the gapless modes with $p\le4$ in Table \ref{tab:multimagnon}
are soft modes giving rise to multipolar TL liquids,
since non-negative excitation energy of $2p$-magnon modes
indicates a repulsive interaction 
between bound $p$-magnons (see the discussion at the end of this section).
We thus find that, as Chubukov first pointed out,\cite{Chubukov1991}
the two-magnon bound state with $k=\pi$ is the soft mode
when $-2.669<J_1/J_2<0$.
Its exact wave function can be easily obtained;
it turns out that the bound-state wave function at $k=\pi$ has amplitudes
only for odd integer values of $r_1$,
which means that the magnons forming a bound pair are on different legs
of the zigzag ladder.
The soft mode signals emergence of a nematic phase
below the saturation field.

\begin{figure}
\begin{center}
\includegraphics[width=70mm]{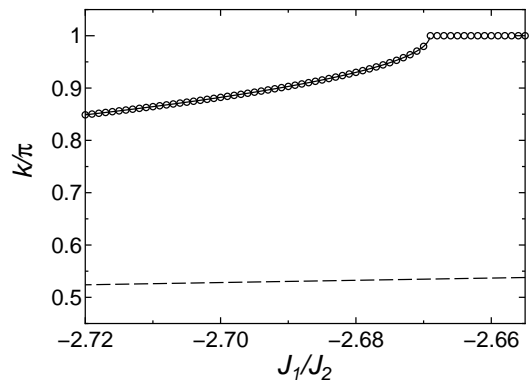}
\caption{
Dependence on $J_1/J_2$ of the center-of-mass momentum $k$
for the two-magnon bound state.
The momentum deviates continuously
from $\pi$ at $J_1/J_2\simeq -2.669$.
The incommensurate momentum $k$
is fitted well to $k/\pi=1-0.67 \sqrt{-2.669-J_1/J_2}$ 
as shown by the solid curve.
The dashed line is the classical estimate $k=2\arccos\,(-J_1/4J_2)$
for two scattering magnons.
}
\label{fig:momentum}
\end{center}
\end{figure}

In the narrow range $-2.720<J_1/J_2<-2.669$,
the soft two-magnon bound state has an incommensurate momentum
$k\ne\pi$.
Our numerical estimate of the commensurate-incommensurate transition
point is consistent with the exact result,
$(J_1/J_2)_c = -2.66908\ldots$.\cite{KuzianD2007}
Beyond the commensurate-incommensurate transition point,
the total momentum $k$ changes continuously
as $k/\pi=1-0.67\sqrt{(J_1/J_2)_c-J_1/J_2}$;
see Fig.~\ref{fig:momentum}.
This suggests continuous nature of the transition
between the commensurate and incommensurate nematic phases
at $h=h_\mathrm{s}$.

As the ratio $J_1/J_2$ is changed towards the end point at $J_1/J_2=-4$,
the magnon number $p$ of the lowest bound-magnon branch increases; 
see Table I.
The total momentum of the bound-magnon mode is always at $k =\pi$,
except in the narrow region of the incommensurate two-magnon bound states
mentioned above.
We expect that many-magnon bound states,
formed by more than seven magnons, should appear
as $J_1/J_2\to-4$.
In our numerical calculation, eight-magnon  
bound states did not come down as the lowest state,
which we suspect was due to finite-size effects coming from
small $r_{\rm max}$.

\begin{figure}
\begin{center}
\includegraphics[width=83mm]{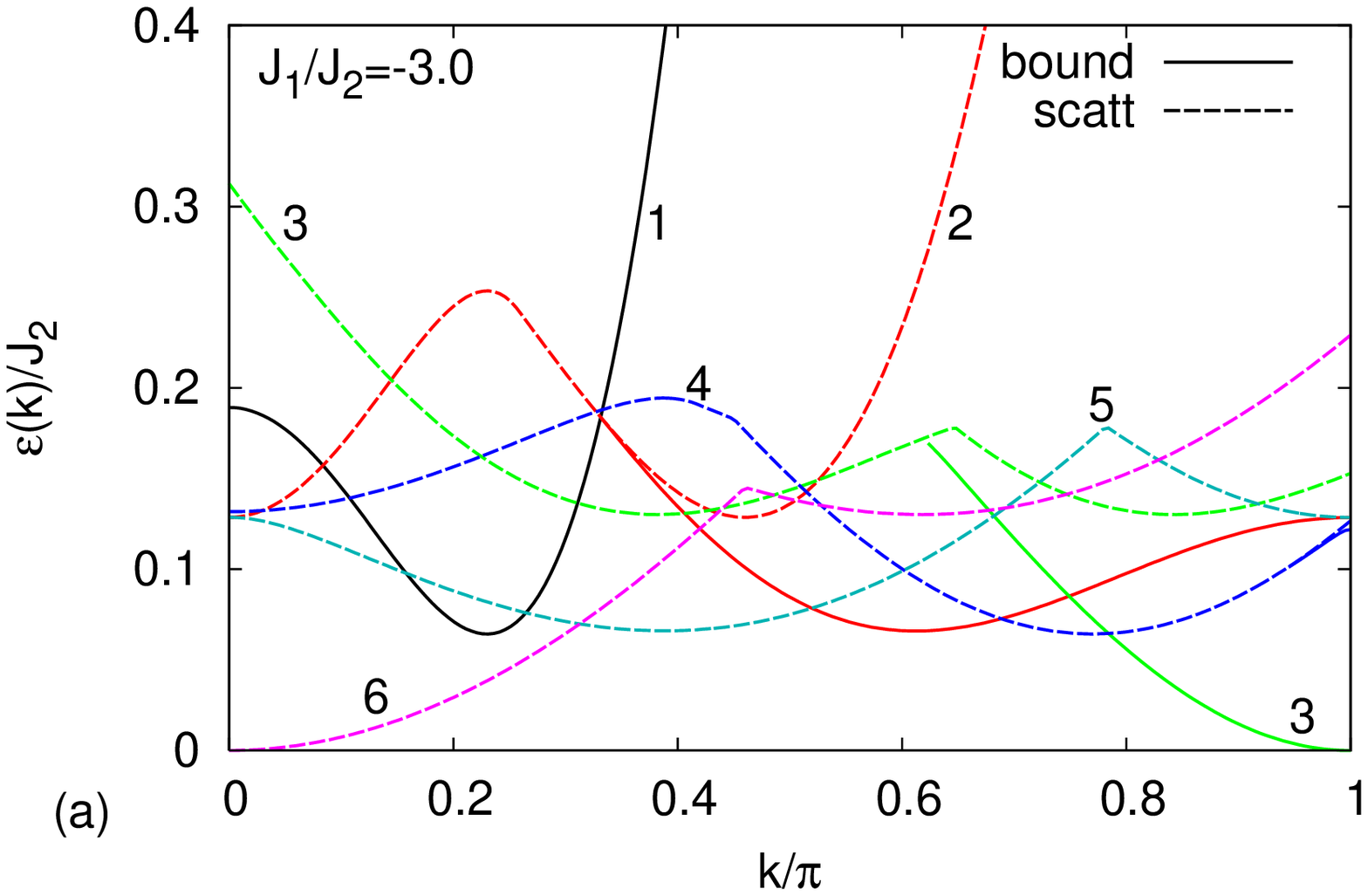}
\includegraphics[width=83mm]{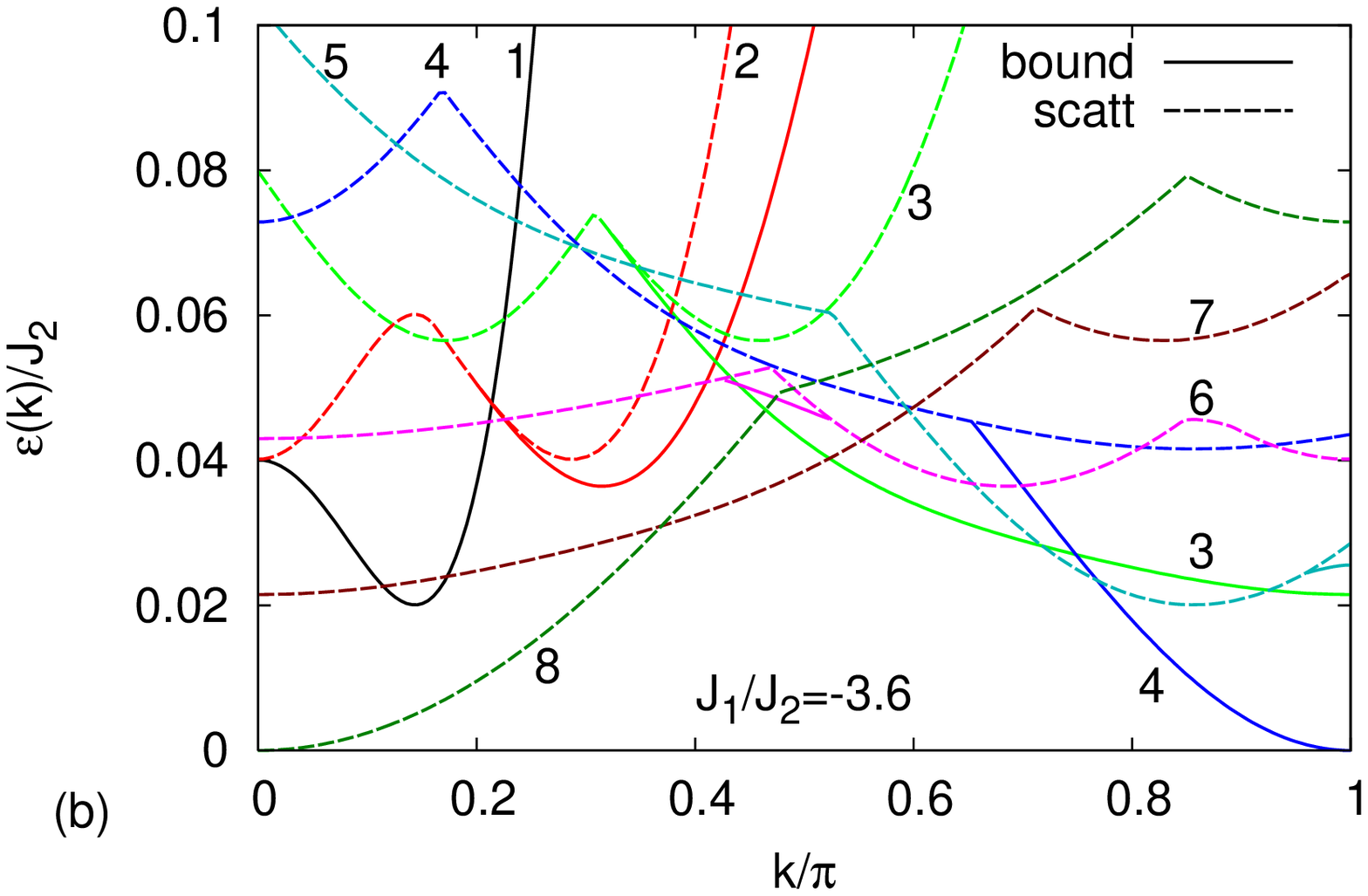}
\caption{
(Color online)
Dispersion curves of multimagnon bands at the saturation field
for (a) $J_1/J_2 = -3.0$ and (b) $J_1/J_2 = -3.6$.
The solid curves are the dispersions of bound states and the
dashed curves show the lower edges of the continuum of scattering states.
For clarity, only the states of up to $6$-magnons are shown in (a)
and up to $8$-magnons in (b).
The numbers printed beside the curves denote the number of magnons.
Bound states inside the scattering continuum are not shown.
}
\label{fig:dispersion}
\end{center}
\end{figure}

To demonstrate stability of the multimagnon bound states,
we show in Fig.\ \ref{fig:dispersion} dispersion curves of bound-magnon
excitations, as well as lower edges of continuous spectra of magnon
scattering states, at the saturation field $h=h_{\rm s}$
for $J_1/J_2=-3.0$ and $-3.6$.
The $p$-magnon scattering states are constructed from
a set of one-, two-, ..., $(p-1)$-magnon (bound) states,
in total of $p$ magnons.
At $J_1/J_2 = -3.0$ [Fig.\ \ref{fig:dispersion}(a)],
the three-magnon bound state is gapless at $k = \pi$.
The branches of one, two, four, and five magnons have finite excitation gaps.
This feature is consistent with a finite binding energy of the three-magnon
bound state.
Furthermore, the state with the lowest energy (at $k = 0$)
in the six-magnon sector belongs to the continuum of
scattering states formed by a pair of
three-magnon bound states.
This indicates that three-magnon bound states are
interacting repulsively with each other.
The repulsive interaction rules out the possibility of
a metamagnetic transition (magnetization jump) at the saturation field,
and instead induces a continuous transition to the triatic phase
which we discuss in more detail in Sec.~\ref{sec:TriQua}.
The multimagnon dispersions for $J_1/J_2 = -3.6$, shown in
Fig.\ \ref{fig:dispersion}(b), can also be understood in the same fashion.
Here it is the four-magnon bound states
that become gapless at the saturation field.
The instability of the fully polarized state
is driven by the four-magnon bound states
with mutual repulsive interactions,
which condense to form a TL liquid with quartic order,
as we show in Sec.\ \ref{sec:TriQua}.

\section{Magnetization curve}\label{sec:magcurve}

\begin{figure}
\begin{center}
\includegraphics[width=86mm]{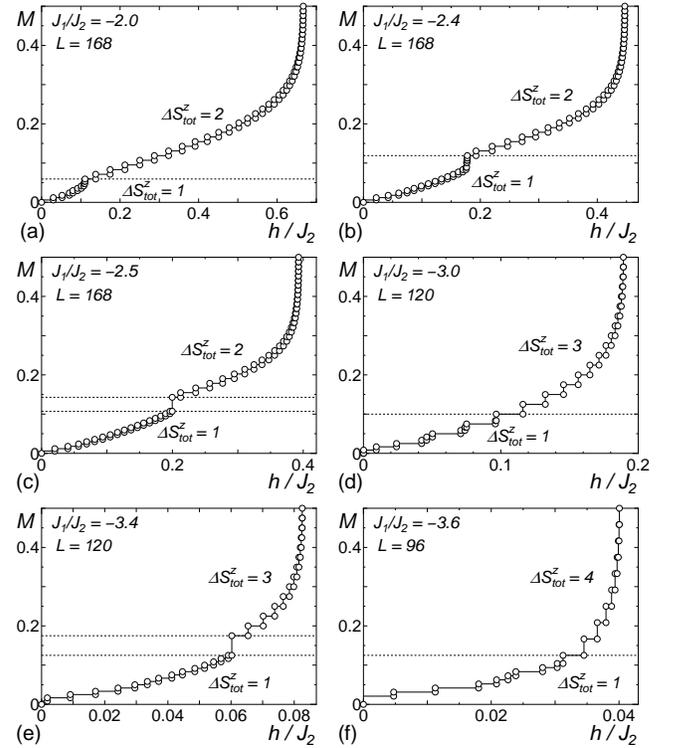}
\caption{
Magnetization curves for (a) $J_1/J_2 = -2.0$, (b) $J_1/J_2 = -2.4$,
(c) $J_1/J_2 = -2.5$, (d) $J_1/J_2 = -3.0$, (e) $J_1/J_2 = -3.4$,
and (f) $J_1/J_2 = -3.6$.
The dotted lines represent the boundaries of the regions of
$\Delta S^z_{\rm tot} = 1$ and $\Delta S^z_{\rm tot} \ge 2$.
}
\label{fig:magcurve}
\end{center}
\end{figure}

Having identified the soft modes at the saturation field $h=h_s$,
we now study magnetization process of
spin chains of finite length, which we obtain numerically
for various values of the coupling ratio $J_1/J_2$.
The numerical results help us to
deduce overall structure of the magnetic phase diagram.
Previous studies\cite{CabraHP2000,HeidrichMHV2006} have found that,
near the saturation field,
the total magnetization $S^z_{\rm tot} = \sum_l s^z_l$ changes
in units of
$\Delta S^z_\mathrm{tot}=2$, 3, and 4
for $J_1/J_2 \gtrsim -2.6$, 
$-3 \le J_1/J_2 \le -2.8$,
and $J_1/J_2 = -3.75$, respectively.
The multispin flip is a natural consequence of the formation
of stable multimagnon bound states.
At lower fields $S^z_\mathrm{tot}$ changes by
$\Delta S^z_\mathrm{tot}=1$.

We obtain magnetization curves from the following procedure.
With the DMRG method we calculate the lowest energy $E_0(M)$
of the model (\ref{eq:Ham}) at $h=0$ in each Hilbert subspace
of magnetization per site $M = S^z_{\rm tot}/L$,
where $L$ is the number of total sites.
The magnetization curve $M(h)$ is then obtained
by finding the magnetization $M$ which minimizes $E_0(M)-hM$
for given $h$.
We have performed the calculation for open chains
of up to $L = 168$ sites.
We kept up to 350 states in our DMRG calculation.

Figure\ \ref{fig:magcurve} shows representative magnetization curves
calculated at various values of $J_1/J_2$.
We clearly see that the total magnetization $S^z_\mathrm{tot}$
changes by $\Delta S^z_{\rm tot} = 1$
at low magnetic fields,
while it shows multispin flip process $\Delta S^z_{\rm tot} \ge 2$
at higher fields.
The magnetization change in the high-field regime is
$\Delta S^z_{\rm tot} = 2$ for $J_1/J_2 \ge -2.7$,
$\Delta S^z_{\rm tot} = 3$ for $-3.4\le J_1/J_2\le-2.75$,
and $\Delta S^z_{\rm tot} = 4$ at $J_1/J_2 = -3.6$.
These features of the magnetization process are
consistent with our finding of stable multimagnon bound
states discussed in Sec.\ \ref{sec:multimagnon}.
The critical field $h_{\rm c}$ and
the critical magnetization $M_\mathrm{c}$,
at which the magnetization step changes from $\Delta S^z=1$
to $\Delta S^z>1$,
are plotted in the phase diagram shown in Fig.\ \ref{fig:phasediagram}.

The magnetic phase diagram (Fig.~\ref{fig:phasediagram}) has four
distinct regions characterized by $\Delta S^z_{\rm tot} = 1, 2, 3,$ and $4$.
These regions correspond to the vector chiral, nematic or SDW$_2$,
triatic or SDW$_3$,
and quartic phases, respectively.
We will discuss each region in detail
in the following sections.
The phase boundary between the region of $\Delta S^z_{\rm tot} = 1$ and
that of $\Delta S^z_{\rm tot}=2$ begins from the critical coupling
$J_1/J_2 = -2.72$ at $h = h_{\rm s}$,
which is the phase boundary between the incommensurate
nematic and triatic phases,\cite{KeckeMF2007}
and appears to go towards small $|J_1|/J_2$ region as
$h$ is decreased.

When $-2.72<J_1/J_2\le-2.5$,
the magnetization curve exhibits a large jump (of order $L^0$)
on the phase boundary between the region of $\Delta S^z_{\rm tot} = 1$
and that of $\Delta S^z_{\rm tot} = 2$, whereas for $J_1/J_2 \ge -2.4$
the magnetization curve appears to become continuous
(i.e., steps are of order $L^{-1}$).
Similarly, we observed a large jump in the magnetization
between the $\Delta S^z_{\rm tot} = 1$
and $\Delta S^z_{\rm tot} = 3$ regions
at $J_1/J_2 = -3.4$,
but not at other values.
However, we find that the magnetization curve at $J_1/J_2 = -2.4$
also develops a sharper change at $M=M_\mathrm{c}$
with increasing the system size $L$,
which turns into almost a discontinuous jump at $L = 168$
[see Fig.\ \ref{fig:magcurve}(b)].
This may suggest that the transition becomes first order
at $L \to \infty$.
Since we do not have a good scheme of extrapolation to $L\to\infty$
for incommensurate values of $M$,
it is difficult to determine the order of the transition
from the numerical calculation alone.
More elaborated treatments, especially analytical ones, would be required
for resolving this issue.

For the parameters calculated, $J_1/J_2 \ge -3.6$, 
we find that at the saturation field $h=h_{\rm s}$ the magnetization curve
approaches $M = 1/2$ continuously in accordance with the previous
studies\cite{CabraHP2000,HeidrichMHV2006,KeckeMF2007}
(see also Note added).
We note that the square-root singularity $1/2-M\propto(h_\mathrm{s}-h)^{1/2}$
is commonly expected for a continuous transition
at the saturation field, where soft excitations are
described as free hard-core bosons or free fermions.

\section{Vector chiral phase}\label{sec:VC}

In this section we take a detailed look at correlation functions
in the vector chiral phase.
We will show that this phase corresponds to the low-field regime
where the magnetization curve has $\Delta S^z_{\rm tot} = 1$ steps,
and the ground state exhibits 
a long-range order of the longitudinal vector chirality,
$
\langle \kappa_l^{(n)} \rangle
= \langle \left( {\bm s}_l \times {\bm s}_{l+n} \right)^z \rangle
\ne 0
$
($n=1,2$).
We first give a brief review of a low-energy field theory for
the vector chiral phase.
We then present numerical DMRG results
and compare them with the theory.

\subsection{Bosonization approach for $|J_1| \ll J_2$}
\label{sec:bosonization-VC}

A field theoretical approach to the vector chiral phase
in the $J_1$-$J_2$ model was developed by Nersesyan
\textit{et al.} in Ref.\ \onlinecite{NersesyanGE1998}, in which
the antiferromagnetic $J_1$-$J_2$ chain with easy-plane anisotropy
was considered.
Kolezhuk and Vekua extended this theory to include effects of
the external magnetic field in Ref.\ \onlinecite{KolezhukV2005}.
Here we follow their approach and apply it to the ferromagnetic
$J_1$-$J_2$ chain.
The theory is based on bosonization of the antiferromagnetic Heisenberg
spin chain and perturbative renormalization-group (RG) analysis valid
for $|J_1| \ll J_2$.

In the limit $|J_1| \ll J_2$, the model (\ref{eq:Ham}) can be regarded
as two antiferromagnetic Heisenberg spin chains which are weakly
coupled by the ferromagnetic interchain interaction $J_1$.
Therefore, we apply the standard bosonization technique
to the two chains separately, treating
the interchain coupling $J_1$ as a weak perturbation.
The low-energy physics of the Heisenberg chains ($n=1,2$) are
described by free bosonic fields $(\phi_n, \theta_n)$
satisfying the equal-time commutation relation
$[\phi_n(x), \partial_y\theta_{n'}(y)] = i\delta(x-y) \delta_{n,n'}$.
The spin operators $\bm{s}_l$ on the site $l=2j+n$
($j\in\mathbb{Z}$)
in the Hamiltonian (\ref{eq:Ham}) are expressed
in terms of
the bosonic fields as
\begin{eqnarray}
s^z_{2j+n} \!\!&=&\!\!
M + \frac{1}{\sqrt{\pi}} \frac{d\phi_n(x_n)}{dx}
\nonumber \\
&&{}\!\!\!
- (-1)^j a \sin[2\pi Mj + \sqrt{4\pi} \phi_n(x_n)] + \cdots,
\label{eq:sz-boson-twochain} \\
s^+_{2j+n} \!\!&=&\!\!
(-1)^j b \, e^{i \sqrt{\pi} \theta_n(x_n)}
\nonumber \\
&&{}\!\!\!
+ b' e^{i \sqrt{\pi} \theta_n(x_n)}
\sin[2\pi Mj + \sqrt{4\pi} \phi_n(x_n)] + \cdots,
\quad
\nonumber \\
\label{eq:s+-boson-twochain}
\end{eqnarray}
where $a$, $b$, and $b'$ are nonuniversal
constants.\cite{HikiharaF2001,HikiharaF2004}
We have introduced the continuous space coordinate $x$, on which
the bosonic fields depend.
On the lattice site $l=2j+n$ the coordinate $x$ takes the
value $x_1=j-1/4$ and $x_2=j+1/4$.
Equations (\ref{eq:sz-boson-twochain}) and (\ref{eq:s+-boson-twochain})
allow us to write the interchain interaction
in terms of the bosonic fields.
The resulting effective Hamiltonian is given by\cite{KolezhukV2005}
\begin{eqnarray}
\widetilde{\mathcal{H}} &=&
\sum_{\nu=\pm} \frac{v_\nu}{2} \int dx
\left[
K_\nu \left( \frac{d\theta_\nu}{dx} \right)^2
+
\frac{1}{K_\nu} \left( \frac{d\phi_\nu}{dx} \right)^2
\right]
\nonumber \\
&&{} + g_1 \int dx \, \sin(\sqrt{8\pi} \phi_- + \pi M)
\nonumber \\
&&{} + g_2 \int dx \, \frac{d\theta_+}{dx} \sin(\sqrt{2\pi} \theta_-)
\label{eq:Ham-boson-twochain}
\end{eqnarray}
with
\begin{equation}
g_1 = J_1 a^2 \sin(\pi M),
\qquad
g_2 = \frac{J_1}{2} \sqrt{2\pi} b^2.
\end{equation}
Here we have introduced bosonic fields
for symmetric $(+)$ and antisymmetric $(-)$ sectors,
$\phi_\pm = (\phi_1 \pm \phi_2)/\sqrt2,
\theta_\pm = (\theta_1 \pm \theta_2)/\sqrt2$.
In lowest order in $J_1$ the TL-liquid parameters $K_\pm$ and 
the renormalized spin velocities $v_\pm$ are
given by\cite{KolezhukV2005}
\begin{eqnarray}
K_\pm&=&K\left(1\mp J_1\frac{K}{\pi v}\right),
\label{eq:K_pm} \\
v_\pm&=&v\left(1\pm J_1 \frac{K}{\pi v}\right),
\end{eqnarray}
where $K$ and $v$ are respectively the TL-liquid parameter 
and the spin velocity of the decoupled antiferromagnetic
Heisenberg spin chains.
The TL-liquid parameter
$K$ is a function of $M$ 
increasing monotonically from $K(M=0)=1/2$ 
to $K(M=1/2)=1$.\cite{Haldane1980,BogoliubovIK1986,CabraHP1998}
In the weak-coupling limit the velocity $v$ is
of order $J_2$,
except near the saturation limit $M\to\frac12$, where
$v\to0$ and the bosonization approach breaks down.

As we can see from Eqs.\ (\ref{eq:sz-boson-twochain}) and
(\ref{eq:s+-boson-twochain}),
the $g_1$ ($g_2$) term in Eq.\ (\ref{eq:Ham-boson-twochain})
originates from the longitudinal (transverse)
part of the interchain exchange coupling.
These coupling constants are renormalized as energy scale
is decreased in RG transformation.
The low-energy physics of the effective Hamiltonian
(\ref{eq:Ham-boson-twochain}) is then determined by
the strongest of the renormalized coupling constants.
In case the $g_1$ term is most relevant,
the $\phi_-$ field is pinned at a value which minimizes
$g_1\sin(\sqrt{8\pi}\phi_-+\pi M)$.
The resulting ground state is in the nematic phase,
as we will discuss in Sec.\ \ref{sec:Nem}.
The vector chiral phase arises when the $g_2$ coupling is
most relevant and renormalized to strong coupling first.

The scaling dimensions of the $g_1$ and $g_2$ terms are
equal to $2K_-$ and $1+(2K_-)^{-1}$, respectively at $J_1=0$.
It is then natural to expect that the $g_2$ term can
dominate over the $g_1$ term
only in high fields
[i.e., for $K_->(1+\sqrt{5})/4$]
when $|J_1|\ll J_2$.\cite{KolezhukV2005}
However, as we discussed in Sec.~\ref{sec:multimagnon}
the two-magnon pairing
is the strongest instability at $h=h_\mathrm{s}$,
which favors the nematic order
near the saturation field
(Fig.~\ref{fig:phasediagram} and Table~\ref{tab:multimagnon}).
In fact, the vector chiral phase is found to be realized
in the weak-field regime where the bare value of the coupling $g_1$
is very small ($M\ll1$)
and where the classical value of the vector chirality is larger
($\theta^c\approx\pi/2$).

In the following discussion
let us assume that the renormalized $g_2$ is the largest coupling.
In this case we may employ the mean-field decoupling
scheme introduced by
Nersesyan \textit{et al.},\cite{NersesyanGE1998}
whose conclusions have been confirmed by numerical
studies.\cite{KaburagiKH1999,HikiharaKK2001,McCulloch2007,Okunishi2008,HikiharaMFK}
In this scheme we assume that
both $d\theta_+/dx$ and $\sin(\sqrt{2\pi} \theta_-)$
acquire finite expectation values so that the $g_2$ term is minimized.
We have essentially two choices:
\begin{subequations}
\label{eq:theta-pin-VC}
\begin{equation}
\langle \theta_- \rangle = + \sqrt{\frac{\pi}{8}},
\qquad
\left\langle \frac{d\theta_+}{dx} \right\rangle
= + \sqrt{\frac{2}{\pi}} (\pi-2Q),
\label{eq:theta-pin-VC(a)}
\end{equation}
and
\begin{equation}
\langle \theta_- \rangle = - \sqrt{\frac{\pi}{8}},
\qquad
\left\langle \frac{d\theta_+}{dx} \right\rangle
= - \sqrt{\frac{2}{\pi}} (\pi-2Q),
\label{eq:theta-pin-VC(b)}
\end{equation}
\end{subequations}
where $Q$ is an incommensurate wave number in the transverse
spin correlation ($0<Q<\pi/2$);
see Eq.\ (\ref{eq:Cspx-VC}).
Note that $\langle \theta_- \rangle$ and
$\langle d\theta_+/dx \rangle$ have the same sign
in the frustrated ferromagnetic chain $J_1 < 0$.
The $Z_2$ symmetry is spontaneously broken when the ground state
selects one of the two choices in Eq.\ (\ref{eq:theta-pin-VC}).

Once the mean-field decoupling is made,
the excitations in the antisymmetric sector $(\phi_-, \theta_-)$
acquire a finite energy gap, and
the field $\phi_-$, which is dual to the pinned field $\theta_-$,
fluctuates strongly.
Therefore the $g_1$ term can be safely ignored.
The symmetric sector $(\phi_+,\theta_+)$ is governed by the
Gaussian model,
\begin{equation}
\mathcal{H}_+=\frac{v_+}{2}\int dx
\left[
K_+\left(\frac{d\theta_+}{dx}\right)^2
+
\frac{1}{K_+}\left(\frac{d\phi_+}{dx}\right)^2
\right],
\label{H_+}
\end{equation}
once we redefine the $\theta_+$ field,
$\theta_+\to\theta_+-\langle d\theta_+/dx\rangle x$,
to absorb the nonvanishing average $\langle d\theta_+/dx\rangle$.
Hence the ground state is a one-component TL liquid.

We are now ready to calculate correlation functions.
Most important of these is the ground-state average
of the vector chirality (\ref{kappa^n_l}),
\begin{subequations}
\label{eq:kappa}
\begin{eqnarray}
\langle \kappa_l^{(1)} \rangle
\!\!&=&\!\!
-b^2 \langle \sin(\sqrt{2\pi} \theta_-) \rangle
= \mp b^2 c_1,
\\
\langle \kappa_l^{(2)} \rangle
\!\!&=&\!\!
- \sqrt{\frac{\pi}{2}} c_2
 \left\langle \frac{d\theta_+}{dx} \right\rangle
= \mp c_2 (\pi-2Q),
\end{eqnarray}
\end{subequations}
where $c_1$ and $c_2$ are positive constants.
These nonvanishing averages indicate that the ground state breaks
a $Z_2$ symmetry and has a vector chiral long-range order.
Since
$\langle \kappa_l^{(1)} \rangle$ and $\langle \kappa_l^{(2)} \rangle$
have the same sign
and satisfy Eq.\ (\ref{eq:bloch relation}),
the spin current $J_n\langle\kappa_l^{(n)}\rangle$
flows as depicted in Fig.~\ref{fig:VCflow}.
The spin current circulates in each triangle in alternating
fashion, and there is no net spin current flow through the whole system.

Two-point correlation functions are also calculated 
using Eqs.\ (\ref{eq:sz-boson-twochain}), (\ref{eq:s+-boson-twochain}),
(\ref{eq:theta-pin-VC}), and (\ref{H_+}).
Here we remind the reader our convention that
the site index $l$ in the original
lattice Hamiltonian (\ref{eq:Ham}) is equal to $2j+n$,
where the integer $j$ is the site index in each antiferromagnetic
Heisenberg chain ($n=1,2$) in the two-chain (zigzag ladder) picture.
Hence $\Delta l=2\Delta j = 2\Delta x$.
The correlation function for the vector chirality is given by
\begin{eqnarray}
\label{eq:kappa_corr}
\langle \kappa_0^{(2)} \kappa_l^{(2)} \rangle
\!\!&=&\!\!
\langle \kappa^{(2)} \rangle^2
\left( 1 - \frac{1}{K_+ [(\pi-2Q)l]^2}\right)\nonumber\\
&&+\frac{(-1)^l A_\kappa}{|l|^{4K_+}}\cos(2\pi M l)+\cdots ,
\end{eqnarray}
where $A_\kappa$ is a constant ($\propto b'^4g_1^2$).
In the two-point function of the chiral operator $\kappa_l^{(1)}$
the uniform $1/l^2$ term in Eq.\ (\ref{eq:kappa_corr}) is replaced
by a $1/l^4$ term.\cite{McCulloch2007}
The transverse and longitudinal spin correlation functions
are obtained as
\begin{eqnarray}
&&
\langle s^x_0 s^x_l \rangle
=
\frac{A}{|l|^{1/4K_+}} \cos(Ql) +\cdots,
\label{eq:Cspx-VC} \\
&&\langle s^z_0 s^z_l \rangle
= M^2 - \frac{K_+}{\pi^2 l^2} + \cdots.
\label{eq:Cspz-VC}
\end{eqnarray}
Lastly the nematic correlation function shows
a faster decay,
\begin{eqnarray}
\langle s^+_0 s^+_1 s^-_l s^-_{l+1} \rangle
= \frac{A'}{|l|^{1/K_+}} \cos(2Q l)
+\cdots.
\label{eq:Cb2m-VC}
\end{eqnarray}
In the above equations 
$A$ and $A'$ are nonuniversal constants, and
we have omitted subleading algebraically-decaying terms 
and exponentially-decaying terms,
such as a short-ranged incommensurate correlation
[$\propto\cos(\pi Ml)$] in Eq.\ (\ref{eq:Cspz-VC}).

We note that
the wave number of the transverse spin correlation function
-- the pitch angle --
is shifted from the commensurate value $\pi/2$ to $Q$.
The vector chiral long-range order and the incommensurate
transverse spin correlation are the hallmark of the vector chiral phase.

\subsection{Numerical results}\label{sec:numerics-VC}

\begin{figure}
\begin{center}
\includegraphics[width=70mm]{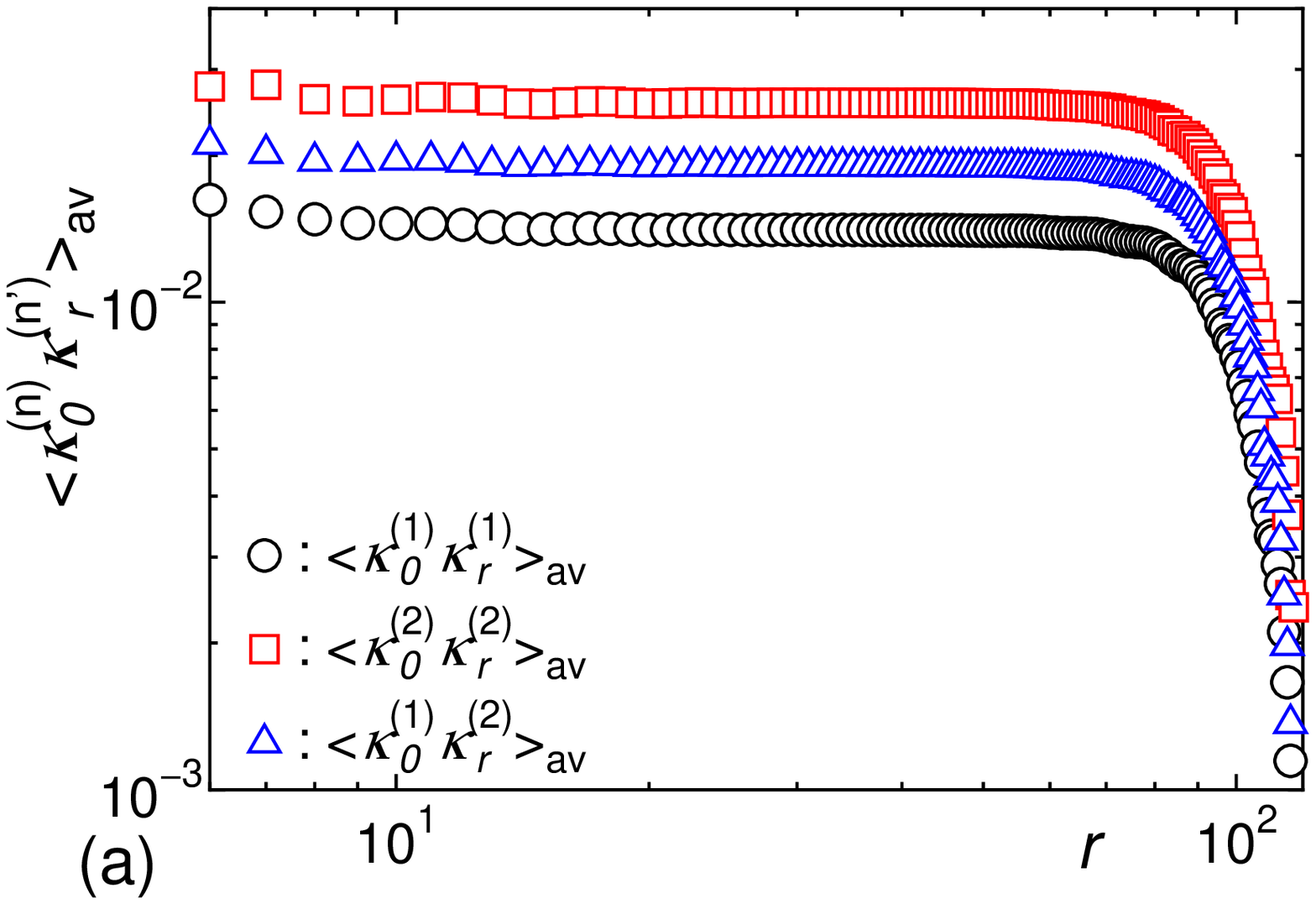}
\includegraphics[width=70mm]{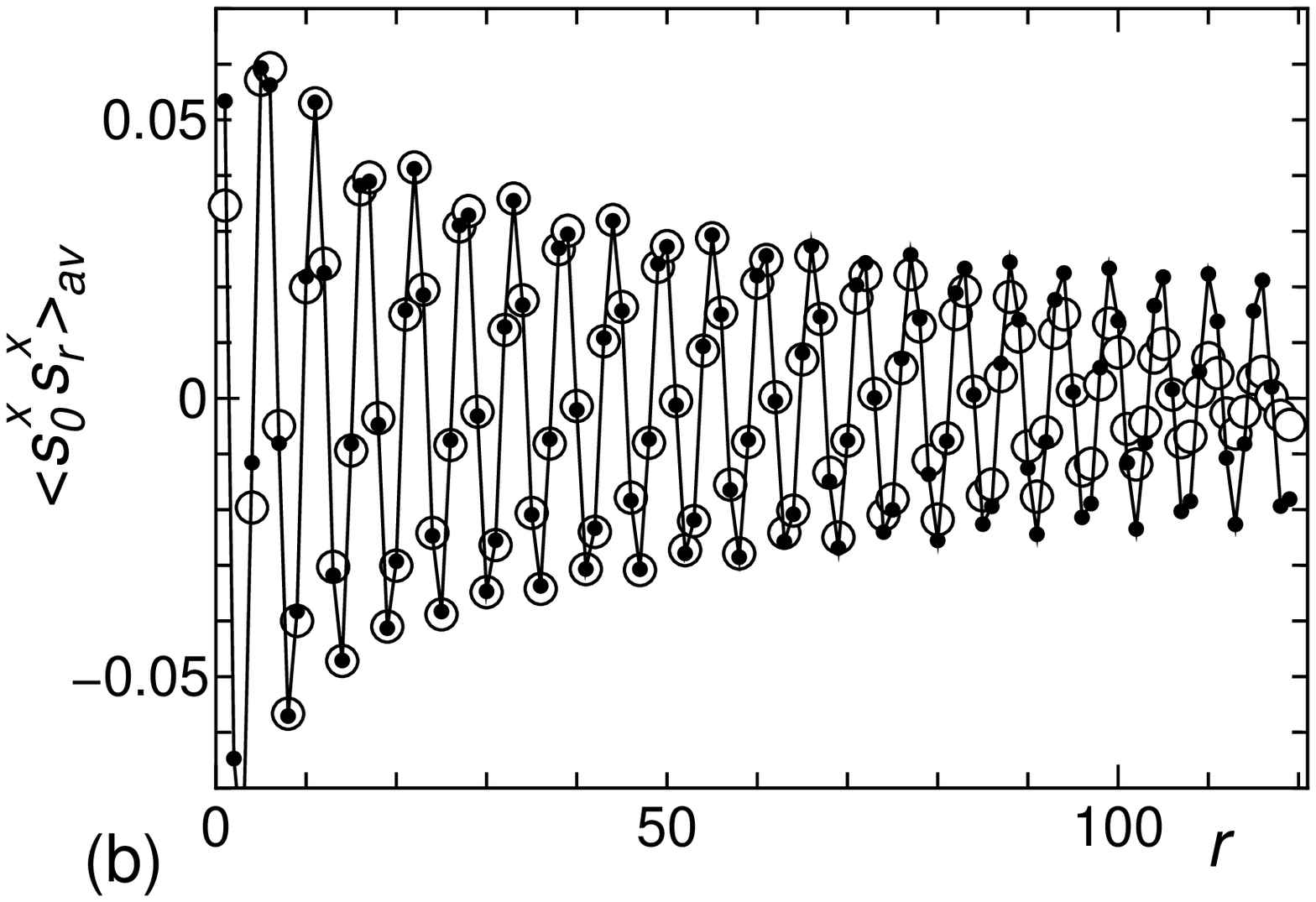}
\includegraphics[width=70mm]{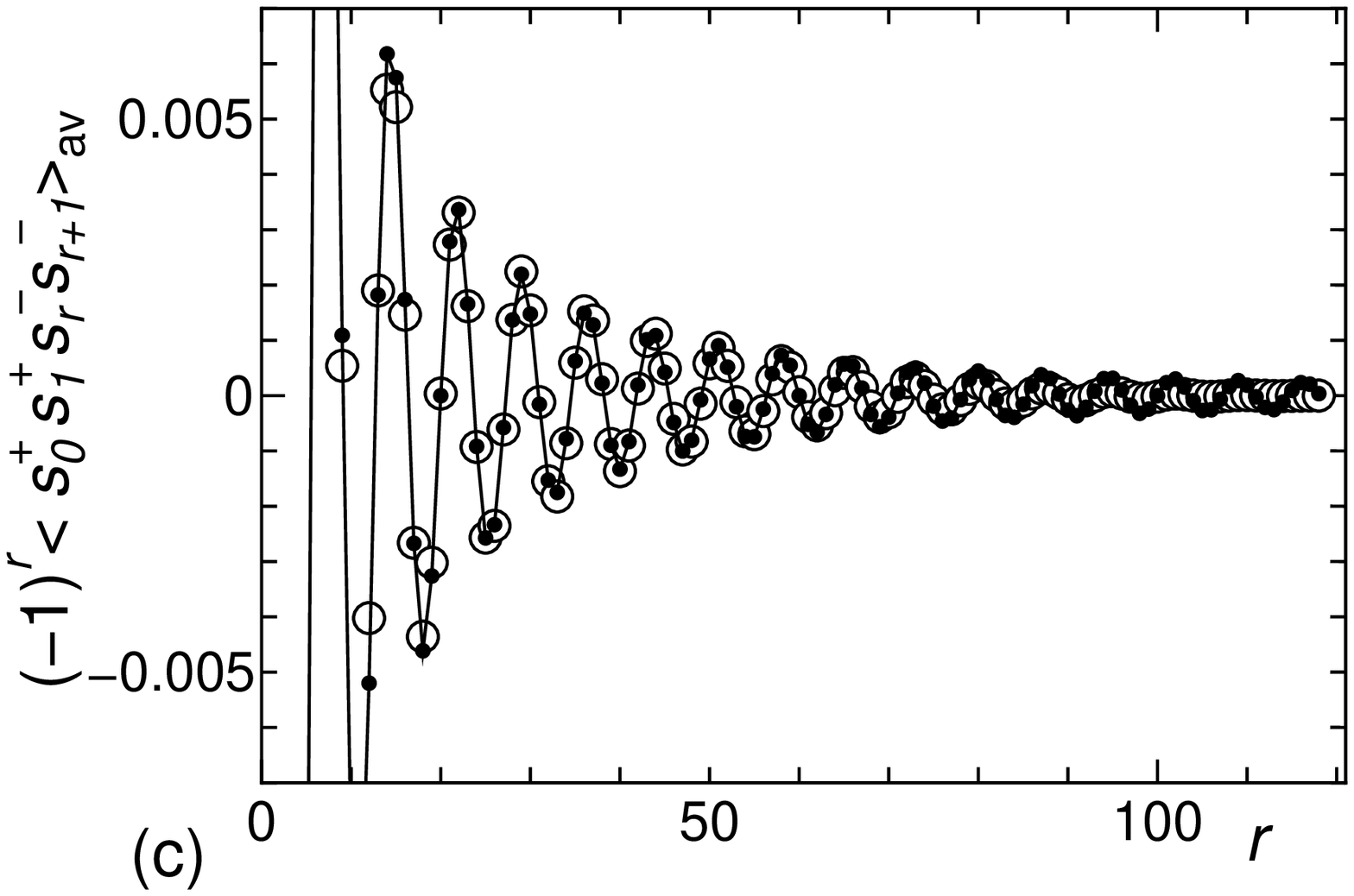}
\caption{
(Color online)
Averaged correlation functions
for $L = 120$ spins with $J_1/J_2 = -2.7$ and $M = 0.1$
in the vector chiral phase;
(a) vector chiral correlation functions
$\langle \kappa^{(1)}_0 \kappa^{(1)}_r \rangle_{\rm av}$,
$\langle \kappa^{(2)}_0 \kappa^{(2)}_r \rangle_{\rm av}$, and
$\langle \kappa^{(1)}_0 \kappa^{(2)}_r \rangle_{\rm av}$,
where $\kappa^{(n)}_r = \left( {\bm s}_r \times {\bm s}_{r+n} \right)^z$
[Eq.\ (\ref{kappa^n_l})],
(b) transverse spin correlation function
$\langle s^x_0 s^x_r \rangle_{\rm av}$,
(c) nematic correlation function
$\langle s^+_0 s^+_1 s^-_r s^-_{r+1} \rangle_{\rm av}$.
Open symbols represent the DMRG data.
Truncation errors are smaller than the size of the symbols.
The solid lines and solid circles in (b) and (c) are fits
to Eqs.\ (\ref{eq:Cspx-VC}) and (\ref{eq:Cb2m-VC}), respectively.
}
\label{fig:correlation-VC}
\end{center}
\end{figure}

Here we present our numerical results, which  support the theory
of the preceding subsection.
The calculation was done for finite open chains
with $L=96$ and $120$ sites, 
unless otherwise mentioned.
In the following, we show mainly the results for $L=120$, 
while we note that the results for $L=96$ exhibits 
essentially the same behaviors as those for $L=120$.
The number of kept DMRG states is up to $350$.
We have performed typically 10-30 DMRG sweeps in the calculation 
and checked the convergence of the results.
Since Eqs.~(\ref{eq:kappa})--(\ref{eq:Cb2m-VC})
are obtained for infinite-length chains,
we need to take care of open-boundary effects in the DMRG data
to make meaningful comparison between the theory and the numerics.
To reduce the boundary effects,
we calculate two-point correlation functions
for several pairs of two sites $(l,l')$ with fixed distance $r = |l-l'|$
selecting the two sites being as close to the center of the chain as possible.
We then take their average for the estimate of the correlation.
We use the notation $\langle \cdots \rangle_{\rm av}$
for the averaged correlation functions below.

Figure\ \ref{fig:correlation-VC}(a) shows a typical $r$ dependence of
the averaged vector chiral correlation functions
in the vector chiral phase ($J_1/J_2=-2.7$ and $M=0.1$).
We clearly see that the vector chirality is
long-range ordered.\cite{boundary-VC}
We also find that not only
$\langle \kappa^{(1)}_0 \kappa^{(1)}_r \rangle_{\rm av}$ and
$\langle \kappa^{(2)}_0 \kappa^{(2)}_r \rangle_{\rm av}$
but also $\langle \kappa^{(1)}_0 \kappa^{(2)}_r \rangle_{\rm av}$
are positive, in agreement with Eqs.\ (\ref{eq:kappa}).
This indicates the ferro-chiral order
as drawn in Fig.~\ref{fig:VCflow}.

\begin{figure}
\begin{center}
\includegraphics[width=86mm]{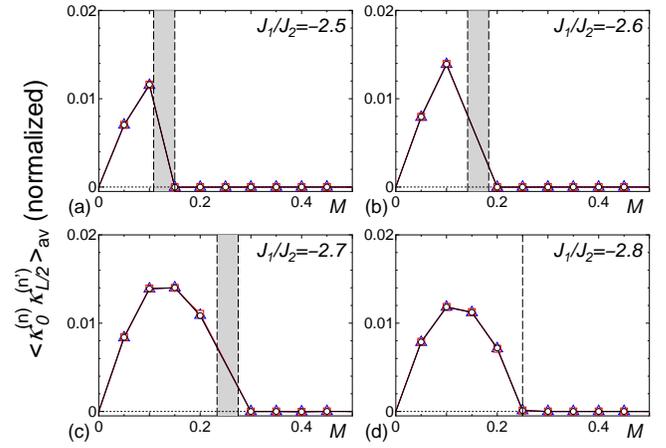}
\caption{
Averaged, normalized vector chiral correlations 
at the distance $r=L/2$ for $L=120$ spin zigzag chain.
Circles, squares, and triangles represent 
$\langle\kappa_0^{(1)}\kappa_{L/2}^{(1)}\rangle$,
$(2J_2/|J_1|)^2\langle\kappa_0^{(2)}\kappa_{L/2}^{(2)}\rangle$,
and
$(2J_2/|J_1|) \langle\kappa_0^{(1)}\kappa_{L/2}^{(2)}\rangle$,
respectively.
(a) $J_1/J_2 = -2.5$, (b) $J_1/J_2 = -2.6$,
(c) $J_1/J_2 = -2.7$, and (d) $J_1/J_2 = -2.8$.
Solid lines are the guide for the eye.
Vertical dashed lines represent the boundaries of the chiral ordered phase
in low magnetization regime
and the chiral disordered phase in the high magnetization regime.
The shaded region corresponds to the magnetization jump 
at the first-order transition.
}
\label{fig:normalized VC}
\end{center}
\end{figure}

In Fig.~\ref{fig:normalized VC} we plot averaged vector chiral
correlations at a distance $r = L/2$ normalized by the coupling ratio,
$\langle\kappa_0^{(1)}\kappa_{L/2}^{(1)}\rangle$,
$(2J_2/|J_1|)\langle\kappa_0^{(1)}\kappa_{L/2}^{(2)}\rangle$,
and
$(2J_2/|J_1|)^2\langle\kappa_0^{(2)}\kappa_{L/2}^{(2)}\rangle$,
as functions of $M$ and $J_1/J_2$.
The three quantities agree, as expected from
Eq.\ (\ref{eq:bloch relation}).
This figure clearly shows where the vector chiral correlation is strong;
The vector chiral order exists in the low-field regime, but disappears
in the high-field regime. 
The shaded regions in Fig.~\ref{fig:normalized VC} correspond to
the magnetization jump discussed in Sec.~\ref{sec:magcurve},
where the transition is clearly first order.
We note that near the boundary of the vector chiral phase 
the vector chiral correlations suffer boundary effects in open chains 
and may underestimate the vector chiral order
(see the discussion at the end of this section).

\begin{figure}
\begin{center}
\includegraphics[width=65mm]{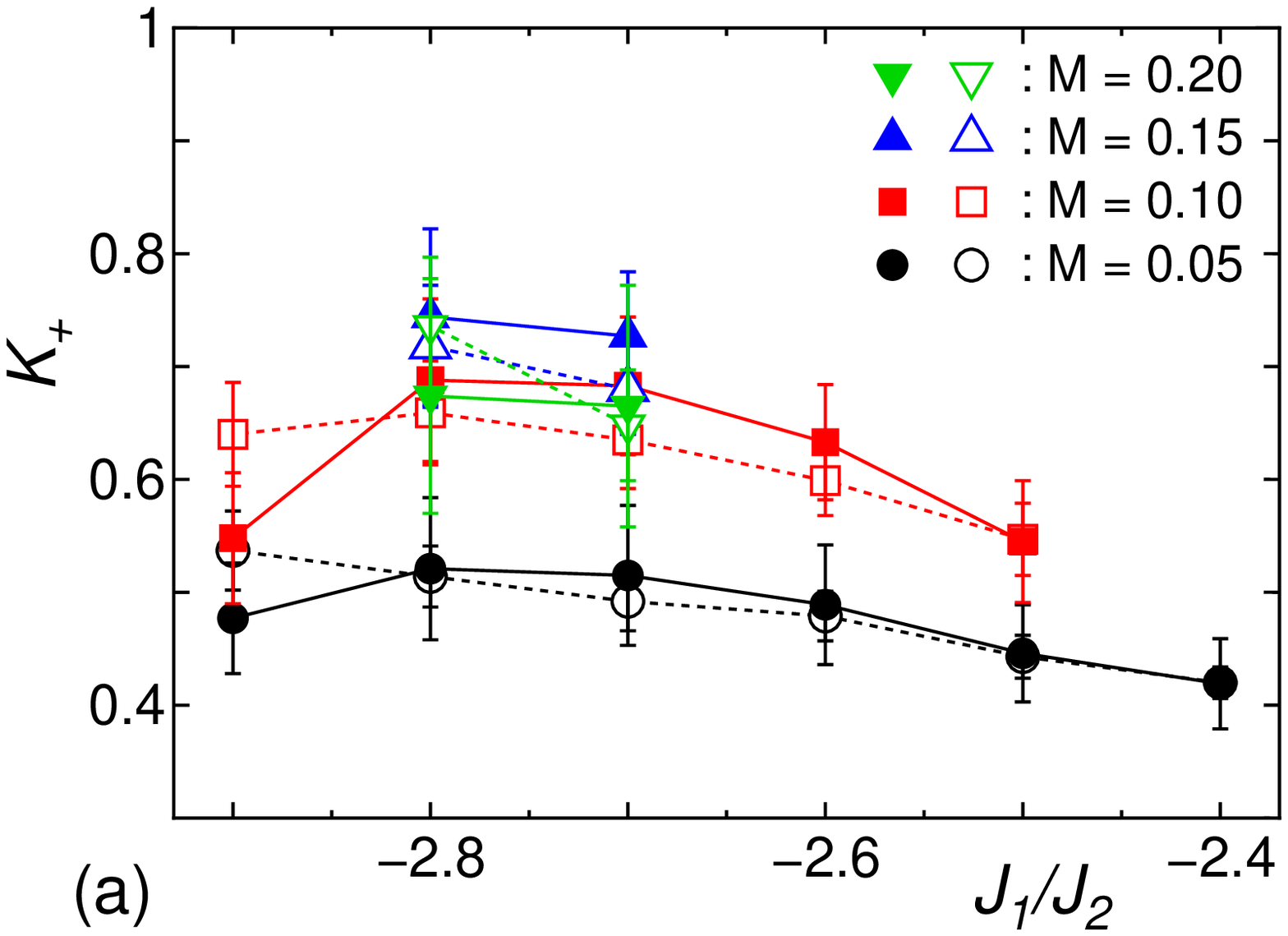}
\includegraphics[width=65mm]{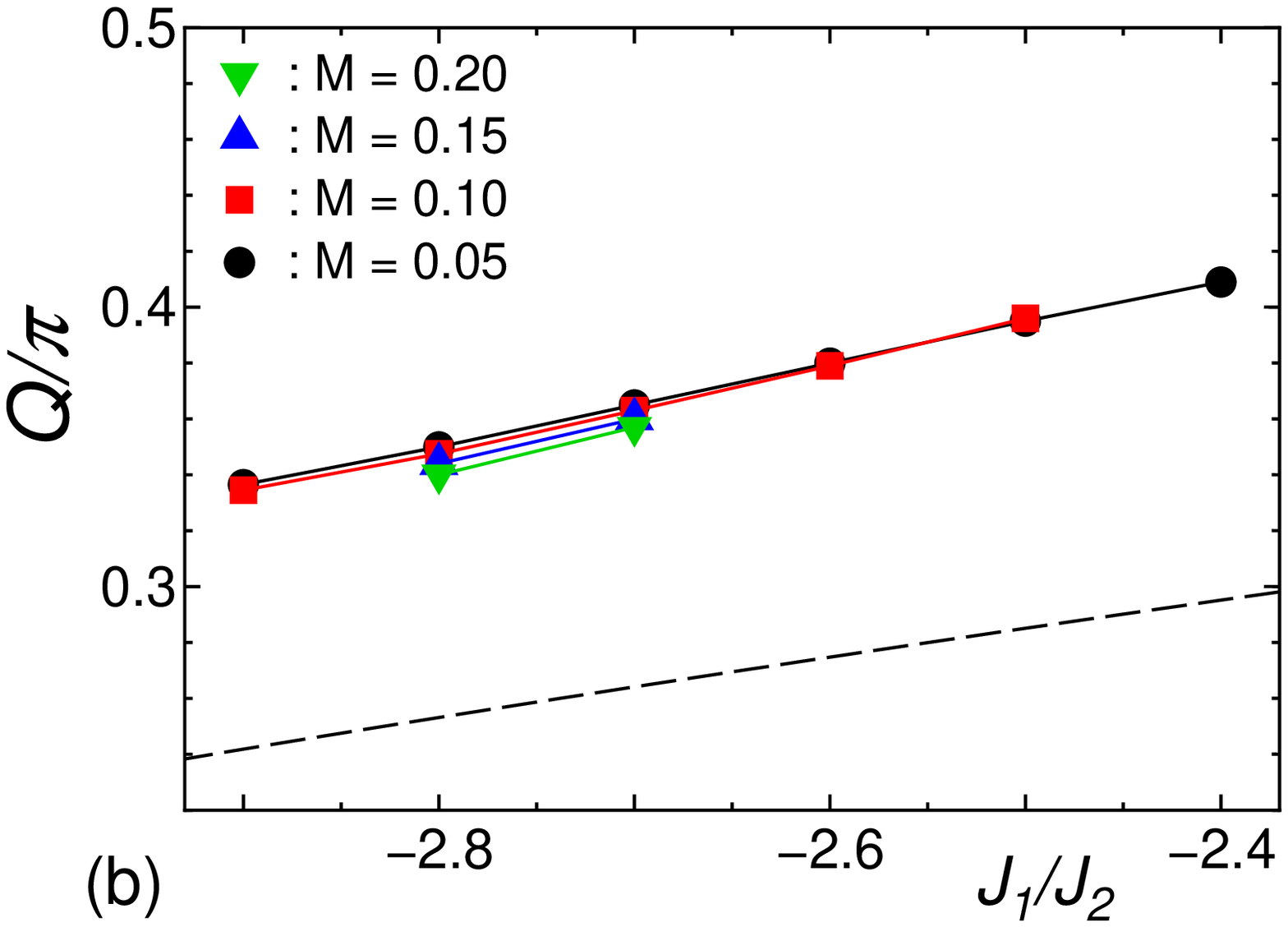}
\caption{
(Color online)
$J_1/J_2$ dependence of (a) the exponent $K_+$
and (b) the incommensurate wave number $Q$ for $L=120$ spin
zigzag chain in the vector chiral phase.
Solid and open symbols in (a) represent the estimates
obtained from the fitting of
$\langle s^x_0 s^x_r \rangle_{\rm av}$
and $\langle s^+_0 s^+_1 s^-_r s^-_{r+1} \rangle_{\rm av}$,
respectively.
The error bars represent the difference of the estimates
obtained from the fitting of the data of different ranges.
In (b), only the results from $\langle s^x_0 s^x_r \rangle_{\rm av}$
are shown since the estimates from $\langle s^x_0 s^x_r \rangle_{\rm av}$
and $\langle s^+_0 s^+_1 s^-_r s^-_{r+1} \rangle_{\rm av}$
are identical to each other within their error bars.
Dashed line in (b) represents the classical pitch angle,
$\phi^c=\arccos(-J_1/4J_2)$.
}
\label{fig:fittingparam-VC}
\end{center}
\end{figure}

We fitted the DMRG data of
the transverse spin and nematic correlation functions
to Eqs.\ (\ref{eq:Cspx-VC}) and (\ref{eq:Cb2m-VC}),
taking $K_+$, $Q$, and the amplitudes $A$ or $A'$ as fitting parameters.
In the fitting procedure, we used the data of the averaged correlation
function $\langle s^x_0 s^x_r \rangle_{\rm av}$
[$\langle s^+_0 s^+_1 s^-_r s^-_{r+1} \rangle_{\rm av}$]
for $L/12 \le r \le L/2$ ($L/6 \le r \le L/2$).
Figures \ref{fig:correlation-VC}(b) and (c)
demonstrate good agreement between numerical data and
the fits to Eqs.\ (\ref{eq:Cspx-VC}) and (\ref{eq:Cb2m-VC}).
Note that both correlation functions are incommensurate.

The exponent $K_+$ and the incommensurate wave number $Q$
obtained from the fitting are shown in Fig.\ \ref{fig:fittingparam-VC}.
The TL-liquid parameter $K_+$ is found to be in the range
$0.4<K_+<0.8$ for the cases we examined numerically, and
the transverse spin correlation function
$\langle s^x_0 s^x_r \rangle_{\rm av}$
is the most slowly decaying correlation function
except for the long-range ordered vector chirality.
This suggests that a magnetic spiral long-range order
in the plane perpendicular to the magnetic field
should be realized in real three-dimensional materials
with additional weak interchain couplings.
The wave number $Q$, which represents the incommensurability
of the transverse spin correlations, shows little dependence on $M$
and decreases as $J_1/J_2$ decreases.
The classical analogue of the wave number $Q$ is
the pitch angle $\phi^c=\arccos(-J_1/4J_2)$,
which shows qualitatively the same feature, but
takes a much smaller value.
We have thus found that the incommensurate wave number is highly renormalized
by quantum fluctuations towards the commensurate value $\pi/2$.

Strictly speaking, the bosonization theory of the previous subsection
is not directly applicable when the interchain ferromagnetic coupling
is strong, $|J_1|/J_2 \gtrsim 1$.
The consistency between the theory and numerics, as demonstrated
by the successful fitting, can be understood, once we postulate that
the vector chiral phase extends 
to the limit $J_1/J_2\to0$, where the bosonization approach is valid,
and that the low-energy physics in this phase is governed by
the same effective theory.

In passing we note that for $J_1/J_2 \lesssim -2.9$ we have not found
clear evidence for long-range order of vector chirality;
although the vector chiral correlation is strong,
it seems to decay slowly at long distances in finite-size systems $L \le 120$.
A possible explanation for this behavior would be
that the energy gap in the antisymmetric sector $(\phi_-, \theta_-)$
is so small that the correlation length becomes very large.
As a result, the bending-down behavior of the vector chiral
correlation functions,\cite{boundary-VC} 
observed near the open boundaries ($r \sim L$) in
Fig.~\ref{fig:correlation-VC}(a), penetrates into the bulk region and
spoils long-range order.
For clarifying the fate of the vector chiral order
for $J_1/J_2\lesssim-2.9$,
calculations for much larger systems are needed.
That is left for future studies.

\section{Nematic phase and spin density wave phase}\label{sec:Nem}

In this section we discuss in detail the nematic phase and
the SDW$_2$ phase,
where two magnons form a bound state with total momentum $k=\pi$.
As we saw in Sec.~\ref{sec:magcurve}
the total magnetization changes in units of
$\Delta S^z_{\rm tot} = 2$ as a result of simultaneous flip
of two spins forming a bound state.
We first review the application of the bosonization theory
described in Sec.~\ref{sec:bosonization-VC} to these phases,
for the sake of completeness of our discussion which partly
complements earlier works.\cite{KolezhukV2005,HeidrichMHV2006,VekuaHMH2007}
We then present an alternative phenomenological
theory\cite{KeckeMF2007} in which a bound magnon pair
is regarded as a hard-core boson.
The theoretical picture developed in these discussions is subsequently
confirmed by numerics.

\subsection{Bosonization theory revisited}

The nematic phase can be described within the bosonization approach
for $|J_1| \ll J_2$.
When the $g_1$ coupling in the effective
Hamiltonian (\ref{eq:Ham-boson-twochain}) is the most relevant,
the field $\phi_-$ is pinned at a value which minimizes the $g_1$ term,
\begin{equation}
\langle\phi_-(x)\rangle
=\sqrt{\frac{\pi}{8}}\left(\frac{1}{2}-M\right).
\label{eq:average phi_-}
\end{equation}
In this case the uniform part of the difference of two neighboring
spins vanishes,
$s^z_l - s^z_{l+1} \sim \sqrt{2/\pi}\partial_x \phi_-=0$,
indicating that two spins are
bound.\cite{CabraHP2000,KolezhukV2005,HeidrichMHV2006,VekuaHMH2007}
The dual field $\theta_-$ is strongly fluctuating, and the $g_2$
is irrelevant.
The fields $\phi_+$ and $\theta_+$ of the symmetric sector remain
gapless and constitute a one-component TL liquid (\ref{H_+}).

The long-distance asymptotic form of correlation functions
can be readily obtained from Eqs.\ (\ref{eq:sz-boson-twochain}),
(\ref{eq:s+-boson-twochain}), (\ref{eq:average phi_-}), and
(\ref{H_+}).
We briefly summarize the results below.

The longitudinal spin correlation has an incommensurate oscillatory
component,
\begin{equation}
\langle s^z_0 s^z_l\rangle
=M^2-\frac{K_+}{\pi^2 l^2}
+\frac{B}{|l|^{K_+}}
\cos\left[\pi l \left(\frac{1}{2}-M\right) \right]
+\cdots,
\label{nematic:szsz}
\end{equation}
where $B$ is a positive constant ($\propto a^2$) and
subleading terms are omitted here and in the equations below.
The third term in Eq.\ (\ref{nematic:szsz}) represents
incommensurate spin density wave correlation.

The transverse spin correlation
$\langle s^+_0s^-_l\rangle$ is short-ranged,
whose correlation length is the inverse of the gap in the
$(\phi_-,\theta_-)$ sector.
Physically, this gap corresponds to the binding energy of
the two-magnon bound state.

The composite operator $s^-_ls^-_{l+1}$ creating a two-magnon
bound state represents the nematic order.
The nematic correlation is alternating and quasi-long-ranged,
\begin{eqnarray}
&&\langle s^+_0s^+_1s^-_ls^-_{l+1}\rangle
\nonumber \\
&&~~~=\frac{B'(-1)^l}{|l|^{1/K_+}}
-\frac{B'' (-1)^l}{|l|^{K_+ + 1/K_+}}
\cos\left[\pi l \left(\frac{1}{2}-M\right) \right]
+ \cdots
\nonumber \\
&&\label{nematic:nematic}
\end{eqnarray}
with $B'$ and $B''$ are positive constants.
Here we note that two down spins are created
and annihilated at neighboring sites in Eq.~(\ref{nematic:nematic}).
In fact, the algebraic decay with the same exponent can be obtained
as long as both distances between the created down spins and
between the annihilated ones are odd integers.
However, when these separations are even, the nematic
correlation functions,
$\langle s^+_0s^+_{2n}s^-_l s^-_{l+2n'}\rangle$ ($n,n'\ll l$),
are expected to be weaker,
as they involve the gapped $\theta_-$ field in lowest order.
This is in accordance with the observation we made in
Sec.~\ref{sec:multimagnon} that the wave function of the two-magnon
bound state with $k=\pi$ at a saturation field is a linear combination
of the states in which the distance between two down spins is
restricted to odd integers.

Comparing Eqs.~(\ref{nematic:szsz}) and (\ref{nematic:nematic}),
we find that the nematic correlation (\ref{nematic:nematic}) is
the most dominant correlation if $K_+>1$.
This phase is called nematic phase.
On the other hand, if $K_+<1$, the most dominant correlation
is the longitudinal spin correlation (\ref{nematic:szsz}).
In this case we have the SDW$_2$ phase.

The vector chiral correlation $\kappa^{(2)}$ shows a power-law decay,
\begin{equation}
\langle\kappa_0^{(2)}\kappa_l^{(2)}\rangle
=-\frac{c_2^2}{K_+ l^2}+\cdots.
\end{equation}
The same $1/l^2$ decay (with a different prefactor) is expected for
$\langle\kappa^{(1)}_0\kappa^{(1)}_l\rangle$,
as the operator product of $\sin(\sqrt{2\pi}\theta_-)$ and the irrelevant
$g_2$ term in the Hamiltonian $\widetilde{\mathcal{H}}$ generates
the $d\theta_+/dx$ operator.

In his pioneering paper,\cite{Chubukov1991} Chubukov suggested
that the nematic phase should have spontaneous dimerization,
$\langle s^z_l(s^z_{l+1}-s^z_{l-1})\rangle\propto(-1)^l$.
However, in the bosonization theory the dimerization operator
is proportional to $\cos(\sqrt{8\pi}\phi_-+\pi M)$, whose average
vanishes because of Eq.\ (\ref{eq:average phi_-}).
We thus conclude that the nematic phase does not have a
spontaneous dimerization.

\subsection{Hard-core Bose gas of bound magnons}\label{sec:bosegas-Nem}

As we discussed in Sec.~\ref{sec:multimagnon},
when $-2.7\lesssim J_1/J_2<0$,
the fully polarized state becomes unstable as a result of formation
of two-magnon bound states
at the saturation field $h_\mathrm{s}$.\cite{Chubukov1991,CabraHP2000,HeidrichMHV2006,KeckeMF2007,VekuaHMH2007}
Below $h_\mathrm{s}$, bound magnon pairs collectively form
a TL liquid with nematic correlation
as well as incommensurate,
longitudinal spin correlation.
Here we develop a phenomenological theory for the phases which
emerge as a result of proliferation of $p$-magnon bound states 
with momentum $k=\pi$,
by assuming that tightly bound $p$-magnons can be treated
as a hard-core boson\cite{Igarashi} and
ignoring internal structure of the bound states.
This is expected to be a good approximation as long as the density
of hard-core bosons is very low, i.e., near the saturation field.
As we see below, for the $p=2$ case, this theory is equivalent to
the $(\phi_+,\theta_+)$ sector of the bosonization theory.

We denote creation and annihilation operators of a hard-core boson by
$b^\dagger_l$ and $b^{}_l$.
Under the assumption that $p$ magnons are tightly bound,
we may relate the creation operator and density operator of bosons
to spin operators,
\begin{eqnarray}
&&
b^\dagger_{\bar l} = (-1)^l s^-_l \cdots s^-_{l+p-1},
\label{eq:bdagger-bosegas} \\
&&
b^\dagger_l b_l^{} = \frac{1}{p}\left( \frac{1}{2}-s^z_l \right),
\label{eq:bdaggerb-bosegas}
\end{eqnarray}
where the $(-1)^l$ factor in Eq.\ (\ref{eq:bdagger-bosegas})
is introduced because the total momentum of bound $p$ magnons
is $k=\pi$; see Sec.~\ref{sec:multimagnon}.
In Eq.~(\ref{eq:bdagger-bosegas}) we identify the site index
$\bar l$ of the boson creation operator $b^\dagger_{\bar l}$
with the center-of-mass coordinate of bound magnons,
$\bar l=l+(p-1)/2$.
From Eq.~(\ref{eq:bdaggerb-bosegas}) we find the density of bosons,
\begin{equation}
\rho=\frac{1}{p}\left(\frac{1}{2}-M\right).
\label{eq:density-bosegas}
\end{equation}
At small but finite density $0<\rho\ll 1$ the hard-core bosons
are a TL liquid at low energy.\cite{giamarchi_book}
Its low-energy effective theory is again a free field theory,
\begin{equation}
\mathcal{H}_0=\frac{v}{2}\int^\infty_{-\infty} d{\bar x}
\left[
K\left(\frac{d\theta}{d{\bar x}}\right)^2
+
\frac{1}{K}\left(\frac{d\phi}{d{\bar x}}\right)^2
\right],
\label{H_0}
\end{equation}
where the bosonic fields $(\phi,\theta)$ play the same
role as the $(\phi_+,\theta_+)$ fields in the $p=2$ case.
Here, we take the lattice spacing between $l$-th and $(l+1)$-th sites 
to be unity and identify $l$ (or $\bar{l}$) with ${\bar x}$.
The TL-liquid parameter $K$ depends on
interactions that work between bosons in addition to the short-range
hard-core repulsion.
When hard-core bosons are free, $K=1$.
This should be the case in the low-density
limit $M\to\frac{1}{2}^-$.
In the continuum limit the operators $b_l^\dagger$
and $b_l^\dagger b_l^{}$ are written as\cite{giamarchi_book}
\begin{eqnarray}
&&
b_l^\dagger=\sqrt{\rho} \, e^{i\sqrt{\pi}\theta}
\sum_{n=-\infty}^\infty e^{2in(\pi\rho {\bar x}+\sqrt{\pi}\phi)},
\label{bosonization1}
\\
&&
b^\dagger_l b_l^{}=\rho + \frac{1}{\sqrt\pi}\frac{d\phi}{d{\bar x}}
+ \rho\cos(2\pi\rho {\bar x} + \sqrt{4\pi}\phi)+\cdots.
\quad
\label{bosonization2}
\end{eqnarray}
Using these bosonization formulas, it is straightforward to
calculate correlation functions of hard-core bosons.
Equations (\ref{eq:bdagger-bosegas}) and
(\ref{eq:bdaggerb-bosegas}) allow us to express these correlation
functions with the original spins $\bm{s}_l$;
we thus obtain the longitudinal-spin and $p$-magnon (multipolar)
correlation functions, $\langle s^z_l s^z_{l'} \rangle$
and $\langle s^+_l \cdots s^+_{l+p-1} s^-_{l'} \cdots s^-_{l'+p-1} \rangle$,
from the density-density correlation function and the propagator
of the bosons, respectively.
In the thermodynamic limit $L\to\infty$ we find
\begin{eqnarray}
&&\langle s^z_0 s^z_{l} \rangle =
\left\langle \left( \frac{1}{2}-p b^\dagger_0 b_0^{} \right)
        \left( \frac{1}{2}-p b^\dagger_{l} b_{l}^{} \right) \right\rangle
\nonumber \\
&&~~= M^2 - \frac{p^2 \eta}{4 \pi^2 l^2}
+ \frac{A_z\cos(2\pi \rho l)}{|l|^\eta} + \cdots,
\label{eq:Cspz-bosegas} \\
&&\langle s^+_0 \cdots s^+_{p-1} s^-_{l} \cdots s^-_{l+p-1} \rangle
= (-1)^{l}\langle b_{\bar 0}^{} b^\dagger_{\bar{l}} \rangle
\nonumber \\
&&~~= \frac{A_{\rm m}(-1)^{l}}{|l|^{1/\eta}}
- \frac{\widetilde{A}_{\rm m}(-1)^{l}}{|l|^{\eta+1/\eta}}
\cos(2\pi \rho l)
+ \cdots,
\label{eq:Cbpm-bosegas}
\end{eqnarray}
where $A_z$, $A_\mathrm{m}$, and $\widetilde{A}_\mathrm{m}$ are
positive constants, and the parameter $\eta$ in the exponents
is related to the TL-liquid parameter $K$ by $\eta=2K$.
Since creating less than $p$ magnons costs a finite energy,
we expect that the transverse-spin correlation functions
$\langle s^x_l s^x_{l'} \rangle$
and, more generally,
$\langle s^+_l\cdots s_{l+p'-1}^+s_{l'}^-\cdots s_{l'+p'-1}^-\rangle$
with $p' < p$ should be short-ranged.

When $p=2$,
Eqs.\ (\ref{eq:Cspz-bosegas}) and (\ref{eq:Cbpm-bosegas})
coincide with Eqs.\ (\ref{nematic:szsz}) and (\ref{nematic:nematic}) by
using the relation (\ref{eq:density-bosegas}) and setting the exponent
$\eta = K_+$.
That is, the two theoretical approaches, the weak-coupling bosonization
theory\cite{VekuaHMH2007} for $|J_1|\ll J_2$ and
the phenomenological hard-core boson theory\cite{KeckeMF2007}
for $\frac{1}{2}-M\ll 1$, give a consistent description of the
nematic and SDW$_2$ phases.
This is in fact expected, as the nematic TL liquid extends
from the saturation limit ($M\to\frac12$) to the weak inter-chain
coupling regime $|J_1| \ll J_2$;
see Fig.~\ref{fig:phasediagram}.

To compare the above theoretical results with numerical data
from DMRG calculation,
we need to modify Eqs.\ (\ref{eq:Cspz-bosegas}) and
(\ref{eq:Cbpm-bosegas}) to include finite-size and boundary
effects.
This can be done by calculating the correlation functions
with Dirichlet boundary conditions on $\phi({\bar x})$.
Here we can borrow results of such calculations from
Refs.\ \onlinecite{HikiharaF2001} and \onlinecite{HikiharaF2004},
in which correlation functions of the spin-$\frac12$ XXZ model
in magnetic field are obtained for open spin chains of length $L$,
once we notice the mapping of the hard-core boson system
onto the spin-$\frac12$ XXZ chain
[$S_l^-=(-1)^l b_l$ and $S^z_l=b^\dagger_lb_l^{}-\frac12$].
In this way we obtain local spin polarization $\langle s^z_l\rangle$
in the $J_1$-$J_2$ spin chain of length $L$,
\begin{equation}
\langle s^z_l \rangle
= \frac{1}{2}(1 - p) - p z(l;q),
\label{eq:szl-bosegas-finite}
\end{equation}
where
\begin{eqnarray}
z(l;q) \!\!&=&\!\!
\frac{q}{2\pi} - a \frac{(-1)^l \sin(ql)}{f_{\eta/2}(2l)},
\label{eq:smallz-bosegas-finite} \\
q \!\!&=&\!\! \frac{2\pi L}{L+1} \left(\rho - \frac{1}{2} \right),
\label{eq:q-bosegas-finite} \\
f_\nu(x) \!\!&=&\!\!
\left[ \frac{2(L+1)}{\pi}
\sin\left(\frac{\pi |x|}{2(L+1)}\right)\right]^\nu.
\label{eq:f-bosegas-finite}
\end{eqnarray}
The site dependence of the polarization $\langle s^z_l\rangle$
comes from Friedel oscillations at open boundaries.
Such oscillations are absent under periodic boundary conditions.
The characteristic wave vector ``2$k_F$'' in the Friedel oscillations is found
from Eqs.\ (\ref{eq:szl-bosegas-finite}),
(\ref{eq:smallz-bosegas-finite}), and (\ref{eq:q-bosegas-finite})
for $L\gg1$ to be
\begin{equation}
2k_F=2\pi\rho,
\label{2k_F}
\end{equation}
which is determined by the density of the hard-core bosons,
and is inversely proportional to the number $p$ of magnons
forming a bound state [see the relation (\ref{eq:density-bosegas})].
This result can be easily checked by DMRG calculation.

The longitudinal spin correlation function in the finite
$J_1$-$J_2$ spin chain is given by
\begin{eqnarray}
\langle s^z_l s^z_{l'} \rangle
\!\!&=&\!\!
\frac{(p-1)^2}{4} + \frac{p(p-1)}{2} \left[ z(l;q) + z(l';q) \right]
\nonumber \\
&&+ p^2 Z(l,l';q),
\label{eq:Cspz-bosegas-finite}
\end{eqnarray}
where
\begin{widetext}
\begin{eqnarray}
Z(l,l';q) \!\!&=&\!\!
\left( \frac{q}{2\pi} \right)^2
- \frac{\eta}{4\pi^2}
\left[ \frac{1}{f_{2}(l-l')} + \frac{1}{f_{2}(l+l')} \right]
- \frac{aq}{2\pi} \left[ \frac{(-1)^l \sin(ql)}{f_{\eta/2}(2l)}
+ \frac{(-1)^{l'} \sin(ql')}{f_{\eta/2}(2l')} \right]
\nonumber \\
&&
+ \frac{(-1)^{l-l'} a^2}{2 f_{\eta/2}(2l) f_{\eta/2}(2l')}
\left\{ \cos[q(l-l')] \frac{f_{\eta}(l+l')}{f_{\eta}(l-l')}
\right.
\left.
- \cos[q(l+l')] \frac{f_{\eta}(l-l')}{f_{\eta}(l+l')} \right\}
\nonumber \\
&&
- \frac{a\eta}{2\pi}
\left\{
  \frac{(-1)^l \cos(ql)}{f_{\eta/2}(2l)} [g(l+l') + g(l-l')]
\right.
\left.
+ \frac{(-1)^{l'} \cos(ql')}{f_{\eta/2}(2l')} [g(l+l') - g(l-l')]
\right\}
\label{eq:capitalZ-bosegas-finite}
\end{eqnarray}
\end{widetext}
with
\begin{equation}
g(x) = \frac{\pi}{2(L+1)} \cot\left[ \frac{\pi x}{2(L+1)} \right].
\label{eq:g-bosegas-finite}
\end{equation}
From Eqs.\ (\ref{eq:szl-bosegas-finite}) and (\ref{eq:Cspz-bosegas-finite}),
the correlation of longitudinal spin fluctuations
is obtained as
\begin{equation}
\langle s^z_l s^z_{l'} \rangle
- \langle s^z_l \rangle \langle s^z_{l'} \rangle
= p^2 \left[ Z(l,l';q) - z(l;q) z(l';q) \right].
\label{eq:Cspz-szsz-bosegas-finite}
\end{equation}
We have two unknown parameters,
$a$ and $\eta$,
in these formulas, which can be obtained by fitting
numerical data to these analytical forms.

Similarly, the multipolar correlation is obtained as
\begin{eqnarray}
&&\langle s^+_l \cdots s^+_{l+p-1} s^-_{l'} \cdots s^-_{l'+p-1} \rangle
\nonumber \\
&&\quad
= A_{\rm m} (-1)^{l-l'}
\frac{f_{1/2\eta}(2l+p-1) f_{1/2\eta}(2l'+p-1)}
{f_{1/\eta}(l-l') f_{1/\eta}(l+l'+p-1)},
\nonumber\\&&
\label{eq:Cbm-bosegas-finite}
\end{eqnarray}
which corresponds to the first term in the
right-hand side of Eq.\ (\ref{eq:Cbpm-bosegas}).
Here fitting parameters are $A_\mathrm{m}$ and $\eta$.

Before closing this section,
we note once again that the phenomenological hard-core boson
theory is applicable to any
phase which appears as a result of the formation of
$p$-magnon bound states with $p=2,3,4,\cdots$ and $k=\pi$.
We will show in the subsequent sections that
the phenomenological theory gives a good description of
correlation functions
not only in the nematic and SDW$_2$ ($p = 2$) phases,
but also in the triatic and SDW$_3$ ($p=3$) phases, and quartic ($p=4$) phase,
which appear for larger $|J_1|/J_2$.
Another advantage of the theory is that it gives a clear intuitive
picture of low-energy excitations.
It is however unable to describe correlations that are related to
internal structures of bound states [such as the feature we
discussed below Eq.\ (\ref{nematic:nematic})].

\subsection{Numerical results}
\label{sec:numerics-Nem}

\begin{figure}
\begin{center}
\includegraphics[width=65mm]{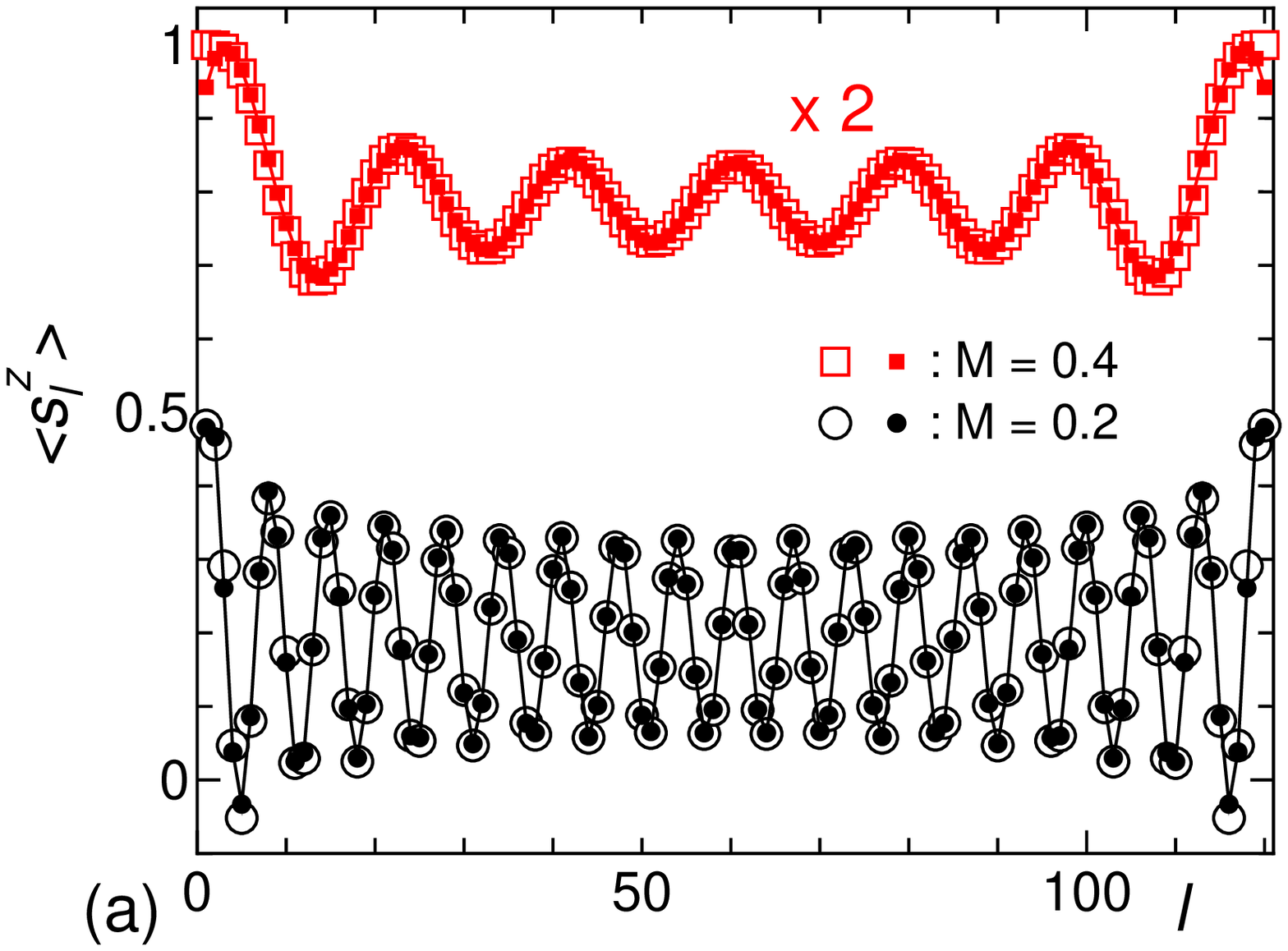}
\includegraphics[width=65mm]{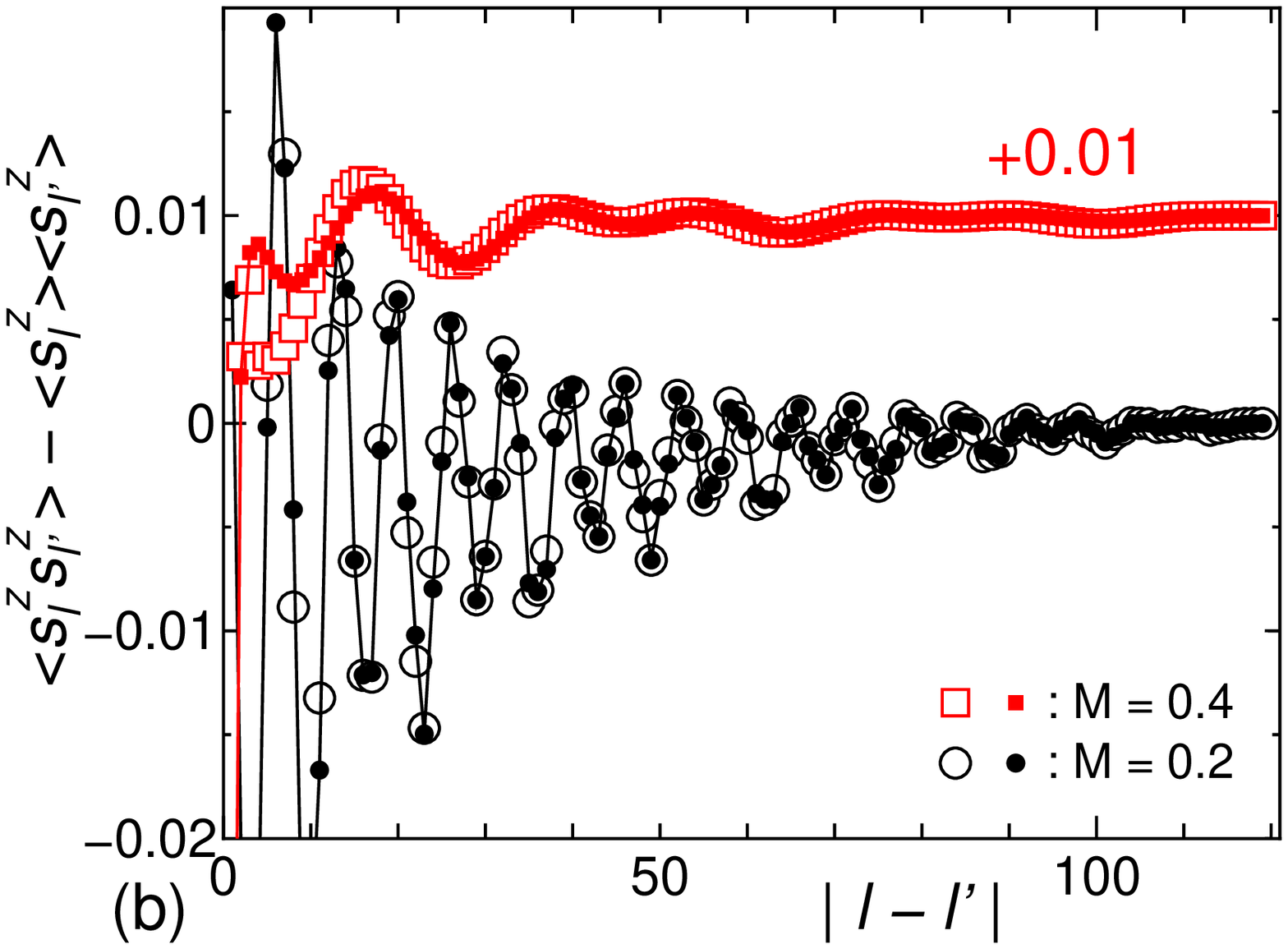}
\includegraphics[width=65mm]{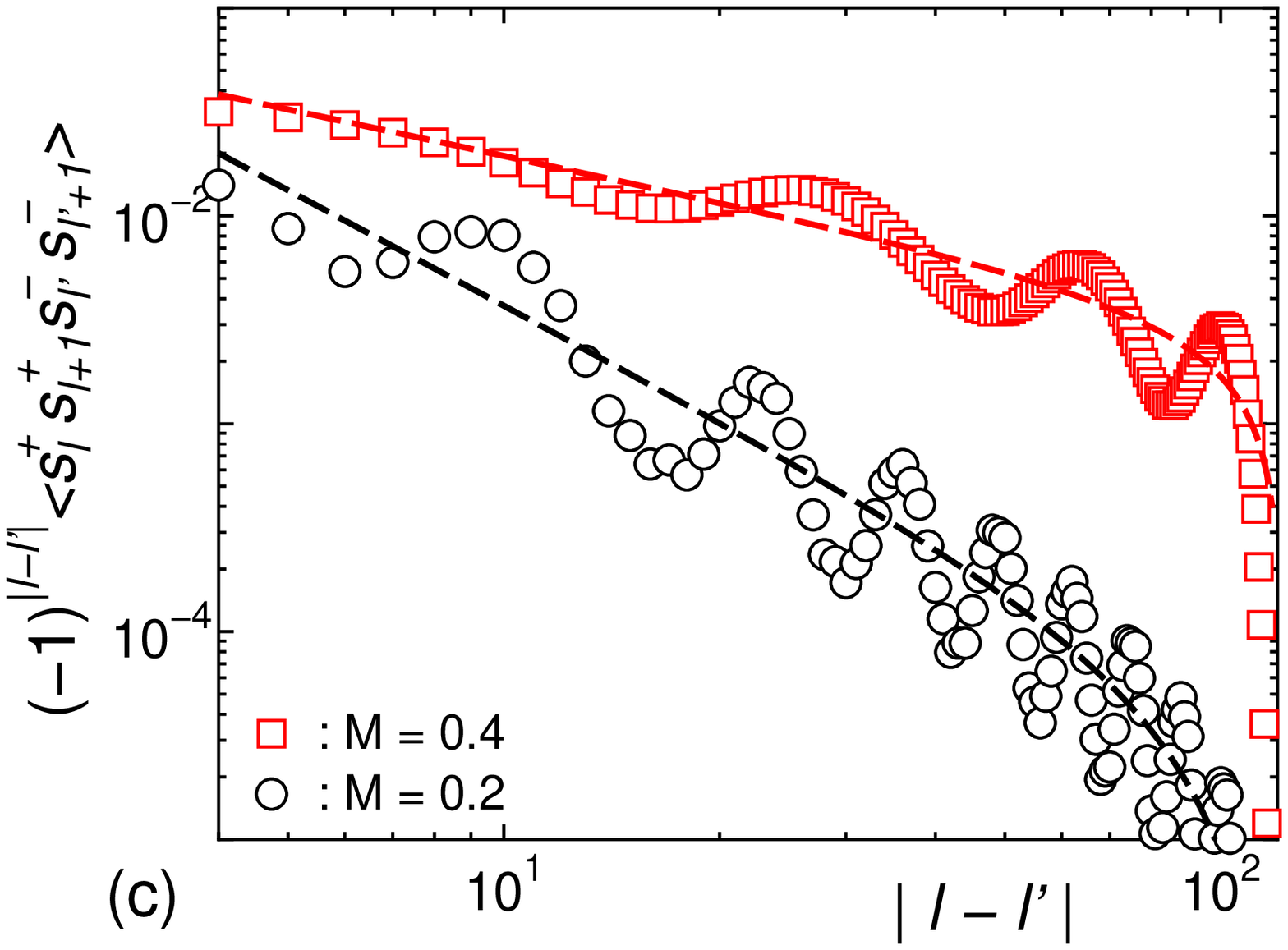}
\includegraphics[width=65mm]{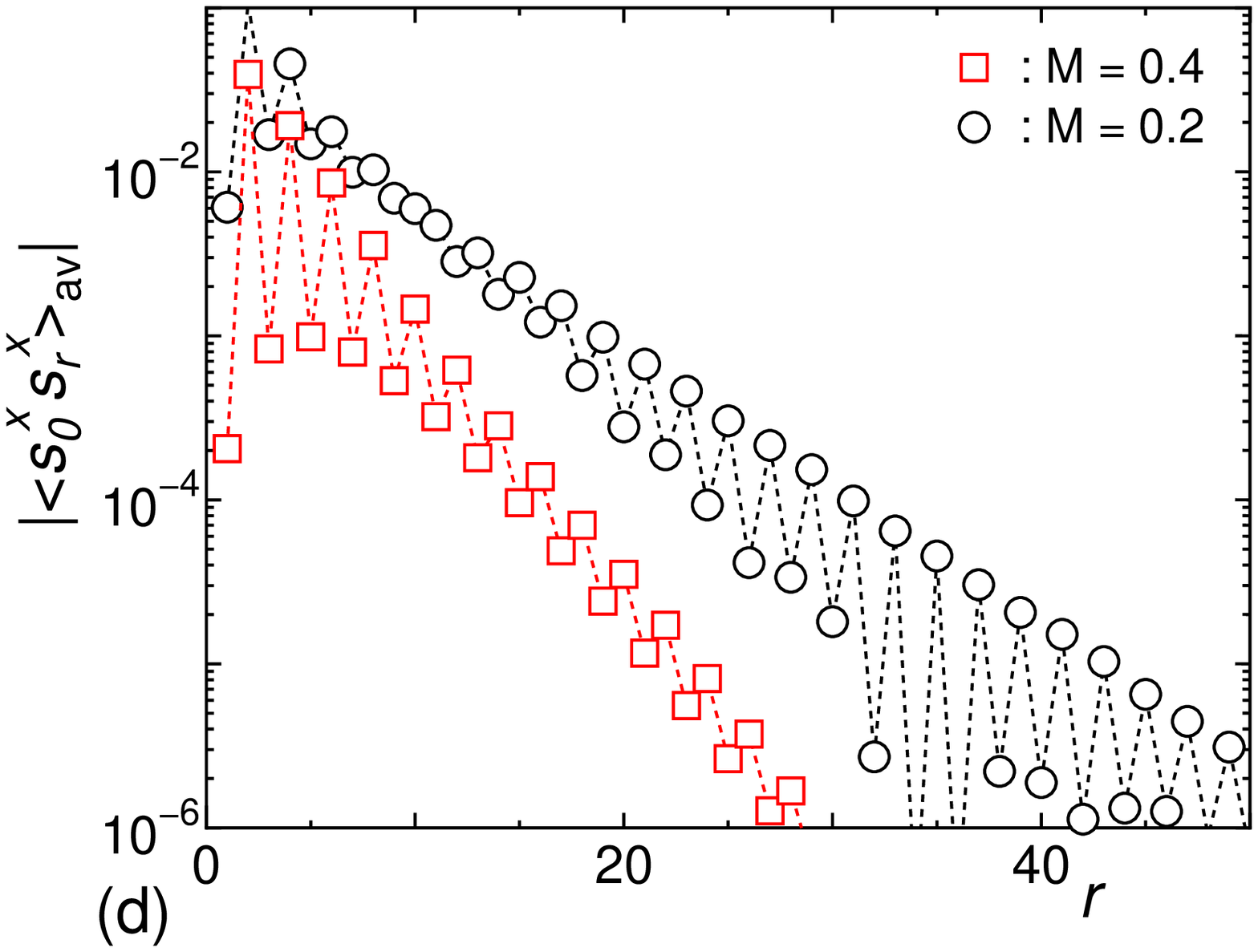}
\caption{
(Color online)
Correlation functions
for $L=120$ spin zigzag chain with $J_1/J_2 = -2.0$, $M=0.2$ and $0.4$,
and their fits to the theory for the nematic and SDW$_2$ phases:
(a) Friedel oscillations in the local spin polarization
$\langle s^z_l \rangle$,
(b) longitudinal spin fluctuation
$\langle s^z_l s^z_{l'} \rangle -
\langle s^z_l \rangle \langle s^z_{l'} \rangle$,
(c) nematic correlation function
$\langle s^+_l s^+_{l+1} s^-_{l'} s^-_{l'+1} \rangle$.
The open symbols represent the DMRG data.
Truncation errors are smaller than the size of the symbols.
In (b) and (c), the data for $l= L/2-[r/2]$
and $l'=L/2 + [(r+1)/2]$ are plotted as a function of $r = |l-l'|$.
The results of the fitting are shown by solid symbols in (a) and (b)
and by dashed curves in (c).
The data for $M=0.4$ are multiplied by a factor 2 in (a)
and shifted by 0.01 in (b).
(d) Absolute values of the averaged transverse-spin
correlation function $\langle s^x_0 s^x_r \rangle_{\rm av}$.
}
\label{fig:correlation-Nem}
\end{center}
\end{figure}

\begin{figure}
\begin{center}
\includegraphics[width=65mm]{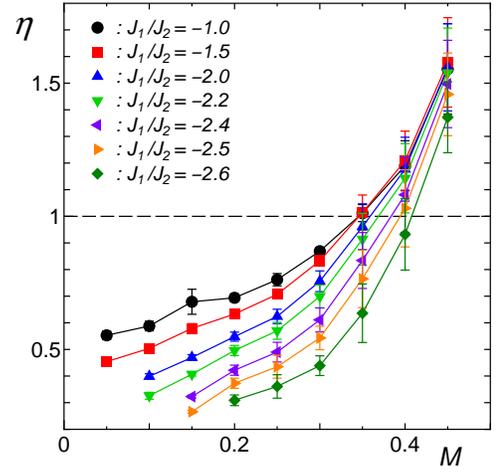}
\caption{
(Color online)
$M$ dependence of the exponent $\eta$ in the nematic phase ($\eta>1$)
and SDW$_2$ phase ($\eta<1$).
The estimates are obtained from the fitting of
$\langle s^z_l s^z_{l'} \rangle$.
The error bars represent the difference of the estimates
obtained from the fitting using the data of different ranges.
}
\label{fig:eta-M-Nem}
\end{center}
\end{figure}

We apply the phenomenological hard-core boson theory to analyze
numerical results in this section.
We use the DMRG method to compute the longitudinal and transverse
spin correlations,
the nematic correlation function, and
the local spin polarization in finite open chains.
We fit the correlation functions to
Eqs.\ (\ref{eq:szl-bosegas-finite})--(\ref{eq:Cbm-bosegas-finite})
with $p=2$,
taking the exponent $\eta$ and the coefficients $a$ and $A_{\rm m}$
as fitting parameters.
Since the formulas already include effects of open boundaries,
here we do not have to take spatial average of
the correlation functions in the fitting procedure.
Finite-size effects are also properly taken
into account in these formulas.
Indeed, we have observed that fitting of numerical results for
$L=96$ and $L=120$ yields the same good quality of agreement
between the numerical data and the fits, and gives essentially
the same estimated values of the fitting parameters.
The results for $L=120$ are shown below.

Figures\ \ref{fig:correlation-Nem}(a) and (b)
show DMRG results of $\langle s^z_l \rangle$ and
$\langle s^z_l s^z_{l'} \rangle -
\langle s^z_l \rangle \langle s^z_{l'} \rangle$
calculated for $J_1/J_2=-2.0$ at $M=0.2$ and $0.4$.
Shown in the same figures are the fits to
Eqs.\ (\ref{eq:szl-bosegas-finite})
and (\ref{eq:Cspz-szsz-bosegas-finite}), respectively, with $p=2$.
We have used numerical data for $6 \le l \le L-5$ to fit
$\langle s^z_l \rangle$ and data for $11 \le |l-l'| \le L-10$
to fit
$\langle s^z_l s^z_{l'} \rangle -
\langle s^z_l \rangle \langle s^z_{l'} \rangle$, and obtained
excellent agreement for both.
(Note that there are only two free parameters $\eta$ and $a$ in the fitting.)
The data of $\langle s^z_l \rangle$ show Friedel oscillations
whose wave length is in good agreement with the theoretical prediction
with $p=2$ (without any fitting parameter).
This is another evidence of the formation of bound magnon pairs.
We observed this consistency in the whole region
of the nematic and SDW$_2$ phases.
The results clearly indicate that
the low-energy physics in this parameter range is indeed described
by the effective theory of hard-core bosons of bound magnon pairs.

For the nematic correlation
$\langle s^+_l s^+_{l+1} s^-_{l'} s^-_{l'+1} \rangle$,
we fit the DMRG result to Eq.\ (\ref{eq:Cbm-bosegas-finite}) with $p=2$,
ignoring the additional oscillating component seen in the DMRG data
which would correspond to
the subleading term in Eq.\ (\ref{eq:Cbm-bosegas-finite}).
We see in Fig.~\ref{fig:correlation-Nem}(c) that
the leading power-law decaying behavior of
$(-1)^{l-l'}\langle s^+_l s^+_{l+1} s^-_{l'} s^-_{l'+1} \rangle$
is fitted rather well by Eq.~(\ref{eq:Cbm-bosegas-finite}).
Figure \ref{fig:correlation-Nem}(d) shows that
the transverse-spin correlation function decays exponentially,
as expected from a finite energy cost for breaking a two-magnon bound state.
These results also support the validity of the effective theory
of hard-core bose gas of bound magnon pairs.

Figure \ref{fig:eta-M-Nem} shows the estimate of $\eta$
obtained from the fitting of $\langle s^z_l s^z_{l'} \rangle$.
As $M$ increases, the exponent $\eta$ increases
across the dashed line $\eta = 1$.
Therefore, the ground state undergoes a crossover from the
low-field SDW$_2$ phase,
where the longitudinal spin correlation function is dominant,
to the high-field nematic phase, where the nematic correlation
dominates.\cite{VekuaHMH2007}
We have found that $\eta$ estimated from the other correlators, 
$\langle s^z_l \rangle$,
$\langle s^z_l s^z_{l'} \rangle 
- \langle s^z_l \rangle \langle s^z_{l'} \rangle$, 
and $\langle s^+_l s^+_{l+1} s^-_{l'} s^-_{l'+1} \rangle$, 
are consistent with Fig.~\ref{fig:eta-M-Nem}.
As $M \to \frac{1}{2}^-$, $\eta$ increases towards $\eta=2$,
in agreement with the theoretical prediction that
the hard-core bosons become free in the dilute limit.

\section{Incommensurate nematic phase}
\label{sec:incommensurate}

\begin{figure}
\begin{center}
\includegraphics[width=65mm]{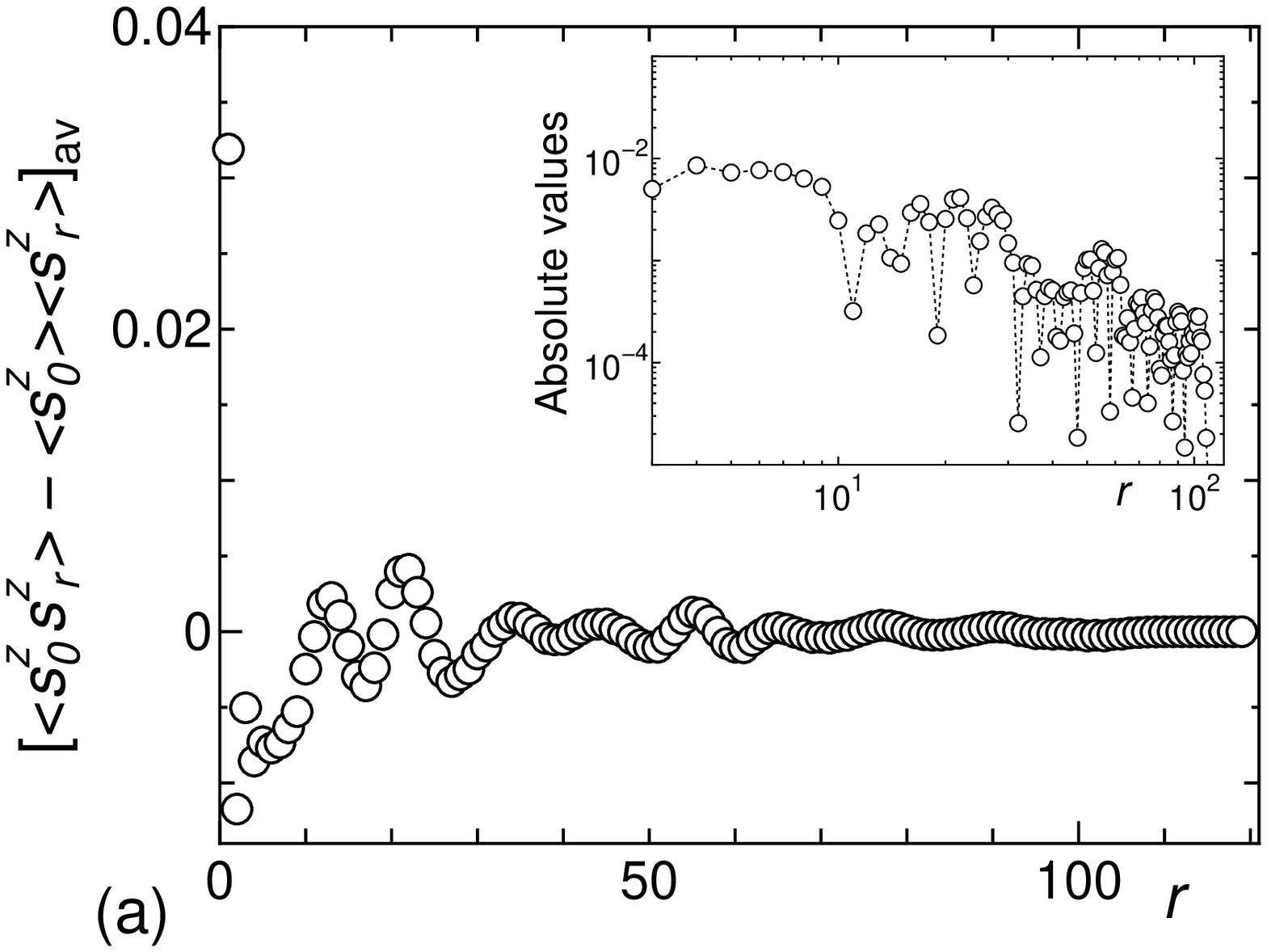}
\includegraphics[width=65mm]{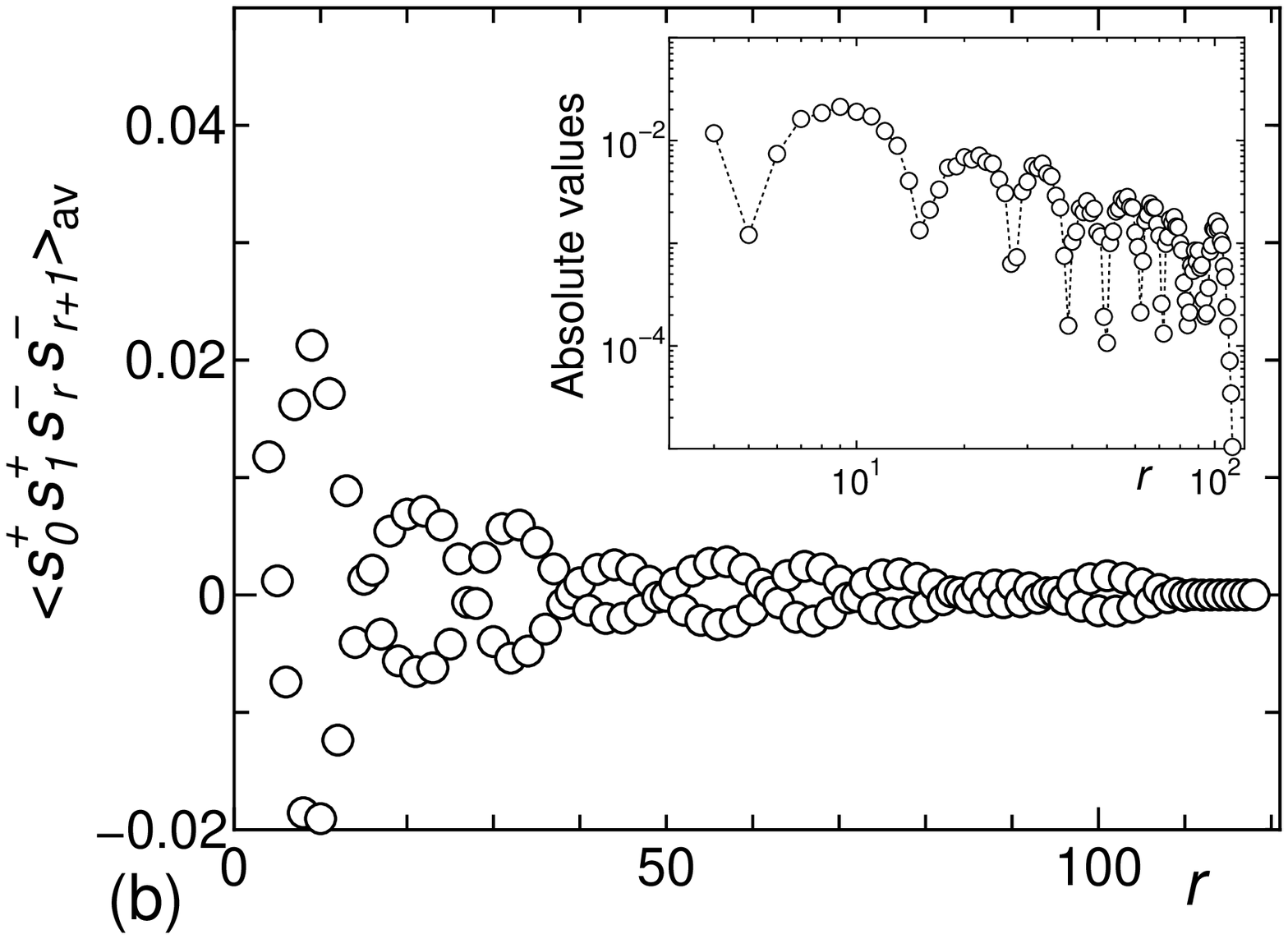}
\includegraphics[width=65mm]{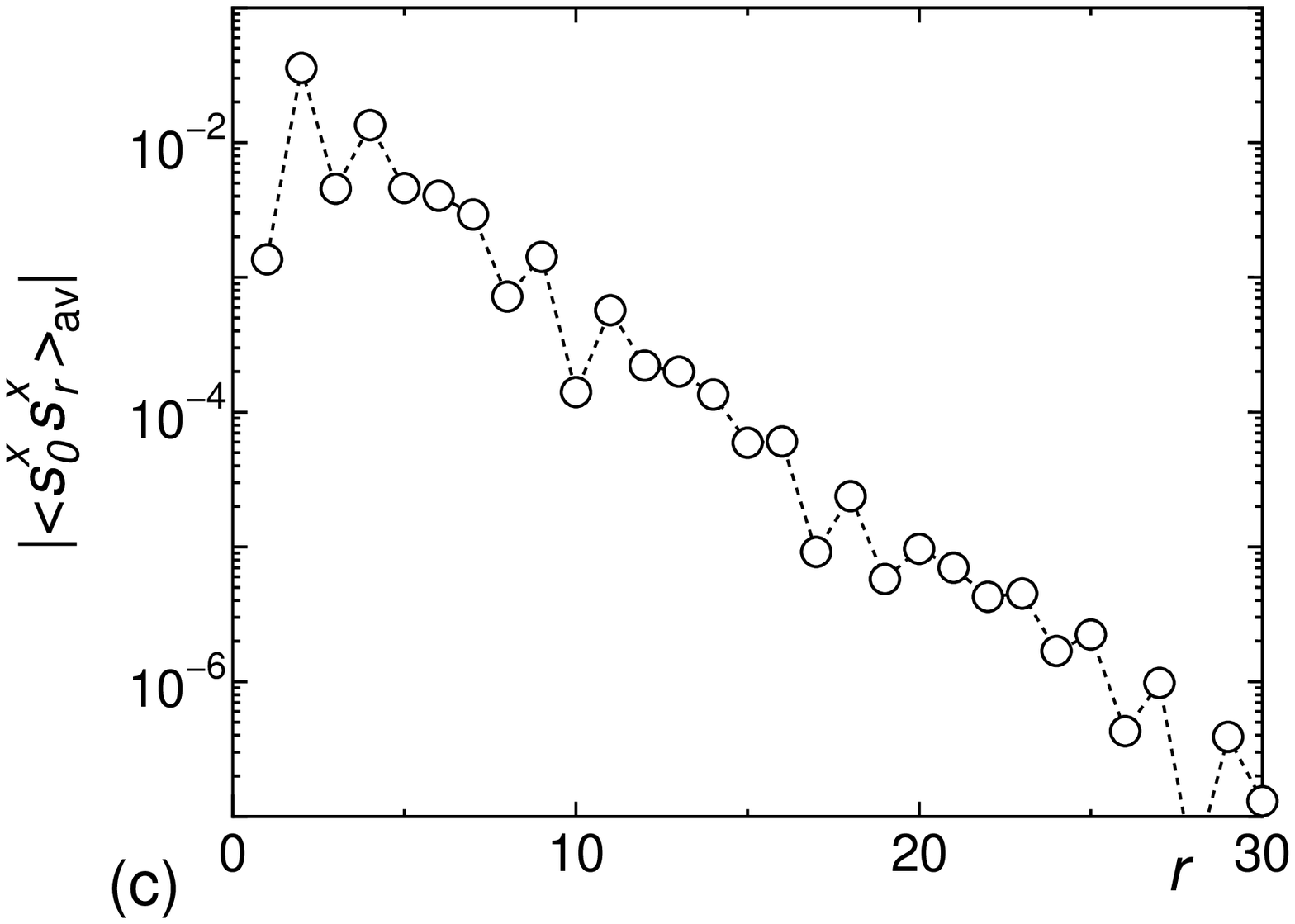}
\includegraphics[width=65mm]{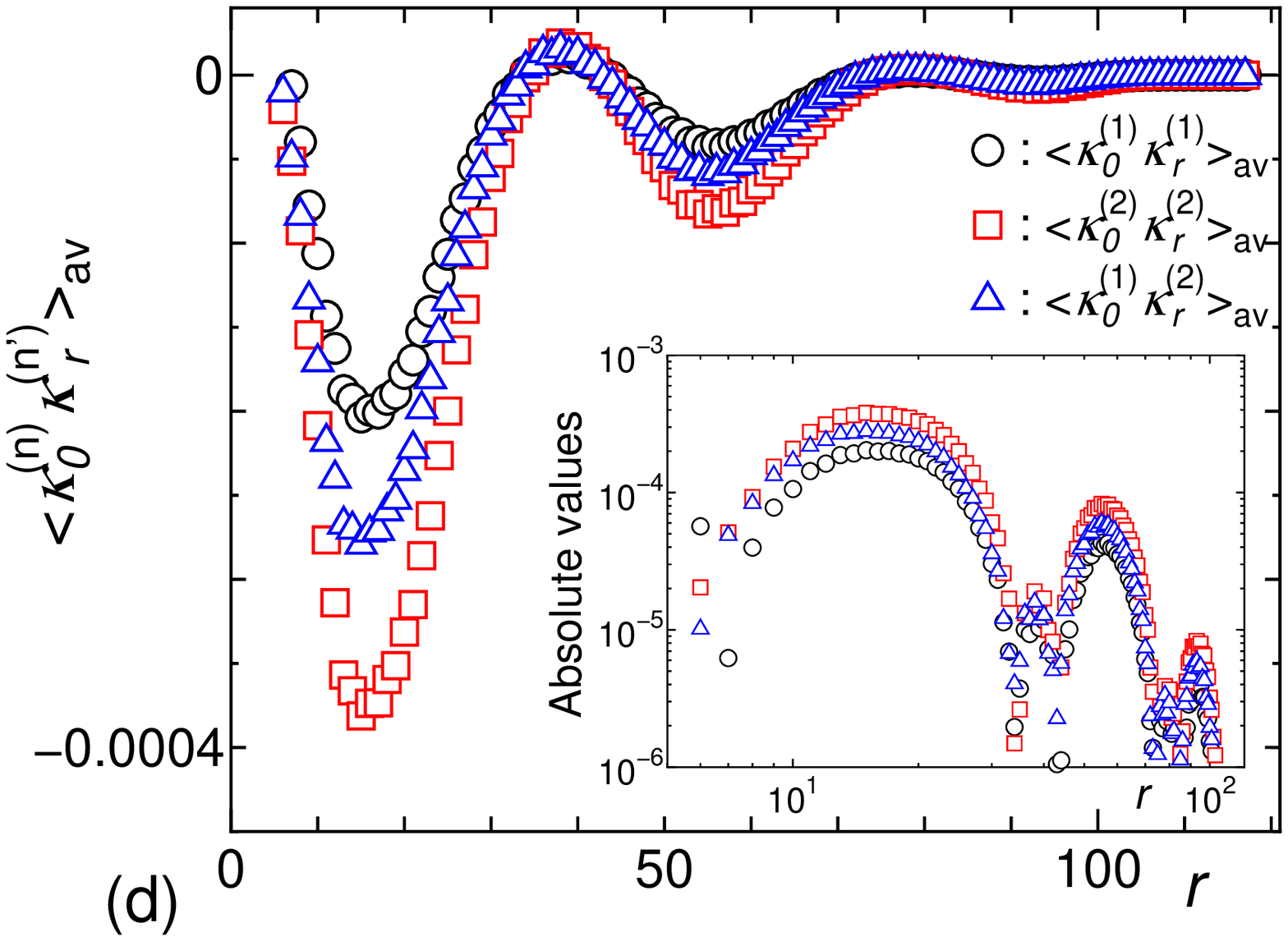}
\caption{
(Color online)
Averaged correlation functions
for $L = 120$ spin zigzag chain with $J_1/J_2 = -2.7$ and $M = 0.4$
in the incommensurate nematic phase;
(a) longitudinal spin fluctuation
$[\langle s^z_0 s^z_r \rangle -
\langle s^z_0 \rangle \langle s^z_r \rangle]_{\rm av}$,
(b) nematic correlation function
$\langle s^+_0 s^+_1 s^-_r s^-_{r+1} \rangle_{\rm av}$,
(c) absolute values of the transverse spin correlation function
$\langle s^x_0 s^x_r \rangle_{\rm av}$, and
(d) vector chiral correlation functions
$\langle \kappa^{(1)}_0 \kappa^{(1)}_r \rangle_{\rm av}$,
$\langle \kappa^{(2)}_0 \kappa^{(2)}_r \rangle_{\rm av}$, and
$\langle \kappa^{(1)}_0 \kappa^{(2)}_r \rangle_{\rm av}$.
Truncation errors are smaller than the size of the symbols.
Insets in (a), (b), and (d) show the absolute values of the data
in a log-log scale.
}
\label{fig:correlation-ICNem}
\end{center}
\end{figure}

\begin{figure}
\begin{center}
\includegraphics[width=75mm]{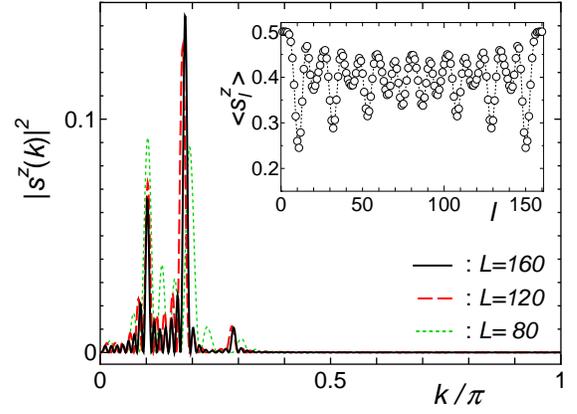}
\caption{
(Color online)
Squared modulus of the Fourier transform, 
$| s^z(k) |^2$,
which is an even function of $k$ and
where 
$s^z(k)=(1/\sqrt{L}) \sum_l e^{ikl} (\langle s^z_l \rangle - M)$, 
for $J_1/J_2 = -2.7$ and $M = 0.4$ in the incommensurate nematic phase.
The solid, dashed, and dotted lines represents the data 
for $L = 160, 120$, and $80$, respectively.
Inset: the local spin polarization $\langle s^z_l \rangle$
for $J_1/J_2 = -2.7$, $M = 0.4$, and $L = 160$.
}
\label{fig:szl-ICNem}
\end{center}
\end{figure}

As we discussed in Sec.~\ref{sec:multimagnon},
for $-2.720 < J_1/J_2 < -2.669$ the fully-polarized state becomes
unstable at the saturation field as a result of formation of
two-magnon bound states with an incommensurate momentum.\cite{KeckeMF2007}
Below the saturation field these bound states are expected to
form a TL liquid with incommensurate nematic correlation.
Such an incommensurate nematic phase was predicted by
Chubukov,\cite{Chubukov1991} who dubbed this phase
the chiral biaxial spin nematic,
as he considered it to have long-range
vector chiral order.
However, our numerical results indicate that the vector chirality
is not long-ranged in the incommensurate nematic phase.

Figure\ \ref{fig:correlation-ICNem} shows our DMRG results
of spatially averaged correlation functions for $J_1/J_2 = -2.7$
and $M = 0.4$.
We see in Fig.~\ref{fig:correlation-ICNem}(a) and (b)
that the correlation of the longitudinal spin fluctuations
$[\langle s^z_0 s^z_r \rangle -
\langle s^z_0 \rangle \langle s^z_r \rangle]_{\rm av}$
and the nematic correlation function
$\langle s^+_0 s^+_1 s^-_r s^-_{r+1} \rangle_{\rm av}$
decay slowly (presumably algebraically)
with an incommensurate modulation.
On the other hand, we find in Fig.~\ref{fig:correlation-ICNem}(c) that
the transverse-spin correlation
$\langle s^x_0 s^x_r \rangle_{\rm av}$ decays exponentially,
indicating the existence of a finite binding energy of
the two-magnon bound pairs.
Finally, Fig.~\ref{fig:correlation-ICNem}(d) shows that
the correlation functions of vector chirality $\kappa_l^{(n)}$
have at most quasi-long-range order with incommensurate
oscillations.

In Fig.\ \ref{fig:szl-ICNem}, we show the local spin polarization 
$\langle s^z_l \rangle$ and its Fourier transform for $L=80, 120$, and $160$.
The Fourier transform exhibits three peaks.
This is in contrast to the cases of the nematic phase in Sec.\ VI and
the triatic and quartic phases (see Sec.\ \ref{sec:TriQua}), 
where the polarization $\langle s^z_l \rangle$ is described
by Eq.\ (\ref{eq:szl-bosegas-finite}) with a single wave number $q$.
The positions of peaks in the Fourier transform are almost independent of $L$, 
suggesting that the incommensurability should not be due to 
finite-size nor open-boundary effects.

We have observed qualitatively the same behaviors of the correlation
functions and spin polarization 
for $J_1/J_2 = -2.7$ and $M > M_{\rm c} \approx 0.35$.
From these results, we conclude that the incommensurate nematic phase
(without chiral long-range order) exists
in the narrow region of the phase diagram;
see Fig.\ \ref{fig:phasediagram}.

Unfortunately, we are not aware of an effective theory
which can give consistent description of these numerical results.
However, in the spirit of the hard-core boson theory in Sec.\ VI,
we may try to treat the two-magnon bound states as hard-core bosons:
\begin{eqnarray}
&&
s_l^-s_{l+1}^-=
e^{i(\pi+\delta)l}b_{1,l}^\dagger
+e^{i(\pi-\delta)l}b_{2,l}^\dagger,
\\
&&
\frac{1}{2}\left(\frac{1}{2}-s^z_l\right)
=b_{1,l}^\dagger b_{1,l}^{}+b_{2,l}^\dagger b_{2,l}^{},
\end{eqnarray}
where $b_{1,l}^\dagger$ and $b_{2,l}^\dagger$ are creation operators
of hard-core bosons with momentum $k=\pi+\delta$ and $\pi-\delta$, 
respectively.
The long-range order of vector chirality would follow if
the average of the boson number difference,
$b_1^\dagger b_1^{}-b_2^\dagger b_2^{}$,
became nonvanishing spontaneously.
Figure~\ref{fig:correlation-ICNem}(d) indicates that this is
not the case, and implies that the low-energy properties in this
phase are determined by two bosonic modes
($b_1^\dagger b_1^{}+b_2^\dagger b_2^{}$ and
 $b_1^\dagger b_1^{}-b_2^\dagger b_2^{}$),
forming a two-flavor TL liquid.

\section{Triatic and Quartic phases}
\label{sec:TriQua}

\begin{figure}
\begin{center}
\includegraphics[width=65mm]{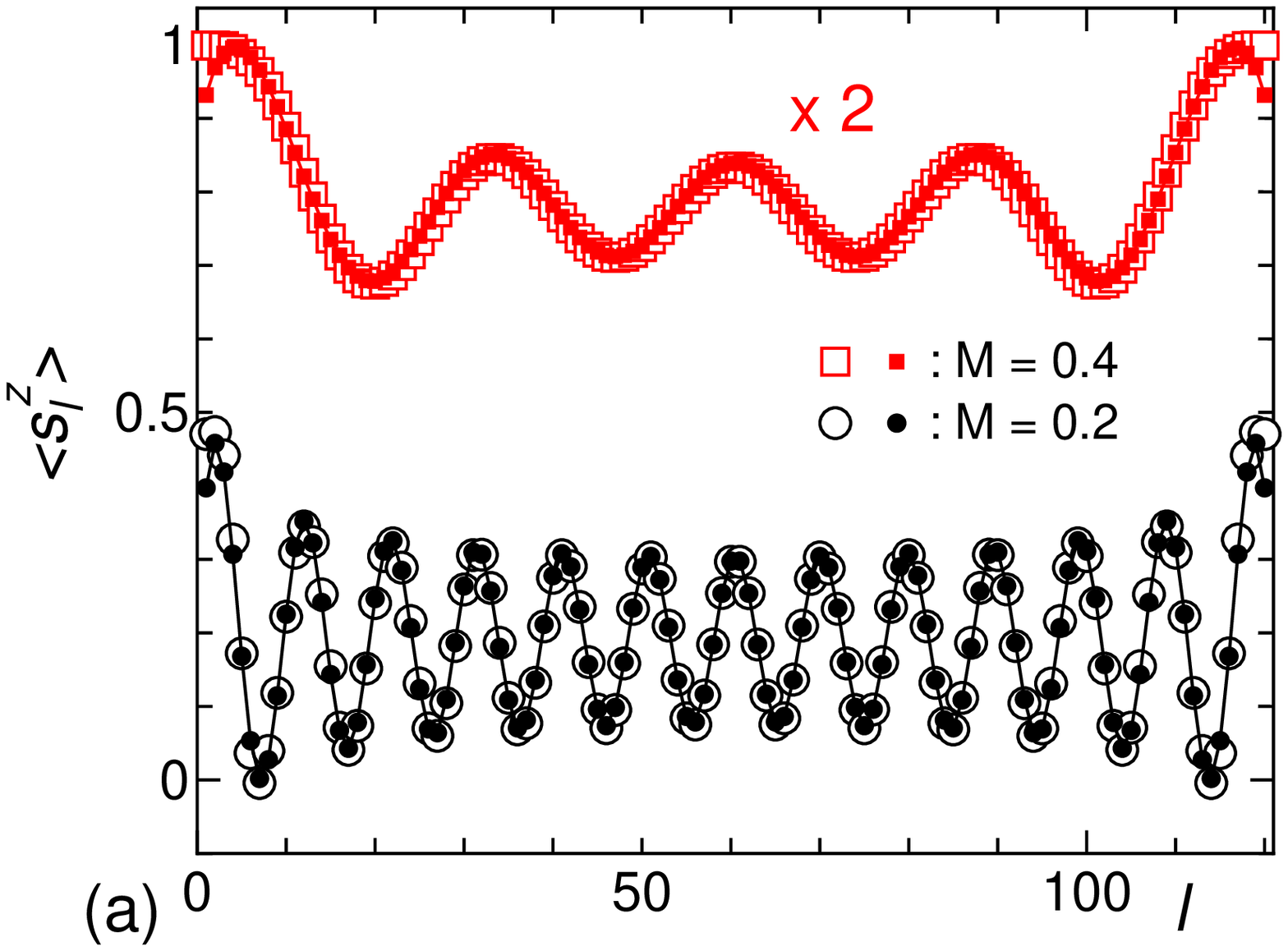}
\includegraphics[width=65mm]{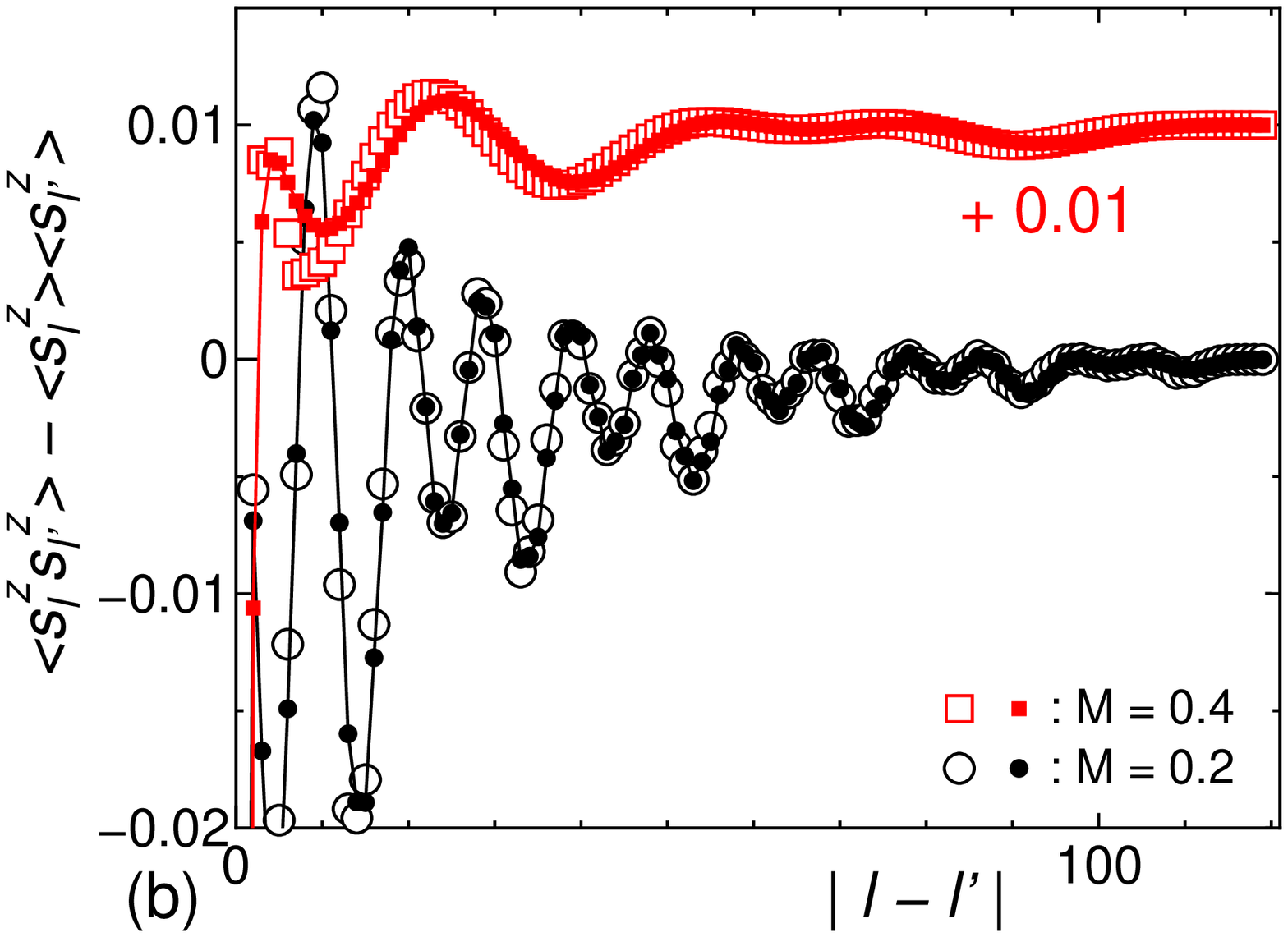}
\includegraphics[width=65mm]{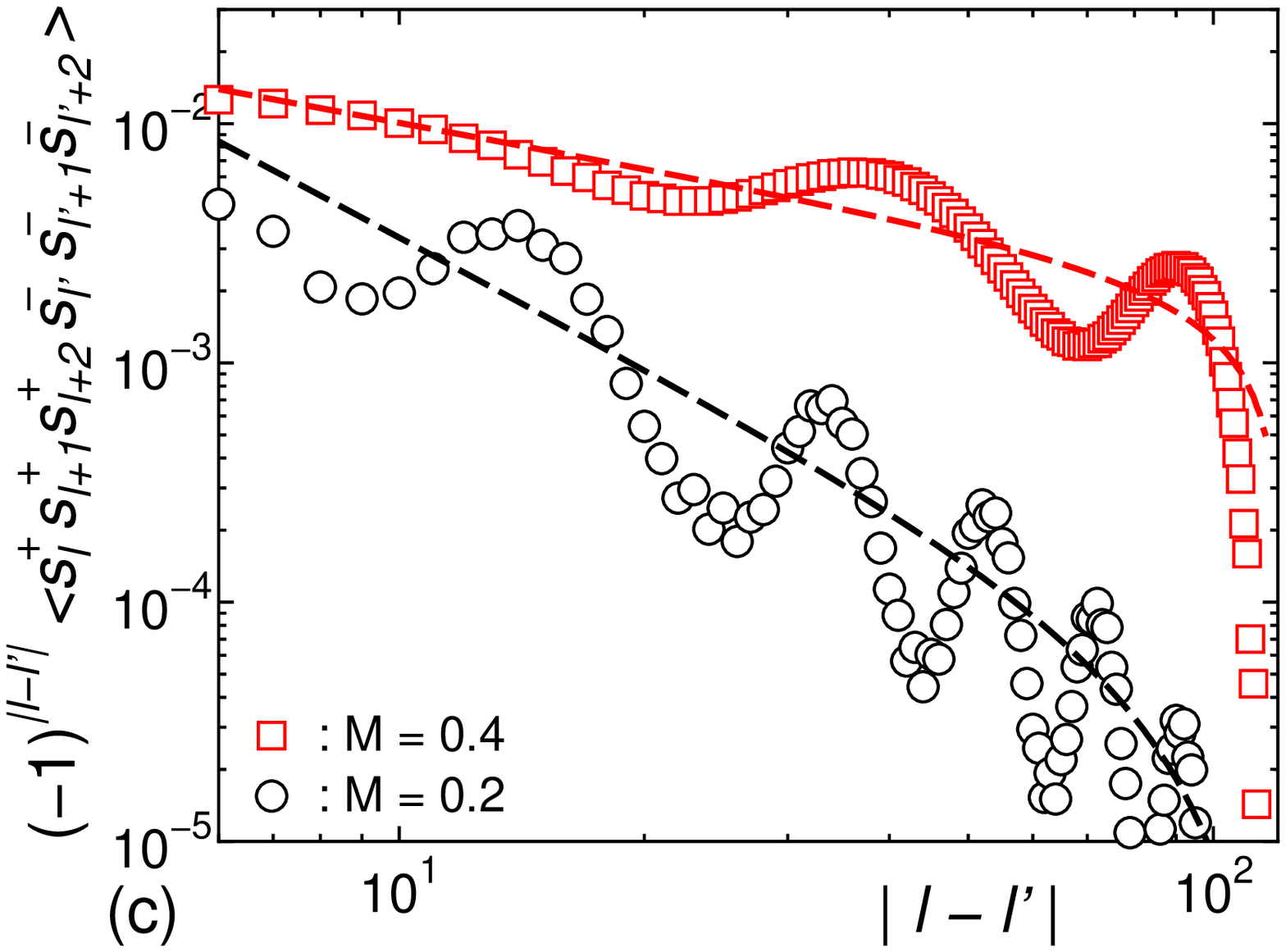}
\includegraphics[width=65mm]{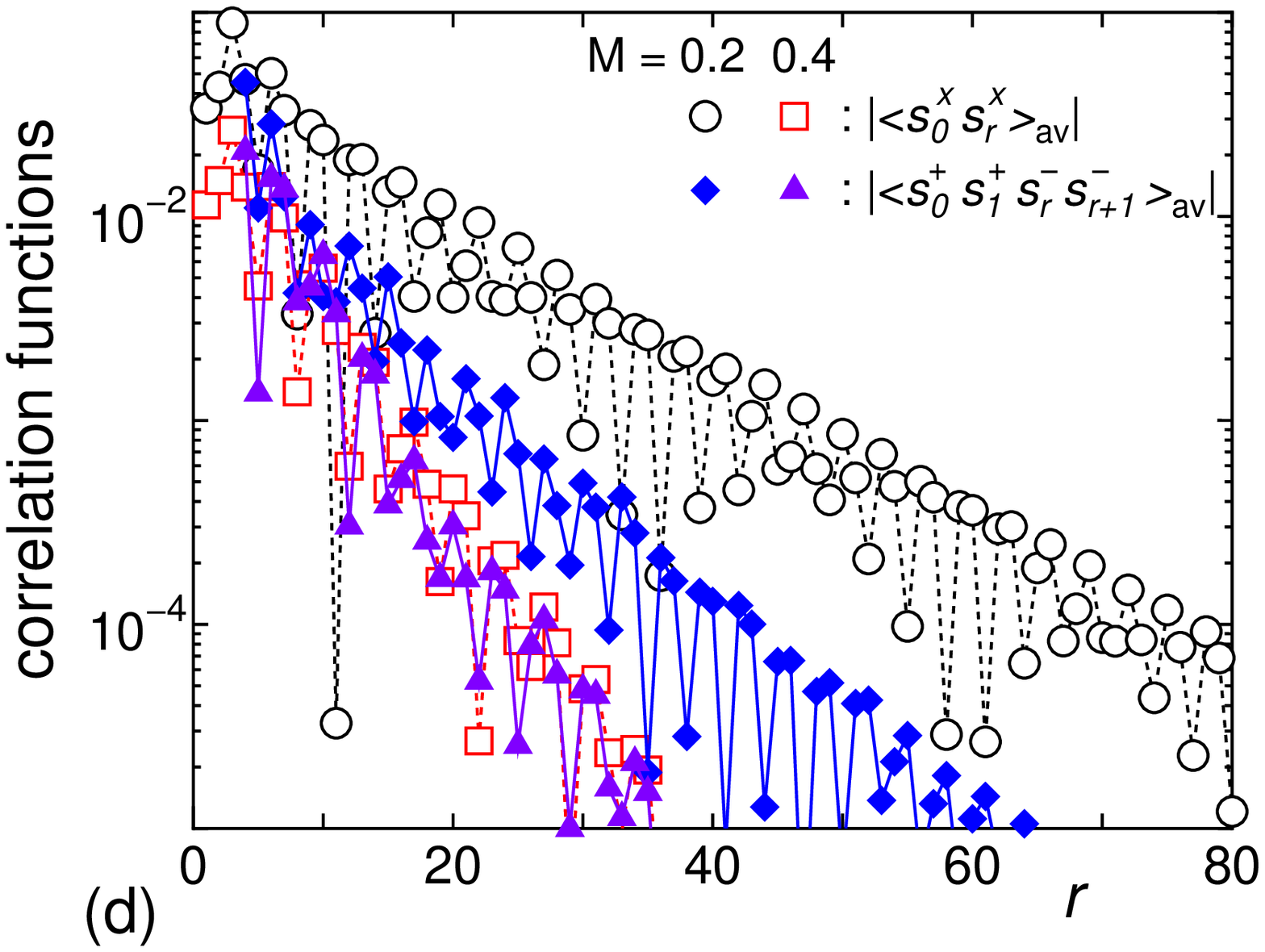}
\caption{
(Color online)
Correlation functions for $L=120$ spin zigzag chain with
$J_1/J_2 = -3.0$, $M=0.2, 0.4$, and their fits to the theory for the triatic phase:
(a) Friedel oscillations in the local spin polarization $\langle s^z_l \rangle$,
(b) longitudinal spin fluctuation
$\langle s^z_l s^z_{l'} \rangle -
\langle s^z_l \rangle \langle s^z_{l'} \rangle$,
(c) triatic correlation function
$\langle s^+_l s^+_{l+1} s^+_{l+2} s^-_{l'} s^-_{l'+1} s^-_{l'+2} \rangle$.
The open symbols represent the DMRG data.
Truncation errors are smaller than the size of the symbols.
In (b) and (c), the data for $l= L/2-[r/2]$
and $l'=L/2 + [(r+1)/2]$ are plotted as a function of $r = |l-l'|$.
The results of the fitting are shown by solid symbols in (a) and (b)
and by dashed curves in (c).
The data for $M=0.4$ are multiplied by a factor 2 in (a)
and shifted by 0.01 in (b).
(d) Absolute values of the averaged transverse-spin and
nematic correlation functions,
$\langle s^x_0 s^x_r \rangle_{\rm av}$
and $\langle s^+_0 s^+_1 s^-_r s^-_{r+1} \rangle_{\rm av}$.
}
\label{fig:correlation-Tri}
\end{center}
\end{figure}

In this section we consider the triatic, SDW$_3$, and quartic phases,
in which
the total magnetization changes by $\Delta S^z_{\rm tot} = 3$ and $4$.
We again apply the hard-core boson theory of Sec.\ \ref{sec:bosegas-Nem}
with $p=3$ and $4$.

Figure\ \ref{fig:correlation-Tri} shows our DMRG results
for the triatic and SDW$_3$ phases.
As a typical example, we chose the coupling ratio $J_1/J_2=-3.0$,
and magnetization per spin $M=0.2$ and 0.4.
We fit the local spin polarization $\langle s^z_l \rangle$
and the longitudinal spin fluctuation correlation
$\langle s^z_l s^z_{l'} \rangle -
\langle s^z_l \rangle \langle s^z_{l'} \rangle$
to Eqs.\ (\ref{eq:szl-bosegas-finite})
and (\ref{eq:Cspz-szsz-bosegas-finite})
with $p=3$, respectively,
taking $\eta$ and $a$ as fitting parameters.
We find that these correlators are fitted
quite well by the formulas.
The fitting of the triatic correlation function
$\langle s^+_l s^+_{l+1} s^+_{l+2} s^-_{l'} s^-_{l'+1} s^-_{l'+2} \rangle$
to Eq.\ (\ref{eq:Cbm-bosegas-finite}) with $p = 3$
also works well, within the approximation that the subleading
oscillating terms are ignored.
The transverse spin and two-magnon (nematic) correlation functions,
$\langle s^x_l s^x_{l'} \rangle$ and
$\langle s^+_l s^+_{l+1} s^-_{l'} s^-_{l'+1} \rangle$,
decay exponentially, in accordance with the theoretical prediction.
All these observations demonstrate the validity
of the bosonic effective theory for the triatic/SDW$_3$ phase.

\begin{figure}
\begin{center}
\includegraphics[width=65mm]{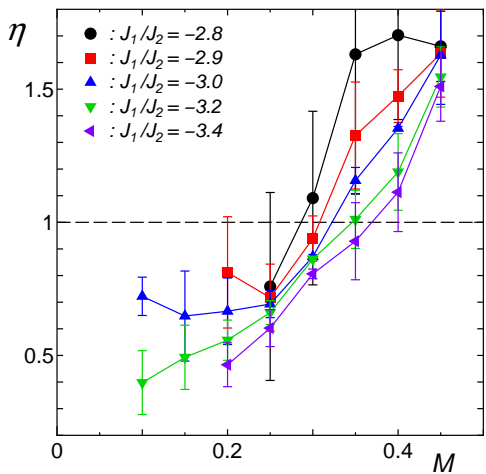}
\caption{
(Color online)
$M$ dependence of the exponent $\eta$
in the triatic phase ($\eta>1$) and SDW$_3$ phase ($\eta<1$).
The estimates are obtained from the fitting of
$\langle s^z_l s^z_{l'} \rangle$ for $J_1/J_2 \ge -3.0$
and $\langle s^z_l \rangle$ for $J_1/J_2 = -3.2, -3.4$.
The error bars represent the difference of the estimates
obtained from the fitting using the data of different ranges.
}
\label{fig:eta-M-Tri}
\end{center}
\end{figure}

Figure\ \ref{fig:eta-M-Tri} shows the exponent $\eta$ 
obtained from the fitting of 
$\langle s^z_l s^z_{l'} \rangle$ for $J_1/J_2 \ge -3.0$ and 
$\langle s^z_l \rangle$ for $J_1/J_2 \le -3.2$.
Although the estimates have rather large error bars,
there is a clear tendency that $\eta$ increases from $\eta < 1$ to
$\eta > 1$ as $M$ increases.
The estimates of $\eta$ obtained from the other correlators, 
including the triatic correlation 
$\langle s^+_l s^+_{l+1} s^+_{l+2} s^-_{l'} s^-_{l'+1} s^-_{l'+2} \rangle$
for $J_1/J_2 \ge -3.0$, exhibit essentially the same feature.
The result indicates that the ground state undergoes
a crossover from the low-field SDW$_3$ phase with
the dominant longitudinal-spin correlation to
the high-field triatic phase where the three-magnon (triatic) correlation
is dominant.
The behavior of $\eta$ at large $M$ is also consistent 
with the theoretical prediction that $\eta\to2$ as $M\to\frac{1}{2}^-$.

\begin{figure}
\begin{center}
\includegraphics[width=65mm]{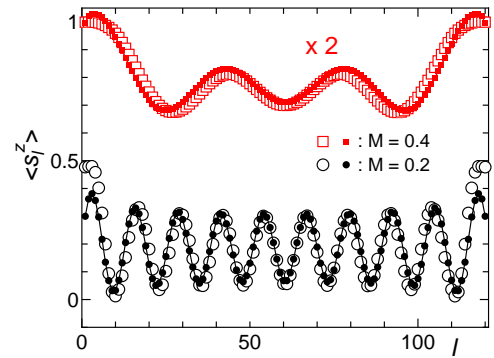}
\caption{
(Color online)
Friedel oscillations in the local spin polarization $\langle s^z_l \rangle$
for $L=120$ spin zigzag chain with $J_1/J_2 = -3.6$ and $M=0.2, 0.4$, and
their fits to the theory for the quartic phase.
The open symbols represent the DMRG data and
solid symbols show the results of the fitting.
Truncation errors of the DMRG data are smaller than the size of the symbols.
The data for $M=0.4$ are multiplied by a factor 2.
}
\label{fig:correlation-Qua}
\end{center}
\end{figure}

For the quartic phase, we show the local spin polarization
$\langle s^z_l \rangle$ calculated at $J_1/J_2 = -3.6$
and $M=0.2$ and 0.4.
(The numerical data of other correlation functions
are not available for $J_1/J_2 < -3$
because of slow convergence of DMRG calculation.\cite{convergence})
We clearly see in Fig. \ref{fig:correlation-Qua} that
the numerical data of $\langle s^z_l \rangle$
are fitted well by Eq.\ (\ref{eq:szl-bosegas-finite}) with $p = 4$.
This gives strong support for the presence of the quartic phase
for these parameters from the following reason.
As we emphasize below Eq.\ (\ref{2k_F}),
the period of the Friedel oscillations in
$\langle s^z_l\rangle$ is directly related to the density $\rho$
of hard-core bosons and, in particular, the number $p$ of magnons 
forming a bound state.
Indeed, if we compare Figs.\ \ref{fig:correlation-Nem}(a),
\ref{fig:correlation-Tri}(a), and \ref{fig:correlation-Qua}
for the same magnetization $M$, we find that
the number of nodes in $\langle s^z_l \rangle$
change as $\propto 1/p$ for
the nematic/SDW$_2$, triatic/SDW$_3$, and quartic phases.
Therefore, the successful fitting of $\langle s^z_l \rangle$
to Eq.\ (\ref{eq:szl-bosegas-finite}) with a certain $p$
can be considered as a strong evidence for the formation of $p$-magnon
bound states.

One may naturally expect that for the quartic phase a spin-density-wave 
(SDW$_4$) regime with dominant longitudinal spin correlation 
should also appear at low magnetic field.
Indeed, the exponent $\eta$ obtained from the fitting of 
$\langle s^z_l \rangle$ for $J_1/J_2 = -3.6$ (not shown here) 
exhibits a tendency that $\eta$ changes from $\eta < 1$ to $\eta > 1$ 
with increasing $M$.
Meanwhile, the estimates of $\eta$ have large error bars, 
which prevent us from determining accurately the crossover point between 
the quartic and SDW$_4$ regions, unfortunately.
We therefore tentatively call the region
of $\Delta S^z_{\rm tot} = 4$ 
the quartic phase, keeping it in mind that 
the low-field part of the phase 
probably includes the SDW$_4$ regime.

\section{Concluding remarks}\label{sec:conclusion}

We have determined the magnetic phase diagram of the spin-1/2 $J_1$-$J_2$
zigzag spin chain
with ferromagnetic $J_1$ and competing antiferromagnetic $J_2$ interactions
under magnetic field.
Asymptotic behaviors of correlation functions have been derived for
various phases, using bosonization approach for weak $J_1$ and
the effective theory for hard-core bosons of bound multi-magnons.
By fitting numerical data of the correlation functions 
obtained by the DMRG method to the analytic forms,
we have successfully identified the vector chiral phase,
nematic/SDW$_2$ phases, triatic/SDW$_3$ phases, and quartic phase.

At low magnetic field, we have found the vector chiral phase,
marked by long-range vector chiral order and
algebraically decaying incommensurate transverse-spin correlations.
The vector chiral state is the quantum counterpart of
the classical helical state.
In the classical $J_1$-$J_2$ model the helical state appears
as the ground state in the whole magnetization region,
whereas in the spin-1/2 model the vector chiral phase
appears only in the low-magnetization regime
(but not at $M=0$).
For larger magnetization, the chiral state is destroyed by the formation of
quantum magnon bound states, and turns into the spin density wave states.

At higher magnetic field, magnons form stable bound states,
caused by ferromagnetic attractive interactions.
The stabilization of magnon bound states is a general feature of
frustrated ferromagnets.
Spin multipolar orders induced by the bound-state formation were discovered
recently in spin-1/2 models on the square lattice,\cite{ShanonMS2006} the
triangular lattice,\cite{MomoiSS2006}
and the two-leg ladder lattice.\cite{HikiharaY2008}
In all of these models, magnon bound states become stable when the
ferromagnetic
state is destroyed by competing antiferromagnetic and/or 
ring-exchange interactions. The unique feature of
the present zigzag chain model is that the number of magnons forming
a bound state
increases consecutively with approaching the ferromagnetic phase boundary 
$J_1/J_2 = -4$. 
This unique feature might relate to the fact that the zigzag spin chain
with $J_1/J_2=-4$ has highly degenerate ground
states.\cite{HamadaKNN1988,TonegawaH1989}

We have found various phases that can be well described
by the effective theory for hard-core bose gas of bound multi-magnons.
Near the saturation field, there appear various multipolar TL liquid phases,
such as the nematic (quadrupolar), triatic (octupolar),
and quartic (hexadecapolar) phases,
which are characterized by the condensation of bound multi-magnons.
With lowering the magnetic field from the saturation field, the increase of
bound-magnon density enhances the effect of repulsion between bound magnons.
This leads to a crossover from a high-field region of the multipolar TL liquids
to a low-field region of spin density wave states, where density waves of bound
multi-magnons dominate.

Lastly we note that our phase diagram appears to be
qualitatively consistent with
the magnetic properties observed in the spin-1/2 chain cuprate LiCuVO$_4$,
whose exchange couplings are estimated as $J_1/J_2\approx-0.4$
with ferromagnetic $J_1$.\cite{Enderle2005}
Recent experiments revealed that, when a magnetic field is applied,
the low-temperature phase undergoes two successive transitions with changing
field.\cite{Banks2007,Buttgen2007}
Besides an usual spin-flop transition
at $h_{\rm c1} \approx 2.5$T, which is presumably
due to spin anisotropy effect,
another magnetic phase transition occurs at $h_{\rm c2} \approx 7.5$T.
Below $h_{\rm c2}$,
(or more precisely, in $h_{\rm c1} < h < h_{\rm c2}$),
the low-temperature phase has spiral spin structure, having
incommensurate spin order in the plane perpendicular to the applied field,
as well as ferroelectlicity.\cite{Naito2007,Yasui2007,Schrettle2007}
In the field above $h_{\rm c2}$, the system has
a modulated magnetic order parallel to the field
while the perpendicular spin components are disordered.\cite{Buttgen2007}
In comparison with the magnetic phase diagram of the $J_1$-$J_2$ model,
it is natural to identify
the magnetic transition at $h = h_{\rm c2}$
with the transition between the vector chiral phase
and the SDW$_2$ phase in our model (\ref{eq:Ham}).
This means that the experimentally observed spin modulated state
in high field can be characterized by a
density wave order of bound magnon pairs.
Furthermore, in the light of our phase diagram,
we predict that the nematic phase, which
has not yet been observed experimentally,
should appear in higher magnetic field.
Further experimental studies on high magnetization
phases would be interesting.
On the theoretical side, it is important to
include effects of interchain couplings, further interactions, 
and spin anisotropy to make more quantitative comparison
between our theoretical results and experiments.
In LiCuVO$_4$, the interchain coupling indeed induces
three-dimensional order at very low temperature $T < T_{\rm N} \approx 2.3$K.
We also note that the critical field $h_{\rm c2}$ is about 0.2
of the saturation field
$h_{\rm s} \approx 41$T and considerably larger than
the value estimated from the one-dimensional $J_1$-$J_2$ model.

{\it Note added:}
Since the submission of this paper, a preprint by
Sudan {\it et al.}\cite{SudanLL2008} has appeared, in which a phase diagram
very similar to ours is obtained independently.
They have found a direct metamagnetic transition from
the vector chiral phase 
to the ferromagnetic phase for large $|J_1|/J_2$ 
where bound states of $p \ge 5$ magnons are formed. 
In Sec.\ \ref{sec:magcurve} we discussed
magnetization curves for $J_1/J_2\ge-3.6$ where only bound magnons
of $p \le 4$ participate in the magnetization process.

\acknowledgments

It is our pleasure to acknowledge stimulating discussions with
Shunsuke Furukawa, Shigeki Onoda, Masahiro Sato, and Oleg Starykh.
This work was supported by Grants-in-Aid for Scientific Research
from the Ministry of Education,
Culture, Sports, Science and Technology (MEXT) of Japan
(Grant No.\ 16GS0219, No.\ 17071011, No.\ 18043003, and No.\ 20046016),
by the Next Generation Super Computing Project, Nanoscience Program,
MEXT, Japan, and
by Japan Society for Promotion of Science (No.\ P06902).
The numerical calculations were performed in part
by using RIKEN Super Combined Cluster (RSCC).

\appendix*

\section{Bloch theorem for spin current}
\label{appendix:bloch}

In this appendix we prove that there is no net spin current flow
in the ground state (even in the vector chiral phase).
This is a variant of Bloch's theorem that there is no net
current flow in the ground state without external 
field.\cite{Bloch,OhashiM,BrayAliN2008}

We begin with defining the spin current.
Since the Hamiltonian (\ref{eq:Ham}) conserves the $z$ component
of the total spin $\sum_l\bm{s}_l$, the $s^z$ current is a well-defined
quantity.
The equation of motion for $s_l^z$ reads
\begin{eqnarray}
\frac{\partial}{\partial t}s_l^z
\!\!&=&\!\!-i[s_l^z,\mathcal{H}]\nonumber\\
&=&\!\!
-J_1\!\left(\kappa_l^{(1)}-\kappa_{l-1}^{(1)}\right)
-J_2\!\left(\kappa_l^{(2)}-\kappa_{l-2}^{(2)}\right),
\quad
\label{ds_l^z/dt}
\end{eqnarray}
from which we deduce the $s^z$ current flowing from the site $l$
to the site $l+n$ is given by $J_l^{(n)}=J_n\kappa_l^{(n)}$ ($n=1,2$).
It also follows from Eq.~(\ref{ds_l^z/dt}) that
\begin{equation}
\frac{\partial}{\partial t}\sum_{m\le l}s_m^z
=-J_1\kappa_l^{(1)}-J_2\left(\kappa_l^{(2)}+\kappa_{l-1}^{(2)}\right),
\end{equation}
implying that the total $s^z$ current flowing through the system
is $J_\mathrm{tot}=J_1\kappa^{(1)}+2J_2\kappa^{(2)}$, where
we have suppressed the site index $l$.

To prove our statement that the net current $J_\mathrm{tot}$ always
vanishes in the ground state,
we consider the $J_1$-$J_2$ spin chain of finite length $L$ with the periodic
boundary condition
and add to its Hamiltonian $\mathcal{H}$ a weak symmetry breaking term
\begin{equation}
\mathcal{H}_y=-y\sum_l J_1\kappa_l^{(1)}
\end{equation}
with the coupling $y>0$.
This term allows us to select a unique ground
state with a broken $Z_2$ symmetry.
Let us assume that the unique ground state $|g\rangle_{L,y}$
has a nonvanishing expectation value of a linear combination
of the vector chiralities,
\begin{equation}
\left\langle J_1\kappa_l^{(1)}+2J_2\kappa_l^{(2)}\right\rangle_{L,y}
>\frac{3y|J_1|}{4},
\label{eq:assumption}
\end{equation}
where
$\langle\cdots\rangle_{L,y}$ is the average
in the ground state $|g\rangle_{L,y}$.

We now introduce the twist operator
\begin{equation}
U_\theta=\exp\left(-i\sum_{l=1}^Ll\theta s_l^z\right)
\label{eq:U_theta}
\end{equation}
with the twist angle $\theta=2\pi/L$.
We then take $U_\theta|g\rangle_{L,y}$ as a trial state, which has a
smaller net spin current than $|g\rangle_{L,y}$, and
compare its energy with the energy of the assumed ground state
$|g\rangle_{L,y}$.
We find the energy difference
\begin{eqnarray}
\Delta E\!\!&=&\!\!
\langle U_\theta^\dagger(\mathcal{H}+\mathcal{H}_y)U_\theta\rangle_{L,y}
-\langle\mathcal{H}+\mathcal{H}_y\rangle_{L,y}
\nonumber\\
&=&\!\!
J_1(\cos\theta-1)\sum_l
\langle s_l^xs_{l+1}^x+s_l^ys_{l+1}^y-y\kappa_l^{(1)}\rangle_{L,y}
\nonumber\\
&&\!\!\!{}
+J_2(\cos2\theta-1)\sum_l
\langle s_l^xs_{l+2}^x+s_l^ys_{l+2}^y \rangle_{L,y}
\nonumber\\
&&\!\!\!{}
-J_1\sin\theta\sum_l
\langle\kappa_l^{(1)}\!+y(s_l^xs_{l+1}^x+s_l^ys_{l+1}^y)\rangle_{L,y}
\nonumber\\
&&\!\!\!{}
-J_2\sin2\theta\sum_l
\langle\kappa_l^{(2)} \rangle_{L,y}.
\nonumber\\&&
\label{eq:DeltaE1}
\end{eqnarray}
Assuming the translation invariance of the ground state, 
we reduce Eq.\ (\ref{eq:DeltaE1}) to
\begin{eqnarray}
\Delta E
\!\!&=&\!\!
-2\pi
\langle J_1\kappa_l^{(1)}+2J_2\kappa_l^{(2)}\rangle_{L,y}
\nonumber\\
&&\!\!{}
-2\pi yJ_1\left\langle s_l^xs_{l+1}^x+s_l^ys_{l+1}^y\right\rangle_{L,y}
+\mathcal{O}(L^{-1})
\quad
\label{Delta E}
\end{eqnarray}
for $L\gg1$.
Using the inequality
$-J_1 \langle s_l^xs_{l+1}^x+s_l^ys_{l+1}^y \rangle_{L,y}<\frac34|J_1|$,
we conclude from Eqs.\ (\ref{eq:assumption}) and (\ref{Delta E})
that $\Delta E<0$.
This is in contradiction with the assumption of
$|g\rangle_{L,y}$ being the ground state.
This means that our assumption (\ref{eq:assumption}) is not valid,
and instead we have
\begin{equation}
\langle J_1\kappa_l^{(1)}+2J_2\kappa_l^{(2)}\rangle_{L,y}
\le\alpha y, 
\end{equation}
where $\alpha$ is a positive constant ($\alpha=3|J_1|/4$).

In the same way, starting from the assumption that the ground state
$|g\rangle_{L,y}$ has a negative net current,
$\langle J_1\kappa_l^{(1)}+2J_2\kappa_l^{(2)}\rangle_{L,y}
<-\frac{3}{4}y|J_1|$,
and using the twisted trial state $U_\theta|g\rangle_{L,y}$ with the angle
$\theta=-2\pi/L$, we can again
show that the trial state has a lower energy than the assumed ground state,
and thereby we have $\langle J_1\kappa_l^{(1)}+2J_2\kappa_l^{(2)}\rangle_{L,y}
\ge-\alpha y$. We thus obtain
\begin{equation}
|\langle J_1\kappa_l^{(1)}+2J_2\kappa_l^{(2)}\rangle_{L,y}|
\le\alpha y.
\end{equation}
We now take the limit $L\to\infty$ and then $y\to0$,
yielding
\begin{equation}
J_1\langle\kappa_l^{(1)}\rangle
+2J_2\langle\kappa_l^{(2)}\rangle=0.
\label{J_tot=0}
\end{equation}
It is straightforward to generalize this identity to 
the case when the ground state breaks translation symmetry 
as well as to other spin Hamiltonians.


\begin{thebibliography}{99}

\bibitem{Chubukov1991}
A.\ V.\ Chubukov, Phys.\ Rev.\ B \textbf{44}, 4693 (1991).

\bibitem{CabraHP2000}
D.\ C.\ Cabra, A.\ Honecker, and P.\ Pujol,
Eur.\ Phys.\ J.\ B \textbf{13}, 55 (2000).

\bibitem{HeidrichMHV2006}
F.\ Heidrich-Meisner, A.\ Honecker, and T.\ Vekua,
Phys.\ Rev.\ B \textbf{74}, 020403(R) (2006).

\bibitem{DmitrievK2006}
D.\ V.\ Dmitriev and V.\ Ya.\ Krivnov,
Phys.\ Rev.\ B \textbf{73}, 024402 (2006).

\bibitem{KuzianD2007}
R.\ O.\ Kuzian and S.-L.\ Drechsler,
Phys.\ Rev.\ B \textbf{75}, 024401 (2007).

\bibitem{KeckeMF2007}
L.\ Kecke, T.\ Momoi, and A.\ Furusaki,
Phys.\ Rev.\ B \textbf{76}, 060407(R) (2007).

\bibitem{VekuaHMH2007}
T.\ Vekua, A.\ Honecker, H.-J.\ Mikeska, and F.\ Heidrich-Meisner,
Phys.\ Rev.\ B \textbf{76}, 174420 (2007).

\bibitem{Furukawa2008}
S. Furukawa, M. Sato, Y. Saiga, and S. Onoda,
arXiv:0802.3256.

\bibitem{Katsura2008}
H. Katsura, S. Onoda, J. H. Han, and N. Nagaosa,
arXiv:0804.0669.

\bibitem{Hase2004}
M.\ Hase, H.\ Kuroe, K.\ Ozawa, O.\ Suzuki, H.\ Kitazawa, G.\ Kido,
and T.\ Sekine, Phys.\ Rev.\ B \textbf{70}, 104426 (2004).

\bibitem{Enderle2005}
M.\ Enderle, C.\ Mukherjee, B.\ F\aa k, R.\ K.\ Kremer, J.-M.\ Broto,
H.\ Rosner, S.-L.\ Drechsler, J.\ Richter, J.\ Malek, A.\ Prokofiev,
W.\ Assmus, S.\ Pujol, J.-L.\ Raggazzoni, H.\ Rakoto, M.\ Rheinst\"{a}dter,
and H.\ M.\ R\o nnow, Europhys. Lett. \textbf{70}, 237 (2005).


\bibitem{Banks2007}
M.\ G.\ Banks, F.\ Heidrich-Meisner, A.\ Honecker, H.\ Rakoto, J.-M.\ Broto,
and R.\ K.\ Kremer, J.\ Phys.: Condens. Matter \textbf{19}, 145227 (2007).

\bibitem{Buttgen2007}
N.\ B\"{u}ttgen, H.-A.\ Krug von Nidda, L.\ E.\ Svistov,
L.\ A.\ Prozorova, A.\ Prokofiev, and W.\ A\ss mus,
Phys.\ Rev.\ B \textbf{76}, 014440 (2007).

\bibitem{Naito2007}
Y.\ Naito, K.\ Sato, Y.\ Yasui, Y.\ Kobayashi, Y.\ Kobayashi,
and M.\ Sato, J.\ Phys.\ Soc.\ Jpn. \textbf{76}, 023708 (2007).

\bibitem{Yasui2007}
Y.\ Yasui, Y.\ Naito, K.\ Sato, T.\ Moyoshi, M.\ Sato, and K.\ Kakurai,
J.\ Phys.\ Soc.\ Jpn.\ \textbf{77}, 023712 (2008).

\bibitem{Schrettle2007}
F.\ Schrettle, S.\ Krohns, P.\ Lunkenheimer, J.\ Hemberger, N.\ B\"{u}ttgen,
H.-A.\ Krug von Nidda, A.\ V.\ Prokofiev, and A.\ Loidl,
Phys.\ Rev.\ B \textbf{77}, 144101 (2008).

\bibitem{MajumdarG1969A}
C.\ K.\ Majumdar and D.\ K.\ Ghosh, J.\ Math.\ Phys. \textbf{10}, 1388 (1969).

\bibitem{MajumdarG1969B}
C.\ K.\ Majumdar and D.\ K.\ Ghosh, J.\ Math.\ Phys. \textbf{10}, 1399 (1969).

\bibitem{Haldane1982}
F. D. M. Haldane, Phys.\ Rev.\ B \textbf{25}, 4925 (1982).

\bibitem{JullienH1983}
R.\ Jullien and F.\ D.\ M.\ Haldane, 
Bull.\ Am.\ Phys.\ Soc.\ \textbf{28}, 344 (1983).

\bibitem{OkamotoN1992}
K.\ Okamoto and K.\ Nomura, Phys.\ Lett.\ A \textbf{169}, 433 (1992).

\bibitem{Eggert1996}
S.\ Eggert, Phys.\ Rev.\ B \textbf{54}, R9612 (1996).

\bibitem{WhiteA1996}
S.\ R.\ White and I.\ Affleck, Phys.\ Rev.\ B \textbf{54}, 9862 (1996).

\bibitem{OkunishiHA1999}
K.\ Okunishi, Y.\ Hieida, and Y.\ Akutsu, 
Phys.\ Rev.\ B \textbf{60}, R6953 (1999).

\bibitem{OkunishiT2003}
K.\ Okunishi and T.\ Tonegawa, J.\ Phys.\ Soc.\ Jpn. \textbf{72}, 479 (2003).

\bibitem{NersesyanGE1998}
A.\ A.\ Nersesyan, A.\ O.\ Gogolin, and F.\ H.\ L.\ E\ss ler,
Phys.\ Rev.\ Lett.\ \textbf{81}, 910 (1998).

\bibitem{KaburagiKH1999}
M.\ Kaburagi, H.\ Kawamura, and T.\ Hikihara,
J.\ Phys.\ Soc.\ Jpn. \textbf{68}, 3185 (1999).

\bibitem{HikiharaKK2001}
T.\ Hikihara, M.\ Kaburagi, and H.\ Kawamura,
Phys.\ Rev.\ B \textbf{63}, 174430 (2001).

\bibitem{KolezhukV2005}
A.\ Kolezhuk and T.\ Vekua, Phys.\ Rev.\ B \textbf{72}, 094424 (2005).

\bibitem{McCulloch2007}
I.\ P.\ McCulloch, R.\ Kube, M.\ Kurz, A.\ Kleine, U.\ Schollw\"{o}ck,
and A.\ K.\ Kolezhuk, Phys.\ Rev.\ B \textbf{77}, 094404 (2008).

\bibitem{Okunishi2008}
K. Okunishi, arXiv:0805.3872.

\bibitem{HikiharaMFK}
T. Hikihara, T. Momoi, A. Furusaki, and H. Kawamura,
unpublished.

\bibitem{White1992}
S.\ R.\ White, Phys.\ Rev.\ Lett.\ \textbf{69}, 2863 (1992).

\bibitem{White1993}
S.\ R.\ White, Phys.\ Rev.\ B \textbf{48}, 10345 (1993).

\bibitem{Schollwock2005}
U.\ Schollw\"{o}ck, Rev.\ Mod.\ Phys.\ \textbf{77}, 259 (2005).

\bibitem{Hallberg2006}
K.\ A.\ Hallberg, Adv.\ Phys.\ \textbf{55}, 477 (2006).

\bibitem{HamadaKNN1988}
T.\ Hamada, J.\ Kane, S.\ Nakagawa, and Y.\ Natsume,
J.\ Phys.\ Soc.\ Jpn. \textbf{57}, 1891 (1988).

\bibitem{TonegawaH1989}
T.\ Tonegawa and I.\ Harada, J.\ Phys.\ Soc.\ Jpn. \textbf{58}, 2902 (1989).

\bibitem{AFcase}
The existence of a vector chiral phase was also established by
recent numerical studies for the
antiferromagnetic $J_1$-$J_2$
spin chain.\cite{McCulloch2007,Okunishi2008,HikiharaMFK}

\bibitem{Andreev1984} A. F. Andreev and I. A. Grishchuk,
Sov.\ Phys.\ JETP \textbf{60}, 267 (1984).

\bibitem{MomoiSS2006}
T.\ Momoi, P.\ Sindzingre, and N.\ Shannon,
Phys.\ Rev.\ Lett.\ \textbf{97}, 257204 (2006).

\bibitem{RappZHH2007}
\'{A}.\ Rapp, G.\ Zar\'{a}nd, C.\ Honerkamp, and W.\ Hofstetter, 
Phys.\ Rev.\ Lett.\ \textbf{98}, 160405 (2007).

\bibitem{CapponiRLABW2008}
S.\ Capponi, G.\ Roux, P.\ Lecheminant, P.\ Azaria,
E.\ Boulat, and S.\ R.\ White,
Phys.\ Rev.\ A \textbf{77}, 013624 (2008).

\bibitem{RouxCLA2008}
G.\ Roux, S.\ Capponi, P.\ Lecheminant, and P.\ Azaria,
arXiv:0807.0412.

\bibitem{HikiharaF2001}
T.\ Hikihara and A.\ Furusaki, Phys.\ Rev.\ B \textbf{63}, 134438 (2001).

\bibitem{HikiharaF2004}
T.\ Hikihara and A.\ Furusaki, Phys.\ Rev.\ B \textbf{69}, 064427 (2004).

\bibitem{Haldane1980}
F. D. M. Haldane, Phys.\ Rev.\ Lett.\ \textbf{45}, 1358 (1980).

\bibitem{BogoliubovIK1986}
N.\ M.\ Bogoliubov, A.\ G.\ Izergin, and V.\ E.\ Korepin,
Nucl.\ Phys.\ B {\bf 275}, 687 (1986).

\bibitem{CabraHP1998} 
D.\ C.\ Cabra, A.\ Honecker, and P.\ Pujol, 
Phys.\ Rev.\ B {\bf 58}, 6241 (1998). 

\bibitem{boundary-VC}
The rapid decrease of the vector chiral correlation functions
at $r \gtrsim 80$ is an open boundary effect.
The bosonic field $\phi_i(x)$ obeys the Dirichlet boundary condition
($\phi=\mathrm{const}$) at the boundaries.
This is not compatible with the vector chiral order which
requires the dual bosonic field $\theta_-(x)$ to be fixed.

\bibitem{Igarashi}
In the Ising limit a similar hard-core boson theory was used
to describe the intermediate phase between the fully polarized
state and (2,2) antiphase state by
J. Igarashi, J. Phys.\ Soc.\ Jpn.\ \textbf{58}, 4600, (1989);
T. Tonegawa, I. Harada, and J. Igarashi,
Prog.\ Theor.\ Phys.\ Suppl.\ \textbf{101}, 513 (1990).

\bibitem{giamarchi_book}
T. Giamarchi, \textit{Quantum Physics in One Dimension,}
  (Clarendon Press, Oxford, 2004).

\bibitem{convergence}
The slow convergence of the DMRG calculation should be ascribed 
to the presence of many low-lying states
in small energy scale due to the strong frustration at $J_1/J_2 \to -4$.

\bibitem{ShanonMS2006}
N.\ Shannon, T.\ Momoi, and P.\ Sindzingre,
Phys.\ Rev.\ Lett.\ \textbf{96}, 027213 (2006).

\bibitem{HikiharaY2008}
T.\ Hikihara and S.\ Yamamoto,
J.\ Phys.\ Soc.\ Jpn. \textbf{77}, 014709 (2008).

\bibitem{SudanLL2008}
J.\ Sudan, A.\ L\"{u}scher, and A.\ M.\ L\"{a}uchli, 
arXiv:0807.1923v1.

\bibitem{Bloch}
D. Bohm, Phys.\ Rev.\ \textbf{75}, 502 (1949).

\bibitem{OhashiM} Y.\ Ohashi and T.\ Momoi,
J.\ Phys.\ Soc.\ Jpn.\ \textbf{65}, 3254 (1996).

\bibitem{BrayAliN2008}
N.\ Bray-Ali and Z.\ Nussinov, arXiv:0803.0984.

\end{thebibliography}
\end{document}